\documentclass[useAMS,usenatbib]{mn2e}
\usepackage{amsmath,astrojournals,fleqn,epsfig,txfonts,rotating}
\usepackage[usenames,dvips]{color} 

{\newif\ifnotend
\notendtrue
\def\veclist{ABCDEFGHIJKLMNOPQRSTUVWXYZabcdefghijklmnopqrstuvwxyz.}
\def\top#1#2.{#1}
\def\tail#1#2.{#2.}
\loop\expandafter\xdef\csname v\expandafter\top\veclist\endcsname%
{{\noexpand\bf\expandafter\top\veclist}}
\edef\veclist{\expandafter\tail\veclist}
\if\veclist.\notendfalse\fi\ifnotend\repeat}

\def\kms{\,{\rm km}\,{\rm s}^{-1}}
\def\mag{\,{\rm mag}}

\def\Gyr{\,{\rm Gyr}}
\def\K{\,{\rm K}} 
\def\pc{\,{\rm pc}}
\def\kpc{\,{\rm kpc}}

\def\Msun{\,{\rm M}_\odot}

\def\feh{\hbox{[Fe/H]}}
\def\meh{\hbox{[Me/H]}}
\def\feh{\hbox{[Fe/H]}}
\def\llg{\log({\rm g})}
\def\logg{\log({\rm g})}

\def\afe{[\alpha/\hbox{Fe}]}
\def\dex{\,{\rm dex}}
\def\figref#1{Fig.~\ref{#1}}

\def\ovcv0{\overline{V_0}}

\def\Teff{T_{\rm eff}}
\def\Mi{M_{\rm init}}
\def\vX{\mathbf{X}}
\def\vO{\mathbf{O}}

\newcommand {\Rsun}{{R_{0}}}

\title[Bayesian spectroscopy]{Fundamental stellar parameters and metallicities from Bayesian spectroscopy: application to low- and high-resolution spectra}
\author[R. Sch\"onrich \& M. Bergemann]
       {Ralph Sch\"onrich$^{1,2}$\thanks{E-mail: ralph.schoenrich@physics.ox.ac.uk},
        and Maria Bergemann$^{3,4}$\\
        $^{1}$ Hubble Fellow, Department of Astronomy, The Ohio State
University, 140 West 8th Avenue, 43210 Columbus, Ohio, USA \\
	$^{2}$ Rudolf-Peierls Centre for Theoretical Physics, University of Oxford, 1 Keble Road, OX1 3NP, Oxford, United Kingdom \\
        $^{3}$ Max-Planck-Institut f\"ur Astrophysik, Karl-Schwarzschild-Str.~1,
        85741 Garching, Germany \\
        $^{4}$ Institute of Astronomy, University of Cambridge, Madingley Road,
        CB3 0HA, Cambridge, United Kingdom}
\date{Draft, \today}
\pagerange{\pageref{firstpage}--\pageref{lastpage}}
\pubyear{2013}

\begin{document}
\maketitle
\label{firstpage}

\begin{abstract}
We present a unified framework to derive fundamental stellar parameters by combining all available observational and theoretical information for a star. The algorithm relies on the method of Bayesian inference, which for the first time directly integrates the spectroscopic analysis pipeline based on the global spectrum synthesis and allows for comprehensive and objective error calculations given the priors.  Arbitrary input datasets can be included into our analysis and other stellar quantities, in addition to stellar age, effective temperature, surface gravity, and metallicity, can be computed on demand.
We lay out the mathematical framework of the method and apply it to several observational datasets, including high- and low-resolution spectra (UVES, NARVAL, HARPS, SDSS/SEGUE). We find that simpler approximations for the spectroscopic PDF, which are inherent to past Bayesian approaches, lead to deviations of several standard deviations and unreliable errors on the same data. 
By its flexibility and the simultaneous analysis of multiple independent measurements for a star, it will be ideal to analyse and cross-calibrate the large ongoing and forthcoming surveys, like Gaia-ESO, SDSS, Gaia and LSST.
\end{abstract}


\section{Introduction}
Observations are a central source of knowledge on almost any entity in astrophysics. Over several centuries of intense research, several principal observational techniques have been developed that are now routinely used to study stars and stellar populations in the Milky Way and other galaxies.
We have information from astrometry, photometry, spectroscopy, but also interferometry, and astroseismology, that give complementary information on the physical parameters of stars (detailed chemical composition, gravities, temperatures, masses and ages) and their kinematics (radial velocities, distances, and orbital characteristics). However, in contrast to e.g. cosmology, where sophisticated Bayesian schemes are well--established \citep[e.g.][]{Drell00, Kitaura08}, stellar parameter determinations are still widely based on best-fit estimates and simple averages between different methods.

The advent of large stellar spectroscopic and photometric surveys like SEGUE/SDSS \citep[][]{Yanny09}, RAVE \citep[][]{Steinmetz06}, APOGEE \citep[][]{Majewski07}, GCS \citep[][]{Nordstroem04}, and the Gaia-ESO survey \citep[][]{Gilmore12}, as well as astroseismic surveys like Kepler \citep[][]{Chaplin11}, makes it necessary to develop fully automated methods for data analysis and determination of stellar parameters. Standard spectroscopic inversion methods are commonly assumed to be accurate, however, they usually involve subjective and hardly reproducible elements, like line fitting and normalisation, or decisions on spectral diagnostic features. Manual analysis of stars is limited to sample sizes of $\sim 1000$ stars, unsuitable for large surveys. Existing automated methods usually suffer from weakly constrained systematics as well as idealised error estimates. So far, most attempts to overcome these problems have concentrated on simple weighted averaging between different methods \citep[e.g.][]{Lee08a, Lee08b}.

The large stellar surveys change stellar astronomy into a precision science, where we cannot limit ourselves to pointing out structures in diagrams, but where knowledge of the error distributions is key to make meaningful model comparisons, e.g. of Galactic evolution and stellar structure. The approach we need must be flexible, objective, applicable to very large datasets, and provide an optimal combination of the different bodies of observational data. The only mathematical apparatus known to permit a systematic combination of different quantities are Bayesian frameworks. The first steps in this direction were made by \cite{Pont04}, \cite{Jorgensen05}, \cite{Shkedy07}, \cite{Bailer10}, \cite{Burnett10}, \cite{Casagrande11}, \cite{Liu12}, \cite{B13}, and \cite{Serenelli13}. The scope and applicability of these studies is limited: they either addressed the problem of fitting a spectrum only \citep[][]{Shkedy07}, partly focussed on the problem of finding the maximum likelihood solution, or rely on simplifications of the observational likelihoods \citep[e.g.][]{Burnett10, Casagrande11}. In part, this problem appears rooted in the introduction of "observables", like effective temperature, \citep[see e.g. Fig. 4 in][]{Rix13}, which have no well-defined place in a Bayesian approach and which are in fact just parameters constrained by another observation.
In contrast, a Bayesian scheme can only fulfil its claim of unbiased information, if a fair account of the observations is given and the full dimensionality of the constraints in parameter space is preserved. 

In the following we will present a new method for the determination of stellar parameters that provides an optimal exploitation of different observational information. The method offers a homogeneous full-scale quantitative recovery of the full probability distributions in parameter space, which are given by the available observations, i.e. photometry, astrometry, spectroscopy, and well-established knowledge from stellar evolution theory and Galaxy structure. The method is objective, computationally efficient, can be readily applied to data from all existing surveys and is robust to missing bits of data, e.g. damaged pixels in a spectrum or low-quality photometry. By embedding spectroscopic analysis directly in scheme, the Bayesian method allows for consideration of all pieces of relevant information at once, thus avoiding unnecessary information loss.

In this first paper of the series, our main goal is to to determine effective temperature, surface gravity, metallicity, mass, age and distances of individual stars. Thus, we limit the input data to spectroscopy, photometry, stellar evolution models and facultative parallax measurements. However, the method can be readily generalised to any number of parameters, such as kinematics or stellar rotation, and include other input information, e.g. astero-seismology and interferometric angular diameters. Furthermore, it is straightforward to analyse star formation history of a whole stellar population, e.g. a young cluster or an old galaxy, using its integrated colours and spectra. Thus the Bayesian method has a very broad scope to applications both in the context of Galactic and extra-galactic research.

The paper is structured as follows. In Sections 2 and 3, we present the details of the algorithm and its implementation, illustrated on two examples. In Section 4 we apply the method to a sample of stars with very high-resolution observations and for a sub-sample of calibration stars from the SDSS/SEGUE catalogue. Section 5 compares to the use of a simplified spectroscopic PDF. Discussion of the algorithm and results and Conclusions are found in the last two Sections.

\section{Method outline}
\subsection{Bayesian scheme}
So far, the majority of observational studies of stars, be it photometric or spectroscopic, have focussed on providing best-fit estimates of stellar parameters. However, accurate comparisons to theoretical models of e.g, galaxy evolution, require the full probability distribution of the derived parameters given the available observations.

This demands a Bayesian formalism. In this context we need to express the probability of a set of parameters $\vX=X_1,\ldots,X_n$ given a set of observations $\vO=O_1,\ldots,O_m$  by the probability that this observation could take place given the set of parameters. By definition the conditional probability $P(\vX | \vO )$, that $\vX$ given $\vO$, derives from the combined probability $P(\vX,\vO )$ as: $P(\vX, \vO) = P(\vX|\vO )P(\vO)$. We can hence write down:
\begin{equation}
P(\vX|\vO) = \frac{P(\vX)}{P(\vO)}P(\vO|\vX) $,$
\end{equation} 
where the {\it posterior} probability $P(\vX|\vO)$ is the conditional probability of the parameter set $\vX$ given $\vO$. $P(\vO|\vX)$, which we call {\it observed likelihood}, is the probability of making the set of observations $\vO$ given the set of parameters $\vX$ and $P(\vX)$ is the prior probability we ascribe to that set of parameters. $P(\vO)$ is the probability that the set of observations was made, which we set to $1$ \citep[][]{Pont04}. This simplifies our problem to
\begin{equation}
P(\vX|\vO) = P(\vX) P(O_1,\ldots,O_m|\vX) $,$
\end{equation} 
where $P(\vX|\vO)$ is the posterior probability distribution function (PDF) on the chosen parameter space.
In our work, observations are conditionally independent given the parameters, i.e. if all parameters are perfectly known, the observations do not provide additional information about each other. Hence we can disentangle the observations by:
\begin{equation}\label{eq:condimp}
P'(\vX) = P(O_1,\ldots,O_m|\vX) = \prod_{(j=1)}^m P(O_j|\vX) $.$
\end{equation}
\subsection{Parameter space}
The parameter set $\vX$ contains all parameters relevant to the problem under investigation and important to the description of a star. This may include surface and interior structure parameters (effective temperature, surface gravity, mean density, etc) as well as any other pieces of information like chemical composition, age, distance, position in the sky, etc. Since we are dealing with a single object, all these parameters are related in some way. However, we can break their dependencies into main groups, using the fact that each type of observations constraints only a sub-set of these parameters, whereas it bears no information on others.

In this work, we define the 'core' parameter space $R_c \equiv (\feh, \Teff, \llg)$ of metallicity (expressed by iron abundance), effective temperature and surface gravity. The parameters in $R_c$ impact all our observations and models. 

Other parameters are constrained by only a subset of observations: e.g. detailed abundances are of importance for spectroscopic observations, while stellar magnitudes in different colour bands $C$ span the space of the photometric parameters. Age $\tau$, initial mass $\Mi$ and present mass $M$ fall into the domain of stellar models. Distance $s$ and parallax $p$ are determined either from direct astrometric observations or via the distance modulus when comparing stellar models with photometry. 

Thus the full parameter space can be disentangled into individual contributions: 
\begin{equation}
R \equiv R_c + R_{sp} + R_{ph} + R_{mod} + R_{others} $,$
\end{equation}
where $R_c$ is the core parameter space the other R$_j$ are the parameters of importance to different types of observations or prior expectations (see Sec. 3.2 to 3.6).
\subsection{Observations} 
In contrast to parameters, which span the n-dimensional space of the posterior probability distribution, the nature of observations is irrelevant. Observations can be anything, from the numbers of electrons on a CCD to a needle on a scale. Each observation puts a constraint on our parameter space, which is its corresponding observed likelihood $P(O|X)$ as a function on parameter space.

Instead of just writing down an observational likelihood, the common approach in astronomy is to "simplify" this by the introduction of "observables". While this term is not well-defined in a Bayesian context, "observables" commonly denote best-fit values for some parameters (like $\Teff$), which appear to be relatively well-constrained by (single) observations. Some studies, like \citep[e.g.][]{Burnett10}, go even further to introduce the errors on those "observables" as further variables in their formalism. From an aesthetic point of view, this results in a rather clumsy and complicated bulk of variables to achieve a simple goal: describing the real observed likelihood. It has two practical consequences. First, "observables" lead to an oversimplification of the observational likelihood, usually with the unjustified (and damaging, see Section~\ref{sec:Gaussian}) approximation as a product of separate Gaussians in each parameter termed "observable". Second, their introduction artificially introduces a "better" class of parameters, raising the wrong suggestion that their values are fixed. This is not true. For any parameter the Bayesian formalism will in general give an estimate different from the best-fit value. 

While selection functions are in most cases essential for understanding observations with theoretical models, this does not apply to the discussed Bayesian schemes. Yet, some studies introduce a selection function in their equations (see a longer discussion in Sec.~\ref{sec:SF} of the Appendix). We refrain from using such a selection function, because only selection criteria based on the parameter space would affect the Bayesian scheme, while a survey selection must be based on random choice or previous observations.

\subsection{Summary of notation}

To facilitate reading the equations we quickly summarize the main notations: We denote the set of observations by $\vO$, the set of parameters by $\vX$, the parameter space by $R$ and all probability distributions by $P$. To cope with the different sources of information, we introduce indices: "ph" for photometry, "sp" for spectroscopy, "astr" for astrometry (parallaxes), in addition we use "mod" for knowledge from stellar models and "pr" for priors. Hence the observational likelihood in full parameter space from spectroscopic observations reads $P(O_{\rm sp} | \vX)$. To facilitate the reading we contract the notation for the conditional probabilities by decorating $P$ with a prime: e.g. $P'_{\rm sp} \equiv P'_{\rm sp}(\vX) \equiv P(O_{\rm sp} | \vX)$. 

Commonly used variables are age $\tau$, stellar mass $M$, solar mass $\Msun$, initial mass $M_{\rm init}$, logarithmic iron abundance $\feh$, general metallicity $\meh$, parallax $p$, distance $s$, and distance modulus $\mu$.

\section{Detailed algorithm}

In this pilot study, we restrict ourselves to the most important basic case: the calculation of stellar parameters, when we have spectroscopic, and/or photometric observations. We will show how to expand this to include parallax measurements. After validating the method on the high-resolution spectroscopic data of nearby stars, we apply it to a sample of low-resolution spectra from SEGUE/SDSS \citep[][]{Yanny09, Allende08} calibration sample.

\subsection{Contributors to the posterior PDF}
With the conditional independence (equation~\ref{eq:condimp}) we simplify the calculation of our posterior:

\begin{eqnarray}{\label{eq:ar}}
P (\vX | O_{\rm sp}, O_{\rm ph}) &\sim& P (O_{\rm sp}| \vX) \cdot P (O_{\rm ph}| \vX) P(O_{\rm astr} | \vX) \cdot P_{\rm mod}(\vX) \cdot P_{\rm pr} (\vX)\notag \\ 
&\sim& P'_{\rm sp} \cdot P'_{\rm ph} \cdot P'_{\rm astr} \cdot P_{\rm mod} \cdot P_{\rm pr} {,} 
\end{eqnarray}
where $O_{\rm sp}, O_{\rm ph}, O_{\rm astr}$ denote the photometric and spectroscopic and astrometric observations, $P_{\rm mod}$ the probability derived from stellar models and $P_{pr}$ the prior probability distribution function. In the second line we abbreviate our observational likelihoods, representing their conditional nature based on the observations in short by a prime. Note that any combination of observational constraints can be dropped from these equations, as well as new observations (e.g. interferometry) can be added by multiplication.

The PDFs from Equ. \ref{eq:ar} have two interesting qualities:
\begin{itemize}
\item some PDFs describe sharp structures in the n-dimensional parameter space, thus lowering the dimensionality of the probability distribution and reducing computational costs (by the multiplication, the combined PDF cannot have higher dimensionality than its components). In other words, the space volume where their PDF is non-negligible has a lower dimensionality than the overall space. For example, stellar models together with model atmospheres map directly from the fundamental stellar parameters ($\Mi, \tau, {\rm \meh})$ to their observed space ($\Teff, \llg, C$).
\item some PDFs constrain only a subset of $j < n$ parameters, i.e. they are flat the other dimensions. Though they can be generalised to the n-dimensional space of the aggregate PDF, most of these dimensions will be redundant, i.e. we have $P(X_1,\ldots, X_n|\vO) \sim P(X_1, \ldots X_j|\vO)$. It can be efficient to merge them in an early step with another PDF that carries more dimensions. An especially valuable case are parameters that are nearly conditionally independent from the other parameters. E.g. the detailed abundances for the majority of chemical elements hardly affect temperature and gravity estimates.
\end{itemize}

The meaning and structure of the single contributors to the posterior $P$ will be examined below.

\subsection{Priors $P_{pr}$}\label{sec:priors}
The priors encode our previous knowledge on the distribution of the examined stellar population in parameter space. The model knowledge $P_{mod}$ will be treated here as a separate prior, though it could be in fact understood as an observation. Appropriate priors are essential to avoid biases in weakly constrained data (see \figref{fig:photmets} for an example). Further, to have set "no prior" means to have adopted a flat prior, which is not fixed under parameter transformations. However, priors must be handled with great caution to obey Cromwell's rule \citep[avoid excluding any outcomes a priori][]{Lindley82} and to avoid overconfidence biases and reproducibility problems. 

How can one cope with uncertainties in a priori parameter distributions? Making a prior just "shallower" or adopting a completely flat prior is not useful, since it actually adopts a worse probability distribution claiming the same certainty on this information as the original prior. However, one typically has estimates on how uncertain the distribution in a set of parameters is. If, within those uncertainties, the precise shape of a prior has important impacts on results, there are two strategies: The uncertainty of a prior can be constrained by demanding consistency with calibration data (data of higher quality, where the prior has less impact) and by imposing internal consistency between different subsamples (e.g. red versus blue stars). The remaining uncertainty in a prior can be covered by hyper-parameters: The shape of the prior can be parametrised and a probability distribution function estimated on these parameters.

%%Our philosophy is to make use of appropriate priors, but to separately flatten priors on each quantity shown in our performance tests\footnote{e.g. when showing the age-metallicity relation, we use a flat prior both on $\meh$ and on $\tau$.}.

There are strong dependencies between most parameters. As the posterior PDF has lower effective dimensionality than the parameter space $R$, we can not set constraints on every single dimension in $R$ without risk of over-constraining the priors. We circumvent this problem by limiting our effective priors to age $\tau$, initial mass $\Mi$, metallicity $\meh$, and distance $s$; all other dimensions are indirectly constrained by these priors and we adopt no additional constraints on them.

Throughout this work we will use the following priors:
\begin{eqnarray}
P_{pr}(\vX) &:=&  p(\tau,M,{\rm  [Fe/H]},s)  \sim \\
&\sim& P(\tau | {\rm  [Fe/H]})  \cdot  P(\Mi) \cdot P(s, l, b) \cdot P(\meh , \feh ) . \notag 
\end{eqnarray}

$P(\meh, \feh)$ is a fixed relation between metallicity (required for the isochrones) and the iron abundance, which we have to introduce, since we do not measure detailed abundances in this paper. The adopted relation is given in Sec. \ref{sec:amprior} of the Appendix.
For $P(M_i)$ we employ a Salpeter IMF \citep[][]{Salpeter55}, with exponent $-2.35$ and independent from metallicity and age. We account for the metallicity-dependent age distribution by adopting a shorter timescale in the star formation history of metal-poor stars. Details are given in the Appendix.

Due to its importance we need to discuss the spatial prior $P(s,l,b)$.\footnote{For simplicity, and to avoid recovering the spatial dependencies our prior invokes, we neglect metallicity dependent structure and separate the spatial prior from the age-metallicity terms.} In general, every sample will cover some fixed angle on the sky (be it so-called pencil beams like in SEGUE or a complete sky coverage), so the actual volume is a cone that covers an effective area $A(s) = k\cdot s^2$, where $s$ is the distance. The constant $k$, given by the sky coverage and selection probabilities, is irrelevant in our context. However, the likelihood to end up in the sample is also proportional to the density of the population in the observed region, so that we obtain:
\begin{equation}
P_s(s) = k s^2 \int\int \rho(s, \omega) d^2\omega
\end{equation}
where we integrate over the sky position $\omega$, and $\rho$ is the spatial density of stars at position $\omega$ and distance $s$. In our formulation, this must be multiplied by another factor of distance $s$, since a fixed magnitude range samples a spatial depth proportional to $s$ (formally this derives from a parameter transformation of the probability density from magnitudes to distance).
Neglect of this prior is not equivalent to a flat distance prior, but one that drops steeply with the third power of the distance, and hence in the absence of parallax measurements would give a strong (and uncertainty dependent) bias towards lower surface gravities.
\subsection{Stellar models $P_{\rm mod}$}\label{sec:mod}

Stellar models describe an effectively three-dimensional constraint in the full parameter space. The corresponding PDF can be represented in the core parameter space $R_c = (\Teff, \logg, \feh)$. In this work, we neglect other dimensions, adopting a simple relationship for alpha enhancement and neglecting initial/present stellar rotation.

The calculation of the PDF is performed by summing the weights of available stellar model points falling into the cells of a dense grid in the target space $R_c$. Throughout this work we fold the models with a Gaussian kernel with widths ${\bf \sigma_c} = (30\K, 0.04 \dex, 0.02 \dex)$ in $(\Teff, \logg, \feh)$.
This error represents the internal uncertainty of the stellar models in parameter space and fills gaps caused by the discrete data representation. In addition, the effective width is augmented by the grid spacing on which the PDF is calculated. 

We use a dense grid of stellar isochrones from the BASTI database
\citep{Pietrinferni04, Pietrinferni06, Pietrinferni09}, kindly provided to us by
S. Cassisi for the stellar parameter determinations in \cite{Casagrande11}. We interpolate the models in the initial mass $\Mi$ to ensure a narrow mass spacing, but do not attempt an interpolation in age $\tau$ or metallicity $\meh$. When summing over the isochrones, we assign to each point $i$ a weight $W_i$ proportional to the parameter space volume it represents:
\begin{equation}
W_i = N_W \Delta \meh \cdot \Delta \tau \cdot \Delta \Mi $,$
\end{equation}
where $N_W$ is the normalisation, $\Delta \Mi = 0.5 (M_{init, i+1} - M_{init, i-1})$ is the average distance to its neighbours in initial mass, $\Delta \meh$ is the average distance in metallicity between the isochrone and its nearest neighbours, and $\Delta \tau$ is the average distance in age $\tau$. On the boundary of the grid we take the distance to the neighbouring point. Note that the approach is identical for stellar tracks instead of isochrones. 
The model probability at each point $\vX$ in our parameter space can then be represented as a weighted sum over all relevant models points $i$:
\begin{equation}
P_{mod}(\vX) = \sum_i W_i \, g((\vX - \vX_i), {\bf \sigma}) $,$
\end{equation}
where $\vX_i$ is the vector in parameter space given by the model grid. Here we represent the uncertainty of the models by an n-dimensional Gaussian $g$ with a dispersion vector $\sigma$. Specifically, we assume $\sigma_c$ on our core parameter space as above, and no additional uncertainties in the other dimensions.

\begin{figure*}
\begin{center}
\epsfig{file=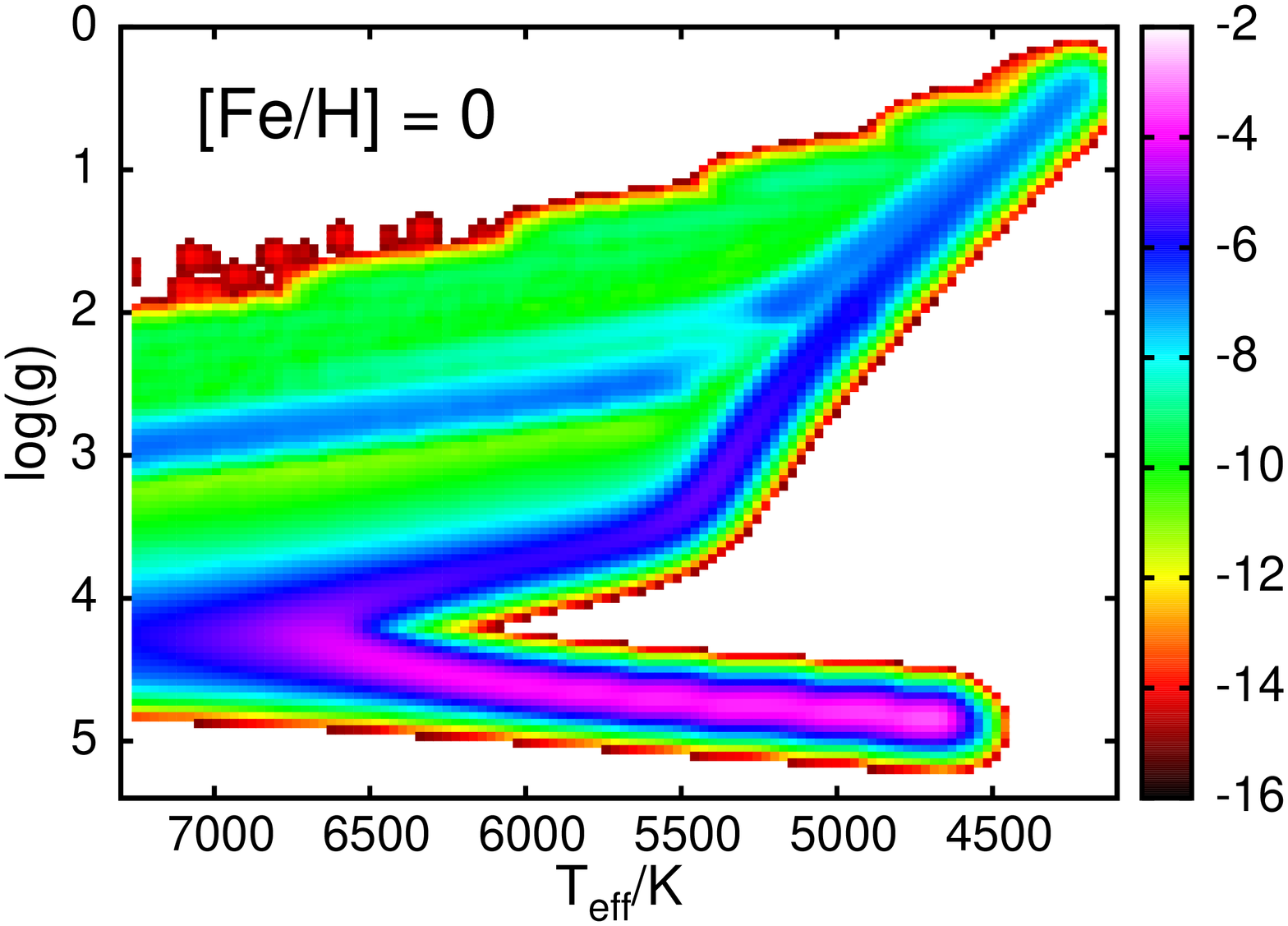,angle=-90,width=0.33\hsize}
\epsfig{file=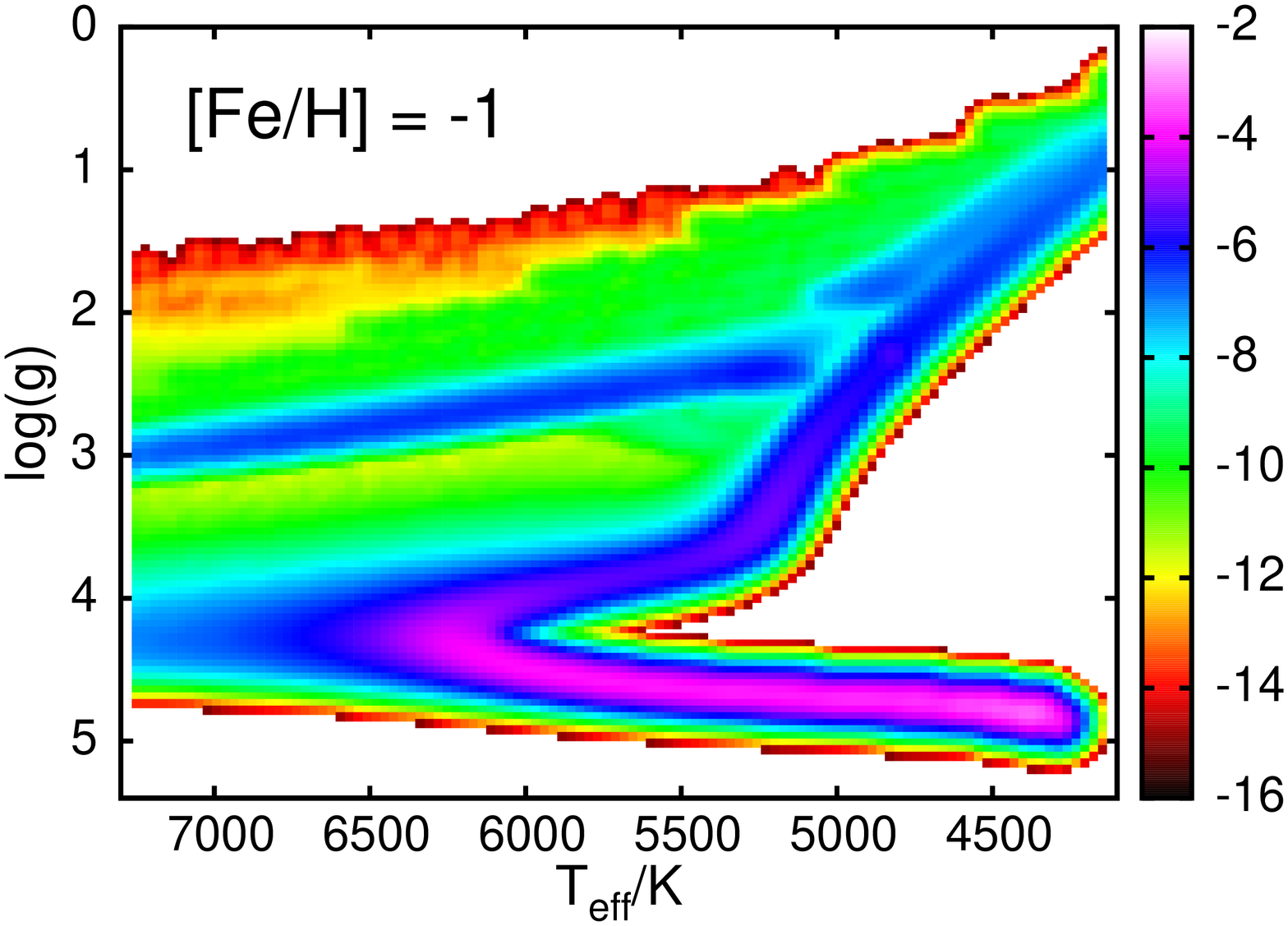,angle=-90,width=0.33\hsize}
\epsfig{file=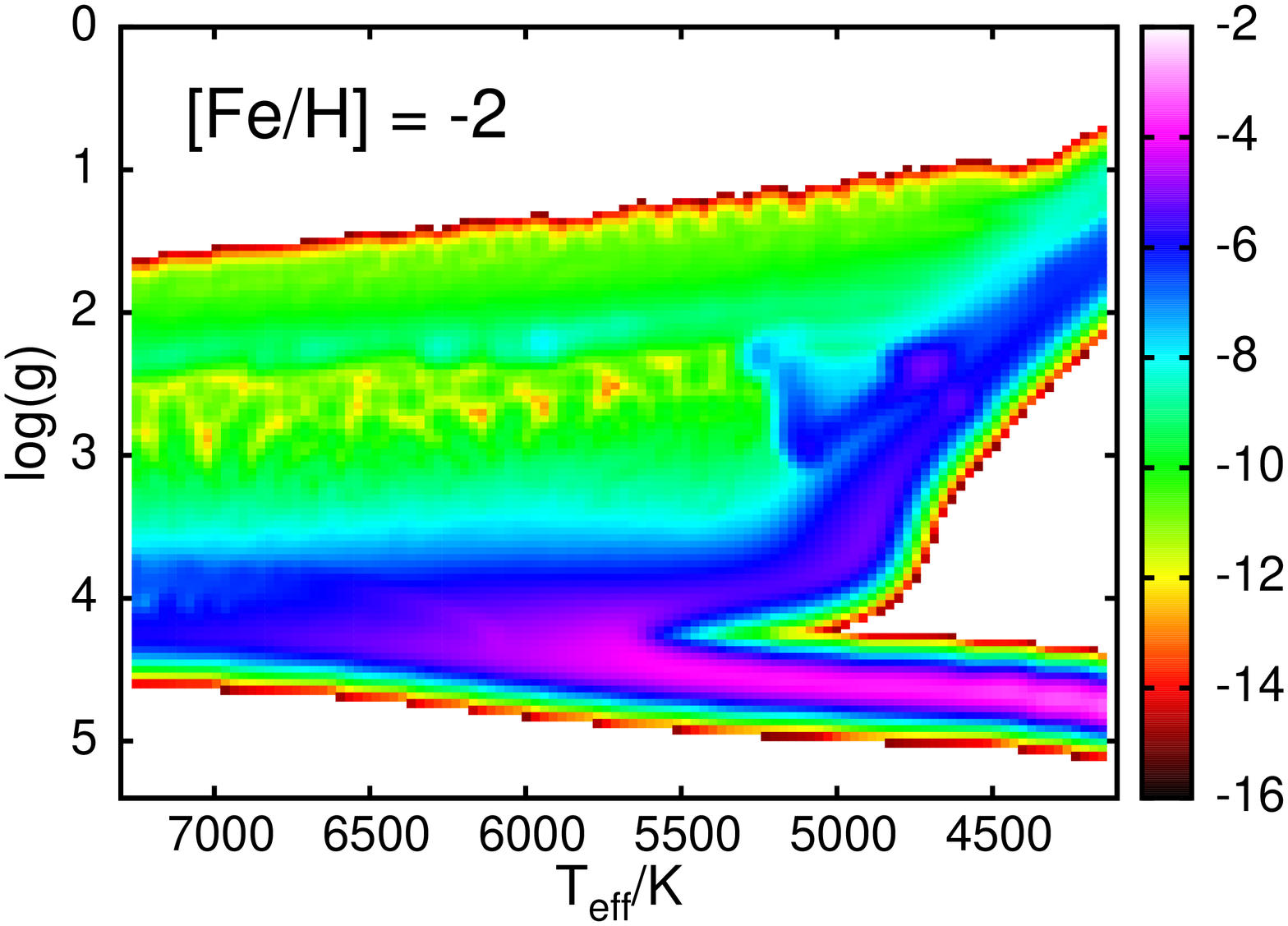,angle=-90,width=0.33\hsize}
\end{center}
\caption{A priori densities from stellar models ($P_{model}$) at metallicities $\feh = -2, -1, 0$ from left to right and accounting for a Salpeter IMF prior, and the metallicity-dependent age prior.}\label{fig:isochvsmet}
\end{figure*}

\begin{figure}
\begin{center}
\epsfig{file=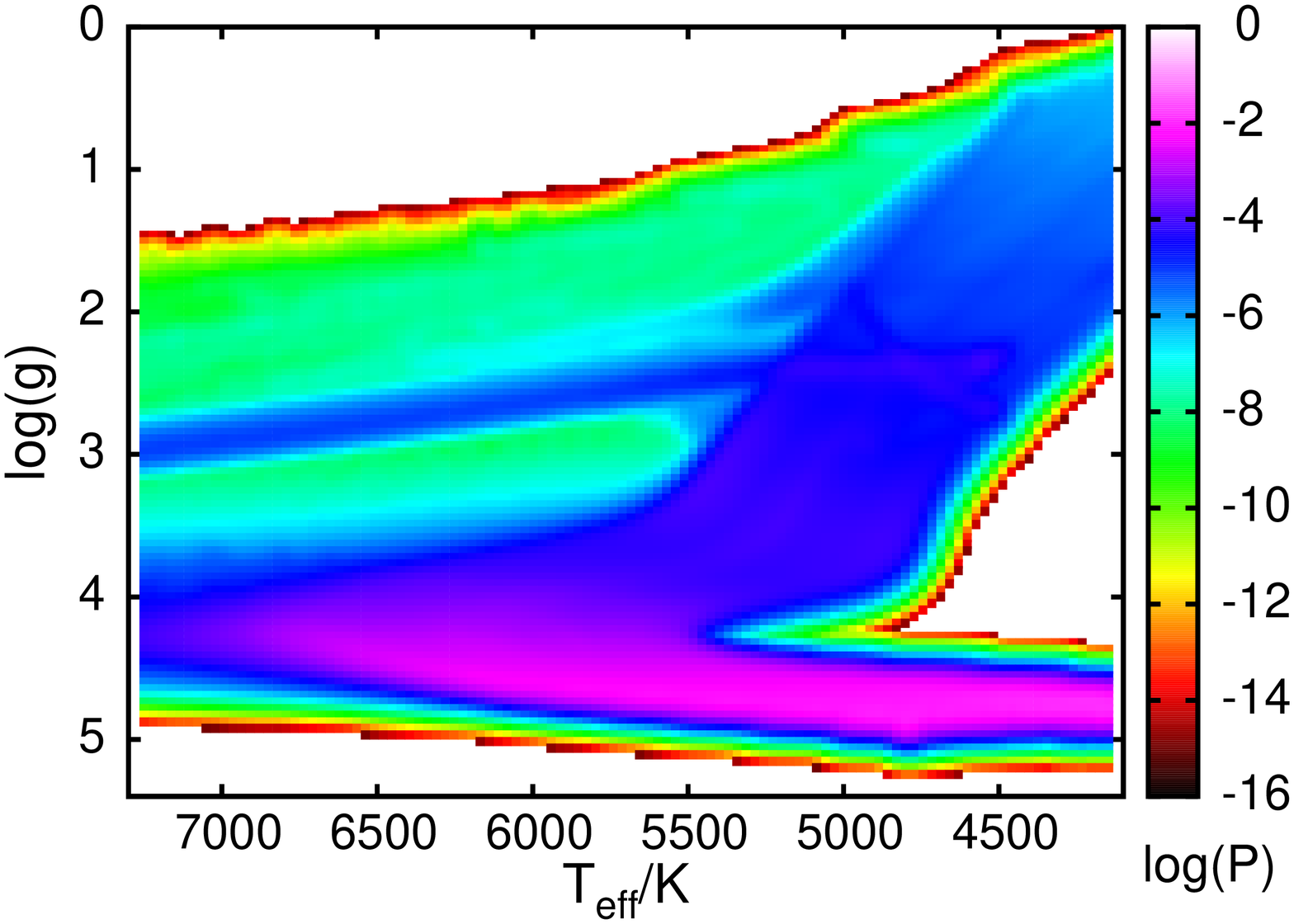,angle=-90,width=\hsize}
\epsfig{file=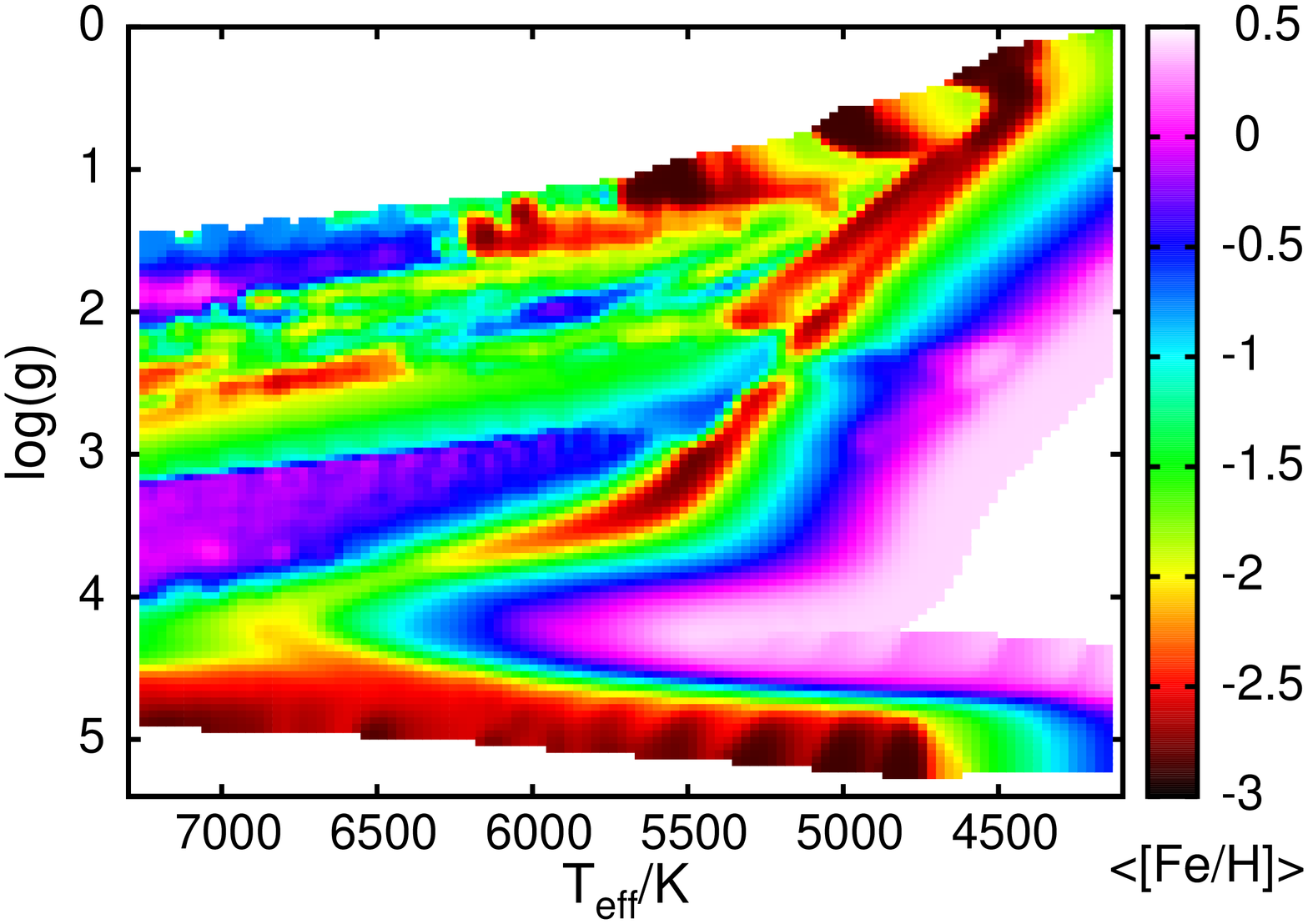,angle=-90,width=\hsize}
\end{center}
\caption{Prior probability densities from stellar models at any metallicity in the $(\Teff, \logg)$-plane (upper panel). We use the Salpeter IMF prior and the metallicity-dependent age prior. The lower panel shows the metallicity expectation value $<\feh>(\Teff, \logg)$ at each point in the $(\Teff, \logg)$-plane. The structures are dominated by the metallicity-dependent $\Teff$-shift and evolutionary differences on the giant branch, e.g. the absence of a horizontal branch for metal-rich stars in exchange for the enhanced red clump.}\label{fig:isochoverview}
\end{figure}
\subsection{Photometric data $P'_{\rm ph}$}
Stellar models couple photometric colours with other parameters \citep[see][for details on the colour calibrations in SDSS filters of BASTI models]{Marconi06}. Thus we best calculate the photometric PDF simultaneously with the stellar models. Denoting the stellar model magnitudes at model point $i$ and colour band $k$ by $C_{i,k}$ and the photometric observation in band $k$ with $O_k$, we have:

\begin{equation}
P_{i}(O_{ph}|{\bf C},s,r) = \prod_k{P(O_k|C_{i,k},s,r)}$,$
\end{equation}
with distance $s$ and reddening $r$.
Lacking sufficient data on the true PDFs, we represent the observational likelihoods of photometric colours, the reddening values and the model uncertainties by a Gaussian $g(x-\mu, \sigma) = exp(-(x-\mu)^2/(2\sigma^2))$, which enables us to combine them into:
\begin{equation}
P(O_k|C_{i,k},\mu(s),r) = g(C_{i,k} + \mu(s) + r\cdot \hat{e}_k - C_k, \sigma_k) $,$
\end{equation}
with the distance modulus $\mu(s) = 5\log(s/10 \pc)$, the reddening strength $r$ multiplied with the reddening vector \citep[for SDSS colours, see][]{Girardi04, An08} in each colour $\hat{e}_k$, and with $\sigma_k^2 = \sigma_{mod}^2 + \sigma_{obs}^2 + \sigma_{r\cdot{e}_k}^2$ as the combined variance/uncertainty of models, observations and reddening. We assume a magnitude uncertainty on stellar models $\sigma_{mod} = 0.01 \mag$ for the high-res sample, neglecting this term for the low-resolution sample. Partly, $\sigma_{mod}$ covers the same uncertainties as our error term {\bf ${\sigma_c}$} on stellar parameters (see Section \ref{sec:mod}). We stress that this uncertainty cannot and should not comprise systematic deviations in stellar models, as those uncertainties should be explored on a larger sample, using hyper-parameters as discussed in Section \ref{sec:priors}.

%%{\bf Magnitude errors on the stellar models derive from uncertainties in the theoretical spectra and filter bandpasses, altered by the calibrations, as well as uncertainties in stellar evolution, and their scope is highly uncertain. Parts . In addition we add $\sigma_{mod} = 0.01 \mag$ in each band for the high-resolution sample.}

The other assumption is the universal reddening vector. This may have to be relaxed when dealing with very different ISM environments. The dust peak and the slope of the reddening spectrum can be shifted, or stars may be individually reddened, e.g. by a circumstellar envelope.

We note that this method can be used to create reddening maps. Since that is beyond the scope of this work, we restrict the sample to stars with relatively low reddening, use reddening values from other sources assuming a fractional reddening error of $10\%$.

\begin{figure}
\begin{center}
\epsfig{file=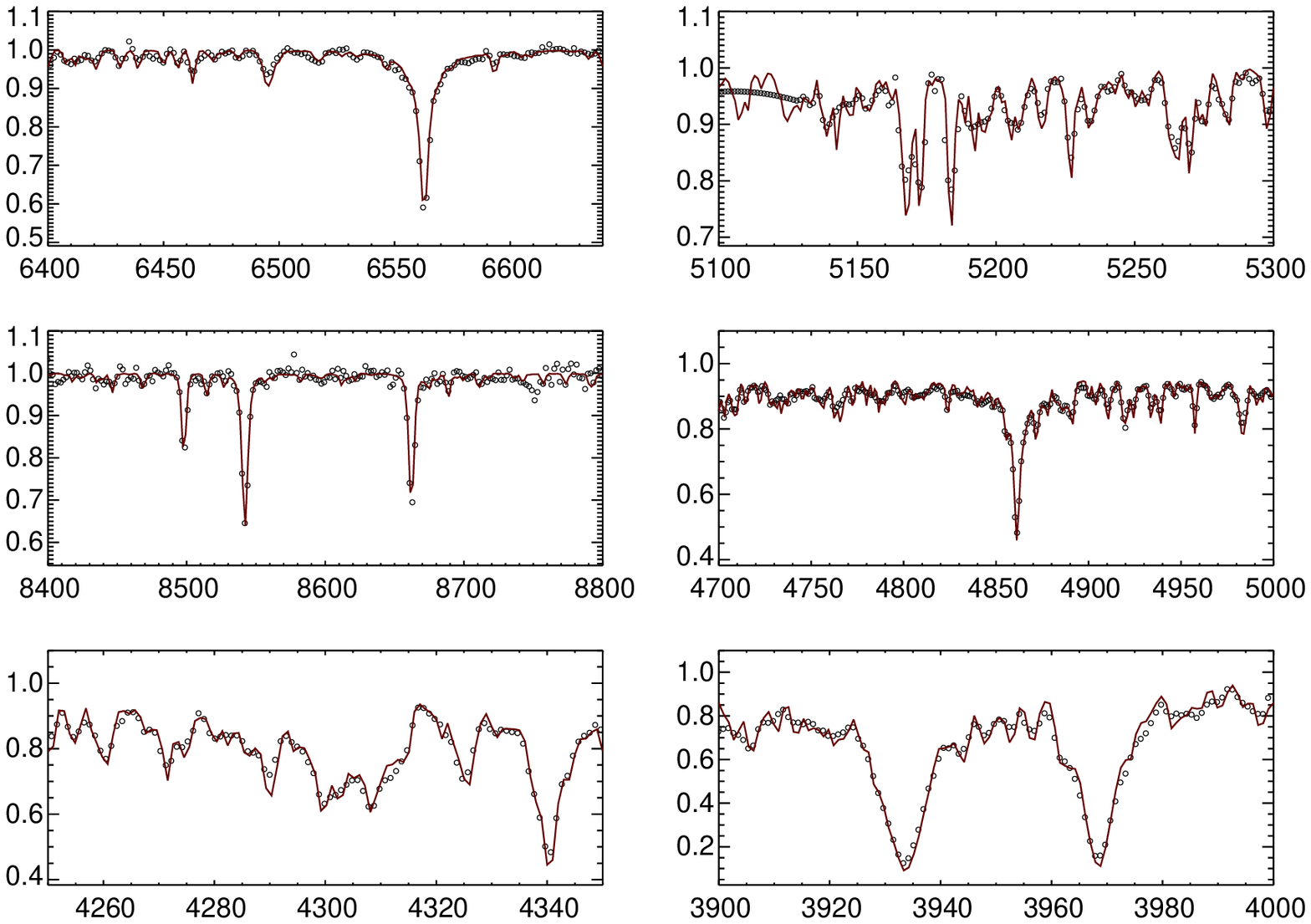,angle=0,width=\hsize}
\epsfig{file=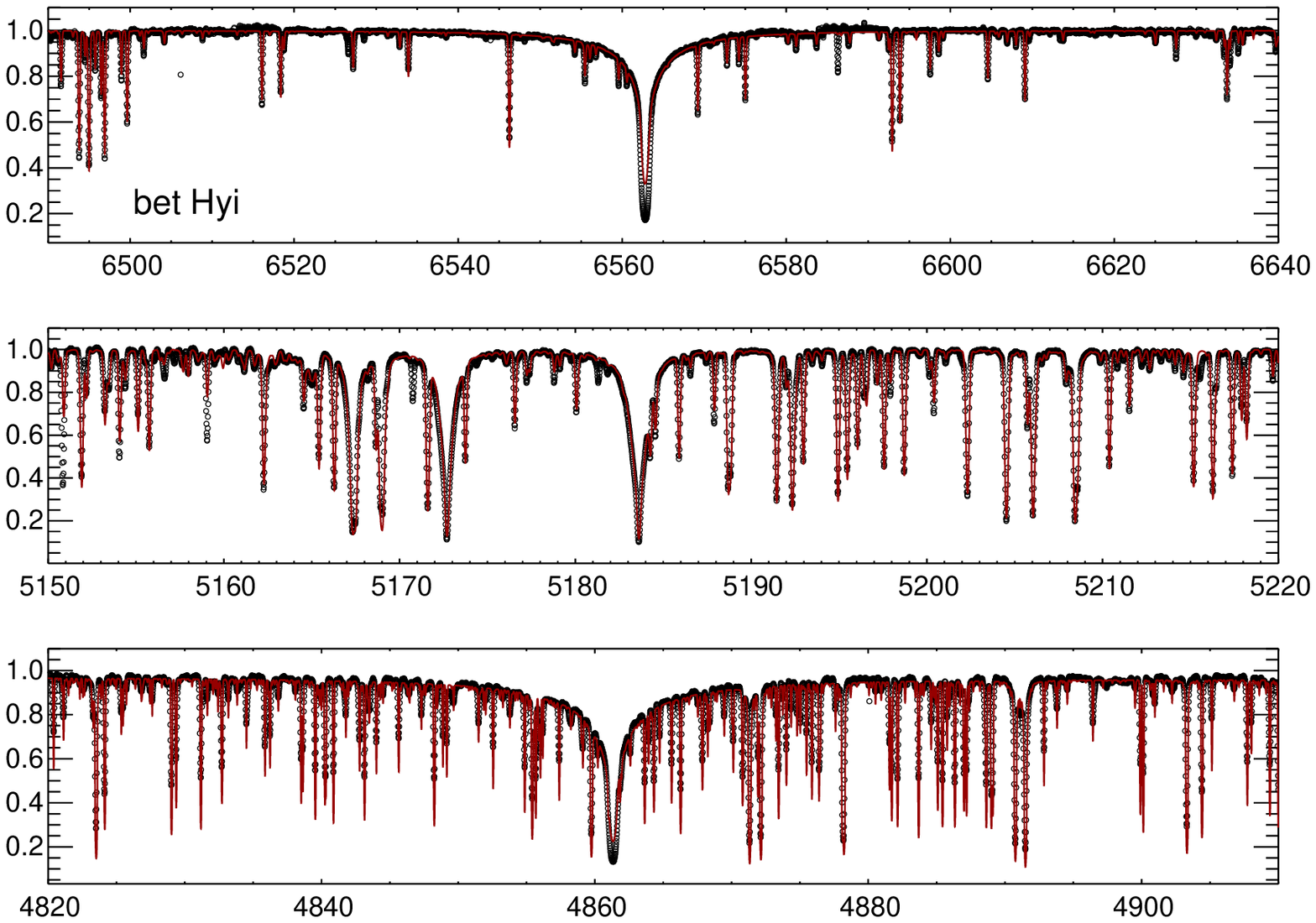,angle=0,width=\hsize}
\end{center}
\caption{Top six graphs: A low-resolution spectrum from the SDSS/SEGUE survey (black, plate no. $2038$, fiber $564$) compared to the best-fitting theoretical model (red). Bottom three graphs: same for a high-resolution spectrum from UVES of $\beta$ Hyi.}
\label{fig:foutr2p2038f564}
\end{figure}

\subsection{Spectroscopic data $P'_{\rm sp}$}

The observational likelihood $P'_{\rm sp}$ in Eq. (\ref{eq:ar}) incorporates all available spectral information. This comprises spectral type, element abundances, rotation, stellar activity (chromospheric emission in cores of strong lines, magnetic effects), inter- and circum-stellar reddening, convection characteristics, etc. 
A spectrum offers by far the largest information content we can obtain on a star. However, this information is limited by incomplete physical knowledge and approximations in modelling the theoretical spectra. At present, calculations of large spectral grids are only possible with 1D hydrostatic codes assuming local thermodynamical equilibrium (LTE), while full hydrodynamic 3D non-LTE calculations are slowly becoming feasible \citep[][]{Bergemann12, Magic13}. Further, the high dimensionality of the problem forbids computing separate grids for all possible chemical compositions.

Here we use the MAFAGS-ODF  \citep{Grupp04a, Grupp04b} grid of model atmospheres designed for late-type (spectral type FGKM) stars, with $4400 < \Teff < 6800$ K, $1.4 < \logg < 4.6$, $-3.9 < \feh < 0.9$. The synthetic spectra are computed with the revised version of the SIU code \citep[][]{Reetz99, Bergemann12}, which has been extensively used during the past two decades for high-precision stellar spectroscopy \citep[see e.g.][]{Korn03, Bergemann08, Onehag11, Shi14}. In comparison to the other three available spectrum analysis codes, MOOG \citep[][]{Sneden73}, SYNTHE \citep{Kurucz05}, and Turbospectrum \citep[][]{Plez12}, SIU already has an implementation of NLTE line formation for any element with pre-computed NLTE level populations. We can thus more easily update it with more realistic physics.

For our grid of theoretical spectra, we use a spacing of ($200 \K, 0.4 \dex, 0.3 \dex$) in ($\Teff, \logg, \feh$) to make linear interpolation between the points reasonable. While the grid covers $4$ values of micro-turbulence from $1$ to $2.5 \kms$, for this pilot study we adopt a fixed micro-turbulence of $2 \kms$ for giants and $1 \kms$ for dwarfs ($\llg \geq 3.5$) (Bergemann et al. in prep), and an $\alpha$-enhancement of $0.4$ dex for $\feh < -0.6$ \citep[e.g.][]{Gehren04}. Such hard cuts and changes in assumed parameters will introduce anomalies in the derived PDF, as exemplified by the micro-turbulence cut at $\llg = 3.5$ in \figref{fig:casestudy1}.  In
total, the three-dimensional ($\meh$, $\Teff$, $\logg$) grid contains $6912$
theoretical spectra and covers the full HRD, as shown in Fig.
\ref{fig:specstuff}. Any other model grid can be easily implemented, with no 
requirement on symmetry or shape, since our code includes a robust interpolation
scheme. Alternatively, one could perform calculations of line formation on the
fly using a grid of model atmospheres. This latter approach is cleaner, however,
it is still computationally too costly. We sample the wavelength windows around
the spectral features important for diagnostic of FGKM stars: $3850-4050$
\AA~(Ca I lines), $4350 - 4450$\AA~(G-band, CN sensitive), $4600 -
4900$\AA~($H_{\beta}$), $5100 - 5300$\AA~(Mg I triplet, main gravity
diagnostics), $6400 - 6640$\AA~($H_{\alpha}$), $8400 - 8800$\AA~(Ca II triplet,
also used in Gaia and in RAVE stellar survey). However, not all pixels in these
intervals are used in the analysis. The high-resolution observed spectra (see
Sec. 4.3.1) do not cover the regions below $4800$ and above $6800$ \AA. We exclude from our statistics all  regions which contain spectral lines of chemical elements other than the temperature- and pressure-sensitive wings of Balmer and Mg I triplet lines, and the Fe I and Fe II lines. Precisely, the weight of all other spectral features, is set to zero. The flat regions
are used for the iterative continuum normalization and are not masked out. To avoid over-confident estimates, we demand that either the temperature uncertainty $\sigma_{\Teff} > 80 \K$ or the metallicity uncertainty $\sigma_{\feh} > 0.08 \dex$, and otherwise flatten the PDF by multiplying the $\chi^2$ distribution with a fixed factor until the condition is met.
Before evaluating the test statistics, the spectra are continuum-normalised
and radial-velocity corrected by cross-correlating with the template theoretical
spectrum for each input combination of stellar parameters.

To obtain the spectroscopic observational likelihood $P(O_{\rm sp}|X_i)$ at each point in parameter space, we resample the synthetic spectrum to the wavelength scale and resolution of the observations and evaluate the goodness-of-fit-statistics $\chi^2$ at each pixel $i$ of the observed spectrum:
\begin{equation}
\chi^2 = \sum\limits_{i=1}^{n} \left(O_i - {S_i}\right)^2/\sigma^2{,}
\end{equation}
where $S$ the template comparison spectrum, $O$ the observed spectrum, $\sigma^{,}$ the weighted observational uncertainty. Noisy and un-informative regions are given less weight using special masks. The final PDF is gained by summing over all pixels within a given segment, and over all segments. 

The original resolution of the synthetic grid is $500\,000$. Thus, the method can be potentially applied to any observed dataset, e.g., low-resolution and high-resolution spectra. For the analysis of the SEGUE spectra, we post-convolved the spectral grids with instrumental resolution, $R = 2000$. A typical fit to a SEGUE spectrum is shown in \figref{fig:foutr2p2038f564}. In the high-resolution mode, we use the resolution of the UVES-instrument ($R = 50\,000$).

\subsection{Parallaxes and other additional data}

The Gaia mission will derive parallax measurements for nearly all stars with spectroscopic information. Parallax measurements only affect the distance $s$ (and distance modulus $\mu$), so that it is straightforward to combine the observational likelihood from parallax measurements $P'_{astr}$ with the photometric and model information.

In the following, we assume a Gaussian parallax error. Cromwell's rule does not apply to mathematical truths, so negative parallaxes are excluded by setting the prior to $0$. This yields:
\begin{equation}
P'_{\rm astr} = P(O_{\rm astr} | p) = N\Theta(p)g(p - p_0, \sigma_{p}) $,$
\end{equation}
where N is a normalisation, $\Theta$ is the Heaviside-function ($1$ for $p \ge 0$ and $0$ for $p < 0$), $g(p-p_0, \sigma_{p})$ is again a Gaussian distribution around the measured parallax $p_0$ (which can be negative) with standard deviation $\sigma$.

\begin{figure}
\begin{center}
\epsfig{file=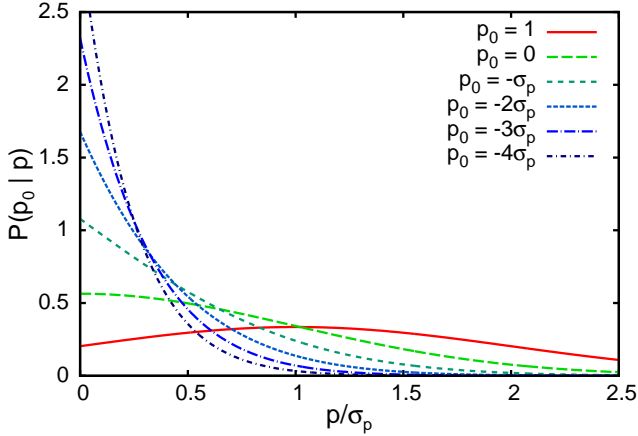,angle=-90,width=\hsize}
\end{center}
\caption{Conditional likelihood of the observation of $p_0 = 0, -\sigma_{p}, \ldots$ under the real parallax $p$, assuming a Gaussian error distribution with standard deviation $\sigma_{p}$. The distributions become significantly narrower around the $0$ for more negative parallax values, reducing the parallax expectation value and raising the expected distance.}\label{fig:parallax}
\end{figure}

\begin{figure}
\begin{center}
\epsfig{file=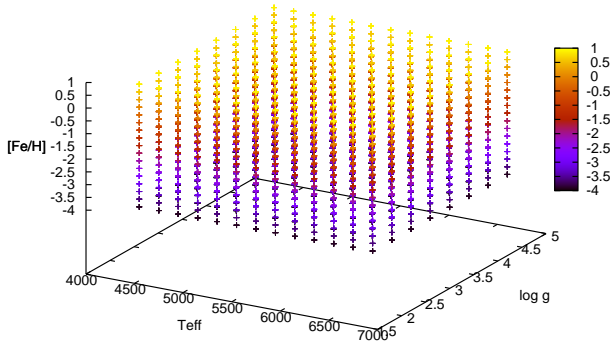,angle=0,width=\hsize}
\end{center}
\caption{The grid of synthetic spectra.}
\label{fig:specstuff}
\end{figure}

It is important not to clip negative values of $p_0$: a small negative value of $p_0$ has still a different information content than a large negative value. In the case of a Gaussian error distribution, the probability ratio between a smaller parallax and a larger parallax rises, the further the measurement is away from zero. Or to use an example: the likelihood ratio between having failed by $3 \sigma$ and by $4 \sigma$ is larger than the likelihood ratio between having failed by $3 \sigma$ and by $2 \sigma$. \footnote{This would only not be true if the error distribution gives constant likelihood ratios for identical distances from the measurement value, i.e. for a declining single exponential.} \figref{fig:parallax} demonstrates how the parallax distributions get more concentrated towards zero, the more negative the measured value is.

To combine $P'_{\rm astr}$ it with the photometric and model PDF, we integrate over the possible distance moduli $\mu$, at each stellar model point $i$:
\begin{equation}
P'_{i, {\rm astr, ph}} = \int{P_{i}(O_{ph}|{\bf C},\mu(s),r)P'_{\rm astr}(\mu) J_{\mu/p}}d\mu $,$
\end{equation}
with the Jacobian $J_{\mu/p} = 20\ln(10)10^{-0.2\mu}$.

In this work, P$_{\rm astr}$ is considered for the stars with high-resolution spectra only (see Sec. \ref{sec:highres}).

\subsection{Combining the PDFs}

Equipped with these results, we can now assemble the combined PDF in equation \ref{eq:ar}. In simple words, the strategy is separate all PDFs into PDFs on the core parameter space $R_c = (\Teff, \logg, \meh)$ and the conditional PDFs on the remaining parameter space given that point in $R_c$. Depending on our needs we can then represent those remaining parameter estimates either as simple moments (expectation value, variance, etc.) at each point in $R_c$, or as full distributions.

Formalising this is a bit tedious, since it involves a conditional probability derived from a conditional probability. To simplify the notation, we use the previous abbreviation of observational dependence with a prime. The combined calculation of photometric and model part yields:
\begin{eqnarray}
P'_{\rm astr,mod,ph,pr} &=& P'_{\rm astr, ph} \cdot P_{\rm mod} \cdot P_{\rm pr} \\
&=& P'_{\rm astr,mod,ph,pr}(\vX_c)P'_{\rm astr,mod,ph,pr}(\va_{\rm ph}|\vX_c) \notag {,}
\end{eqnarray}
where $\vX_c$ is the vector of parameters in our core parameter space $R_c$ and $\va_{\rm ph}$ is the vector of remaining parameters constrained by the photometric and astrometric observations, models and priors, i.e. $\va_{\rm ph} = (M_i, \tau, \vC, r, s, \ldots)$. Similarly, we separate the spectroscopic information:
\begin{equation}
P'_{\rm sp} = P'_{\rm sp}(\vX_c)P'_{\rm sp}(\va_{\rm sp} | \vX_c) $,$
\end{equation}
where $\va_{\rm sp}$ denotes all other parameters constrained by spectroscopic observations, like detailed abundances, or stellar rotation. In this work we do not use this supplementary information, so that we can drop the term $P'_{\rm sp}(\va_{\rm sp} | \vX_c)$. Most of the parameters in $\va_{\rm sp}$ will not coincide with the parameters in $\va_{\rm ph}$, but if they correspond, they must be written into the core parameter space. For example, rotation and stellar activity available from high-quality spectra constrain stellar ages. We will discuss this in a future work.

We can now calculate the final probability distribution function:
\begin{equation}{\label{eq:finprob}}
P(\vX|\vO) = P'_{\rm astr,mod,ph,pr,sp}(\vX_c) P'_{\rm sp}(\va_{\rm sp} | \vX_c) P'_{\rm astr,mod,ph,pr} (\va_{\rm ph} | \vX_c) $,$
\end{equation}
where 
\begin{equation}
P'_{\rm astr,mod,ph,pr,sp}(\vX_c) = P'_{\rm sp}(\vX_c) P'_{\rm astr,mod,ph,pr} (\vX_c)$.$
\end{equation}

\subsection{Calculating projections, central values, and uncertainties}
We can gain the conditional probability distribution in a lower number of parameters by marginalising, i.e. by integrating out the other dimensions in the joint conditional probability distribution function. E.g., to exclude the parameter x$_{j+1}$, we write:
\begin{eqnarray}
P(X_1,\ldots,X_{j},X_{j+2},\ldots,X_n|\vO) = \int{P(X_1,\ldots,X_{n}|\vO)dx_{j+1}}
\\
P(X_j|\vO) = \int\int{P(X_1,\ldots,X_{n}|\vO)}dx_1 \ldots dx_{j-1} dx_{j+1} \ldots
dx_{n} 
\end{eqnarray}
From this we can obtain the moments of the probability distribution in each
variable or group of variables:
\begin{eqnarray}
\left< X_{j} \right> &=& \int x_jP(X_j|\vO) dx_j \\
\left< X_{j}^2 \right> &=& \int x_j^2 P(X_j|\vO) dx_j \\ 
\end{eqnarray}
where $\left< X_{j} \right>$ denotes the expectation value of the parameter
$X_j$, and the standard deviation $\sigma_{X_j} = \sqrt{\left<X_{j}^2 \right> - \left<
X_{j} \right>^2}$.
\subsection{Short recipe of the algorithm}

\begin{figure}
\begin{center}
\epsfig{file=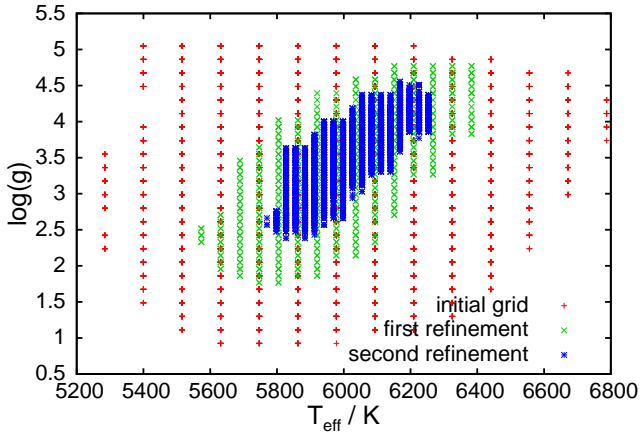,angle=-90,width=\hsize}
\end{center}
\caption{The process of iterative grid refinement in our pipeline for our example SEGUE star (plate $2038$, fiber $564$). Most points in the $(\Teff, \logg)$-plane have several metallicity values.}\label{fig:search}
\end{figure}

In short the steps are as follows:
\begin{itemize}
\item 1) Combine photometric and astrometric information together with the priors and sum over all stellar model points to obtain a preliminary PDF $P_{prel}$ in core parameter space, calculate moments or full PDFs for the remaining dimensions.
\item 2) In regions of parameter space, where the probability is larger than a threshold value\footnote{The threshold should be sufficiently small to ensure coverage of the final PDF. Here we use a generous $10^{-30}$ per bin. Compared to a number of $\sim 2 \cdot 10^{-6}$ bins we hence neglect a negligible fraction of the probability mass.}, calculate a coarse grid of spectroscopic probabilities and approximate the PDF $P_{sp}$ by interpolation.
\item 3) Multiply $P_{prel}$ with $P_{sp}$ to obtain an approximate posterior PDF $P$. Determine a refined grid in parameter space to better sample the spectroscopic PDF and iterate steps 2) + 3) (Fig. \ref{fig:search}).
\end{itemize}

The threshold value on a binned PDF was chosen as $t \ll 1/(N\cdot M)$ where $N$ is the number of stars and $M$ is the effective number of bins, because in a large sample we have to expect the presence of rare objects, which will have low preliminary probabilities. As a different condition one can formulate that the integral of the PDF over parameter space must be
\begin{equation}
N\cdot \int_{P(\vX) < t}P(\vX)d\vX \ll 1 $.$
\end{equation}

\begin{figure*}
\begin{center}
\epsfig{file=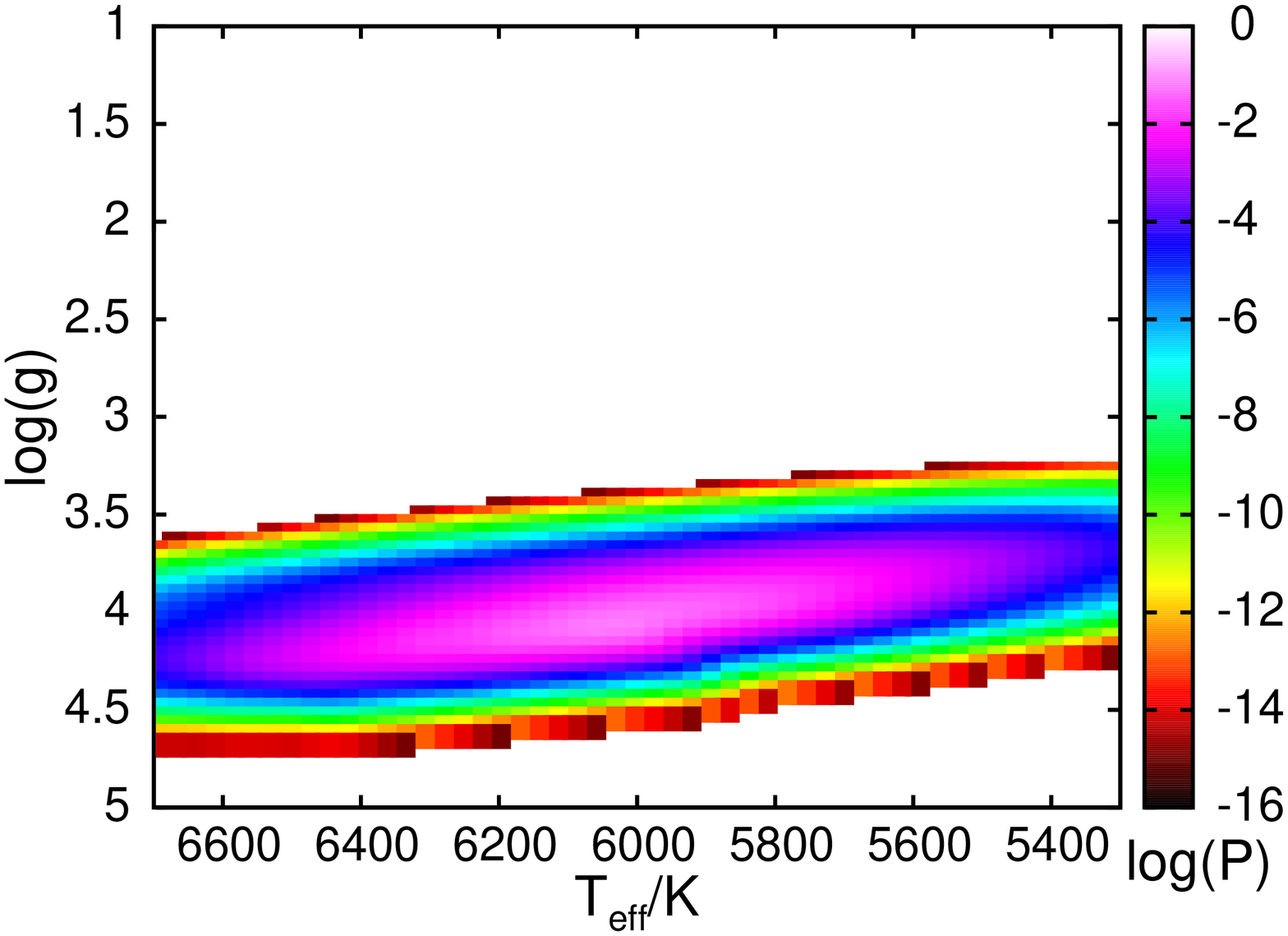,angle=-90,width=0.49\hsize}
\epsfig{file=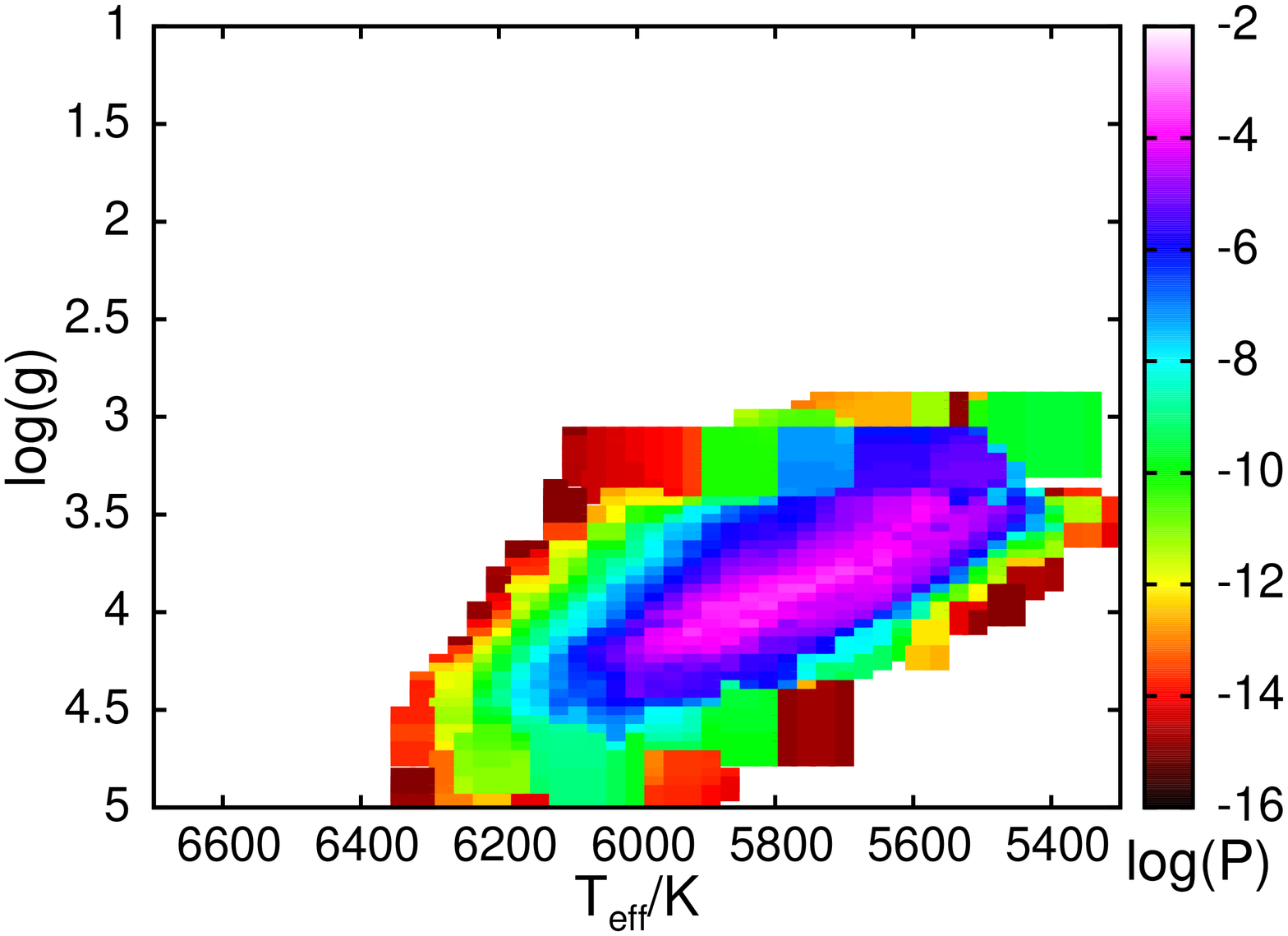,angle=-90,width=0.49\hsize}
\epsfig{file=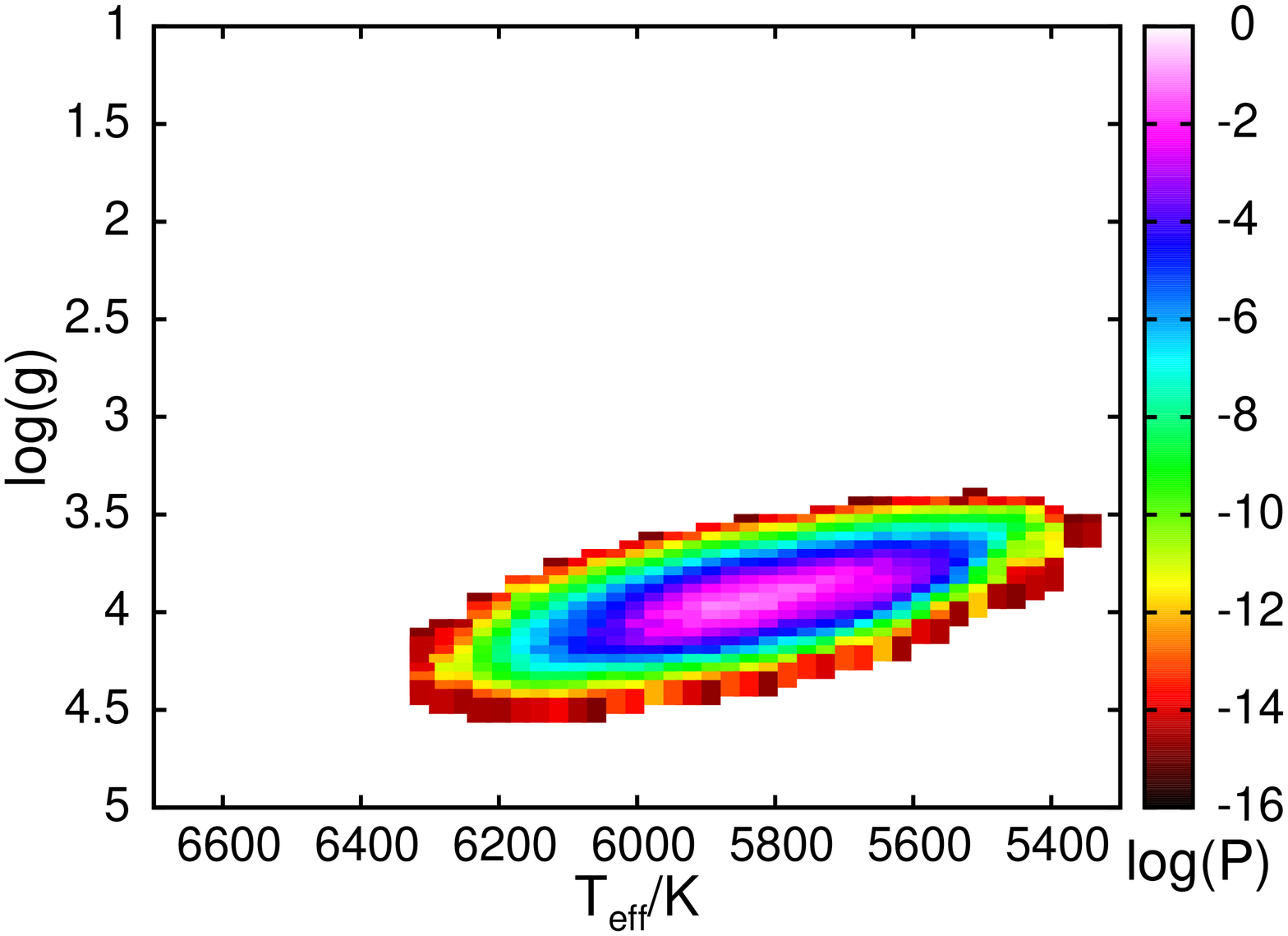,angle=-90,width=0.49\hsize}
\epsfig{file=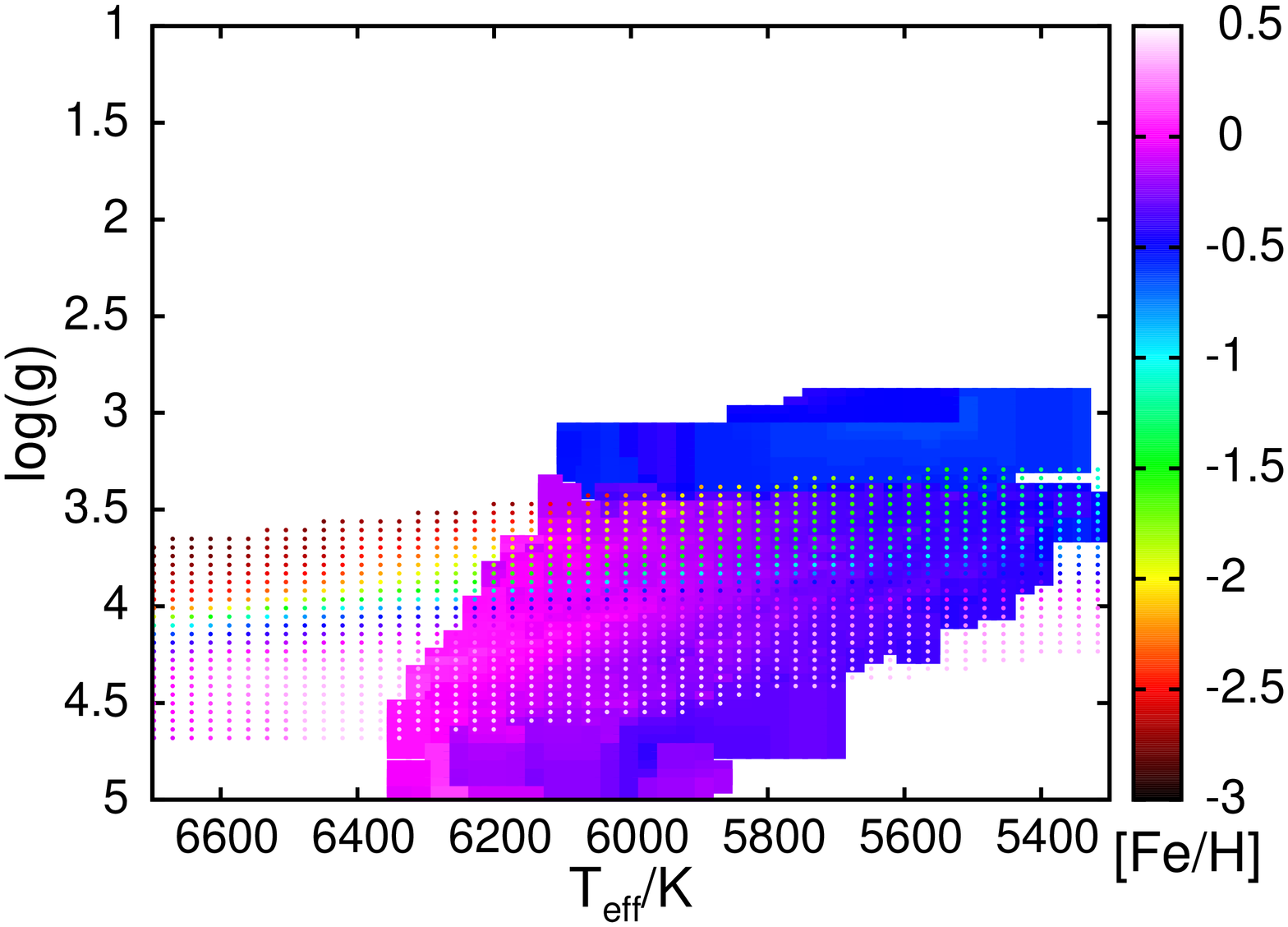,angle=-90,width=0.49\hsize}
\end{center}
\caption{The PDF's for $\beta$ Hyi from the high-res sample. Top left: the combined PDF from photometry-prior-stellar evolution projected onto $(\Teff, logg)$-space; top right: the spectroscopic PDF in photometrically allowed space. These two estimates combine to the overall PDF in the bottom left panel. The bottom right: metallicity expectation values from spectroscopy (coloured area) and the photometric-model part (coloured dots).}
\label{fig:casestudy2}
\end{figure*}

\subsection{Selected examples}

To illustrate the algorithm, we describe here the results for two stars. One is $\beta$ Hyi from our high resolution data sample, for which we have basic Johnson photometry, high resolution spectroscopy and a Hipparcos parallax. The other star, randomly selected from the SEGUE data sample (plate number $2038$ and fiber number $564$), is a turn-off subgiant. In this case we have SDSS photometry and a low-resolution spectrum from SEGUE.

\begin{figure*}
\begin{center}
\epsfig{file=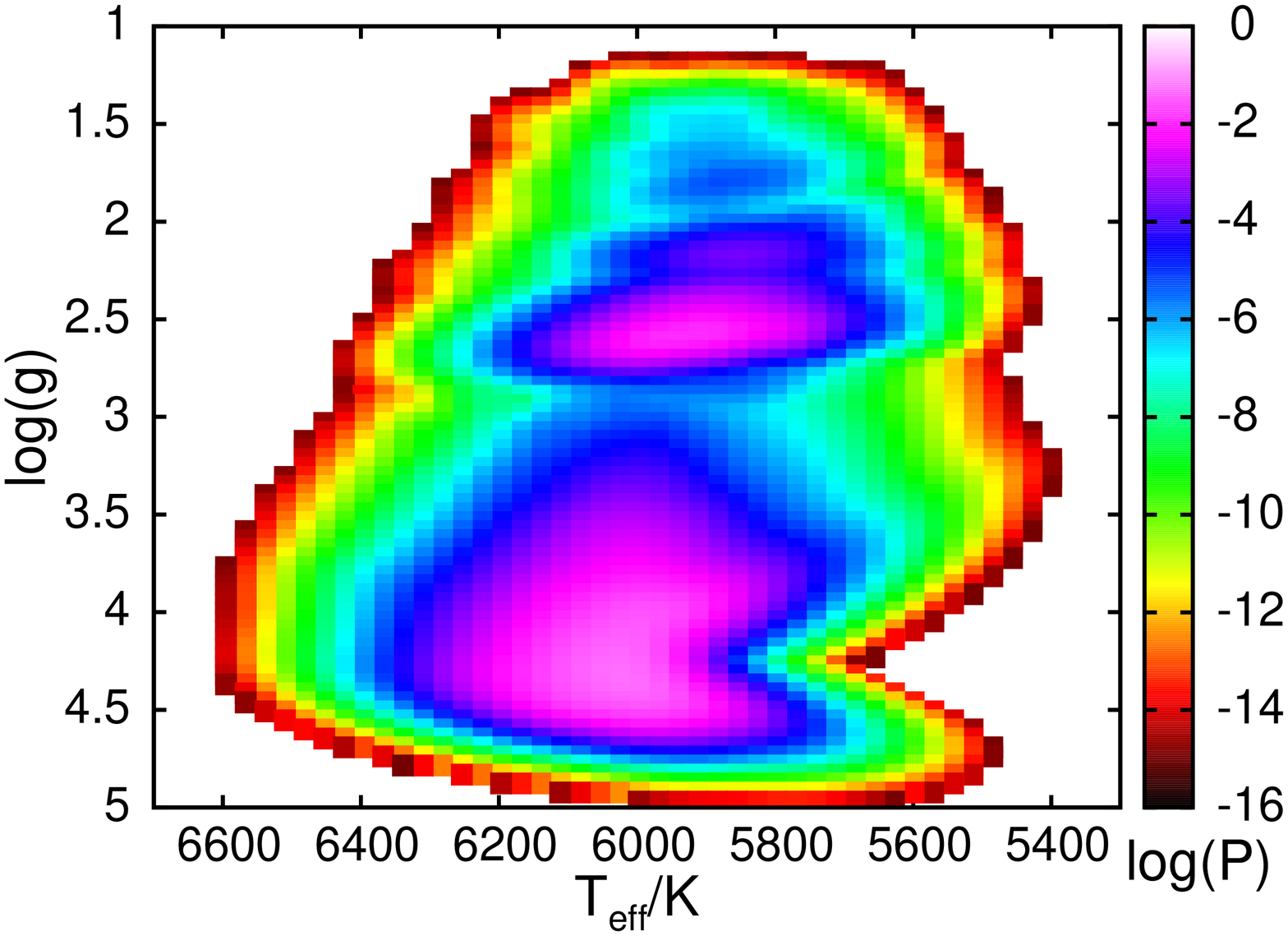,angle=-90,width=0.49\hsize}
\epsfig{file=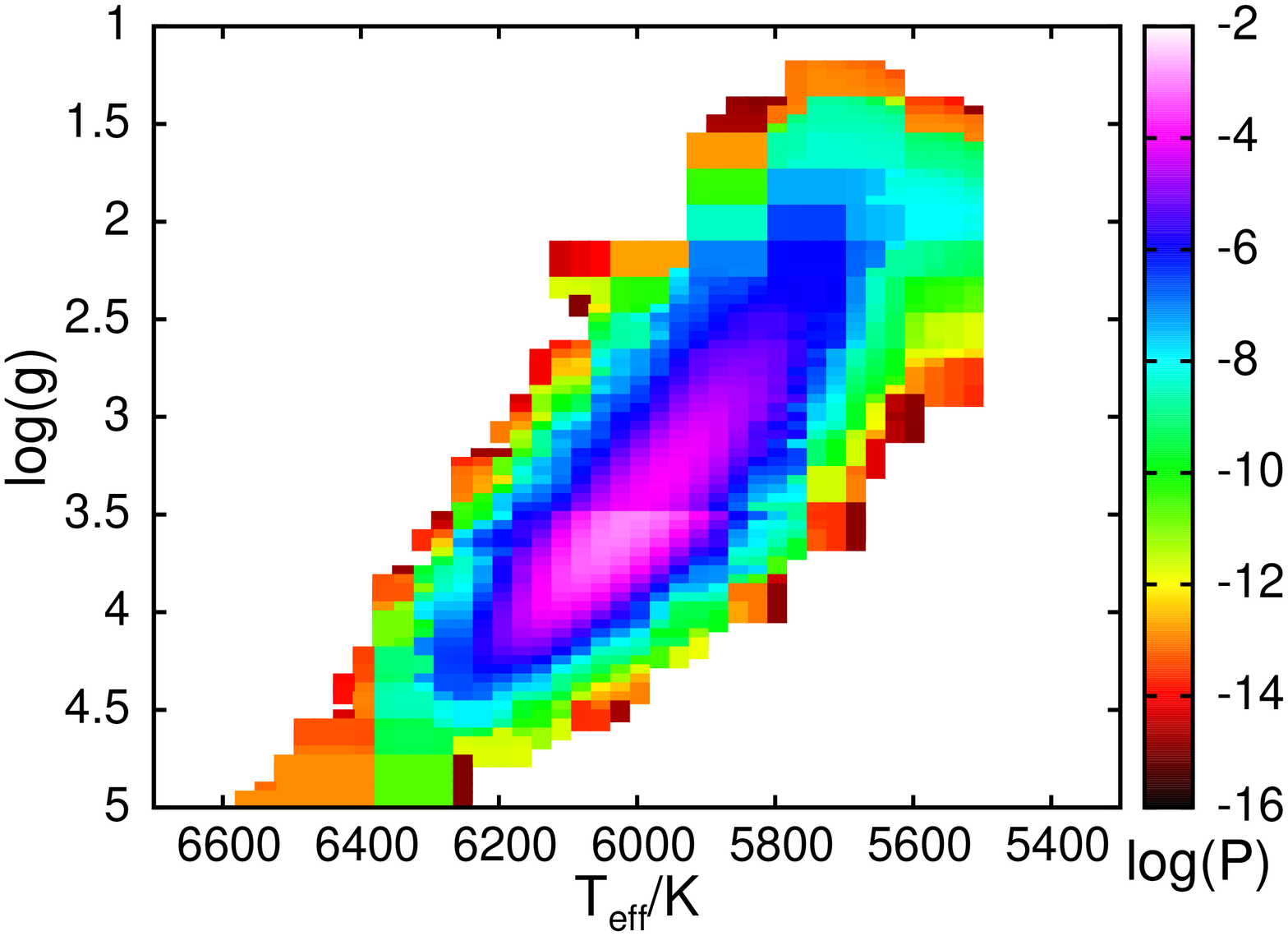,angle=-90,width=0.49\hsize}
\epsfig{file=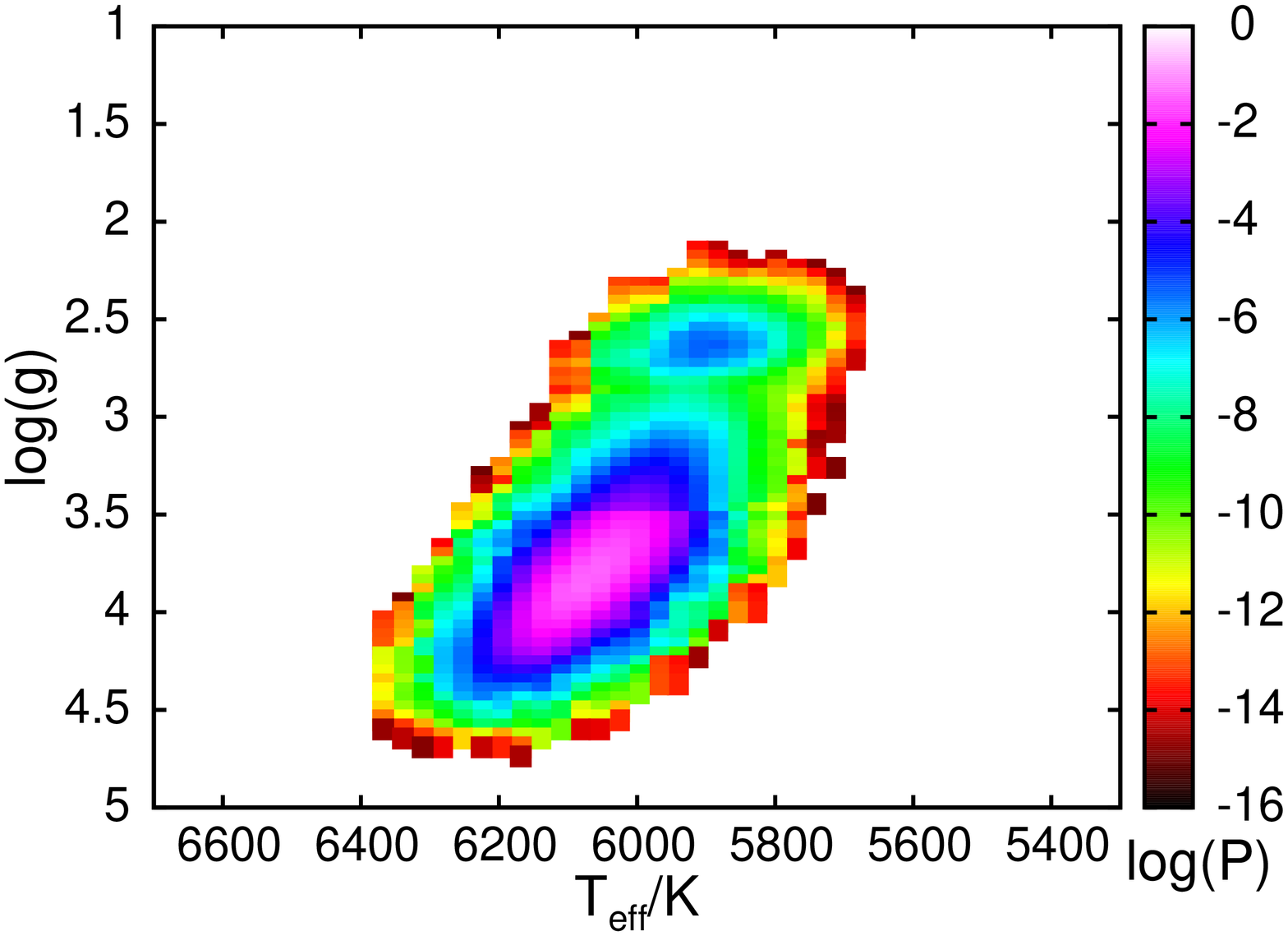,angle=-90,width=0.49\hsize}
\epsfig{file=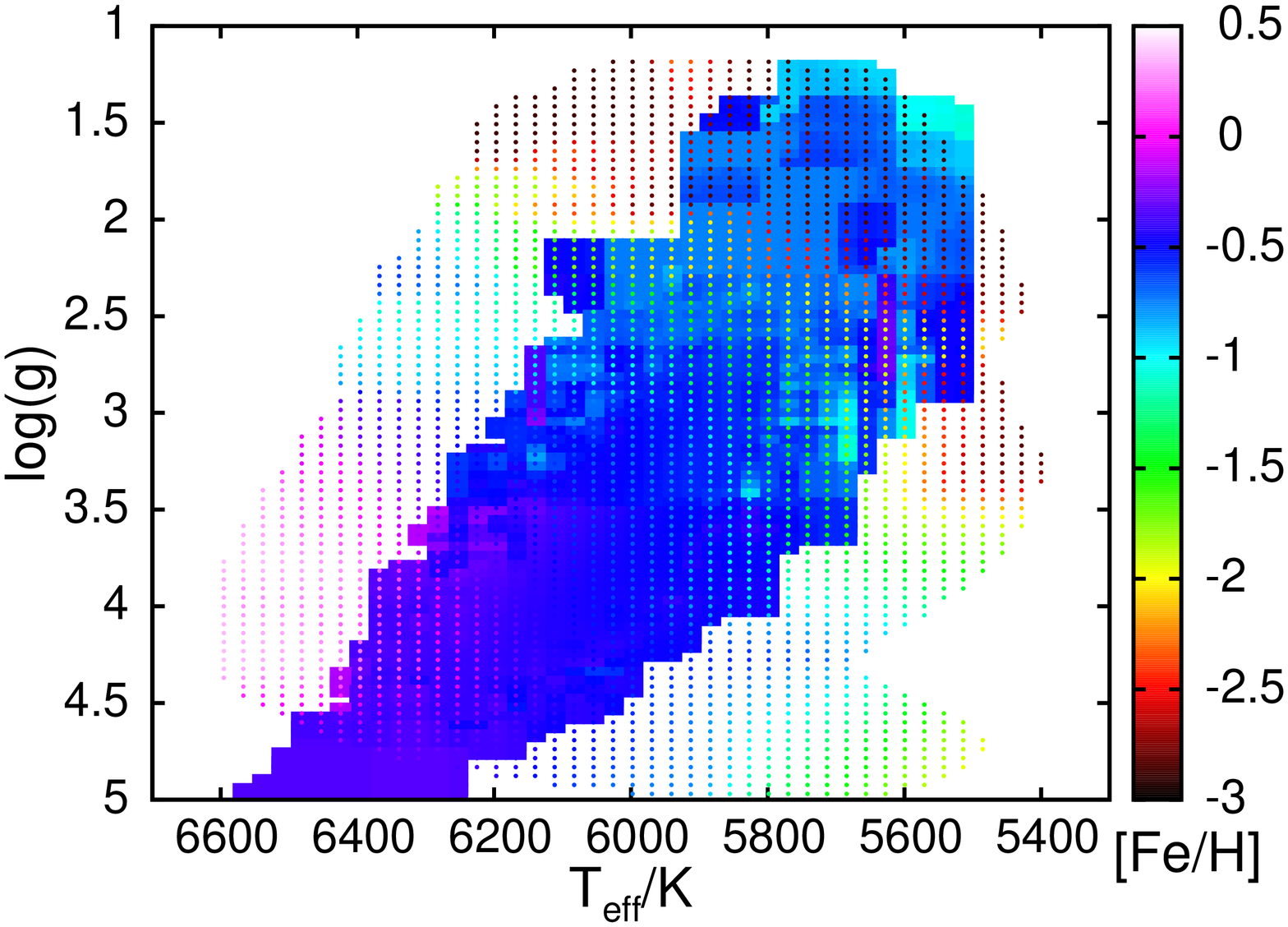,angle=-90,width=0.49\hsize}
\end{center}
\caption{The PDFs for one of the sample stars from the SEGUE sub-sample (see Sec. 3.10). Top left: the combined PDF from photometry-prior-stellar evolution projected onto $(\Teff, logg)$-space; top right: the spectroscopic PDF in photometrically allowed space. These two estimates combine to the overall PDF in the bottom left panel. The bottom right: metallicity expectation values from spectroscopy (coloured area) and the photometric/model part (coloured dots).}
\label{fig:casestudy1}
\end{figure*}

The resulting probability distribution functions in the $(\Teff, \log g)$-plane are shown in \figref{fig:casestudy2} and in \figref{fig:casestudy1}. The top left panel shows the combined PDF from photometry, prior and stellar evolution; the top right panel shows the spectroscopic PDF in photometrically allowed space. These two estimates combine to the final posterior PDF in the bottom left panel. The corresponding metallicities are shown by colour coding in the bottom right panel. The individual probability densities from photometry-stellar evolution and spectroscopy are clearly different in shape and in location. 

The Hipparcos parallax combined with photometry and stellar models puts tight constraints on the surface gravity of $\beta$ Hyi in \figref{fig:casestudy2}. This leads also to a tight correlation between metallicity and gravity as evident from the coloured dots in the bottom right panel. A moderate step in the spectroscopic PDF at $\logg \sim 3.5$ is produced by a step in micro-turbulence in our current grid of theoretical spectra, which will disappear with the improved grids in preparation. The calculation does not cover the full allowed region of the spectroscopic PDF (see the coarse behaviour at smaller gravities in the top right panel), saving computation time, since the joint PDF (lower left panel) is fully represented. The final expectation values and uncertainties are $\Teff = (5837 \pm 72)\K$, $\logg = (3.981 \pm 0.068) \dex$, and $\feh = (-0.196 \pm 0.074) \dex$ versus $\Teff = (5873 \pm 38) \K$, $\logg = (3.98 \pm 0.02) \dex$, and $\feh = (-0.08 \pm 0.02) \dex$ in the reference sample (described in the next Section).

\begin{figure}
\begin{center}
\epsfig{file=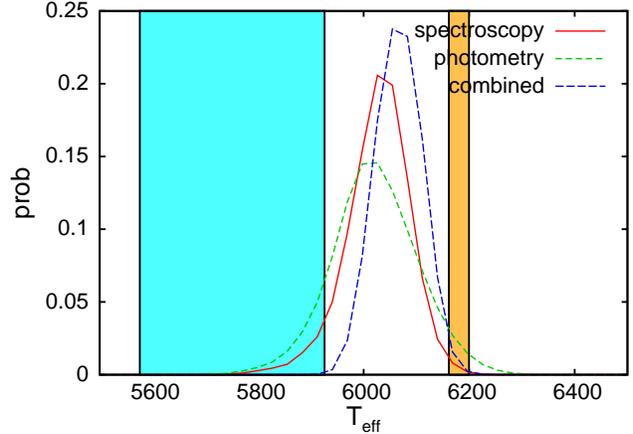,angle=-90,width=\hsize}
\end{center}
\caption{Projected 1D $\Teff$ distributions for the discussed SEGUE star versus the $1\sigma$ intervals from Allende Prieto et al.(2008, light blue) and from SEGUE DR9 (orange). Note how the combined estimate differs from a naive expectation when looking at photometric and spectroscopic information separately.}\label{fig:Teffex}
\end{figure}

While neither the photometric part nor the spectroscopic constraints are very tight for the SEGUE star in \figref{fig:casestudy1}, the combined PDF is very well defined. This shows the benefits of solving the problem in the full parameter space $R_c$. While points in the $\logg-\Teff$ plane may be allowed by both derivations, the corresponding limits on the third dimension $\feh$ are in disagreement, ruling them out. These are the regions in the bottom right panel of \figref{fig:casestudy1}, where the colours are mismatched. To stress this point we show the one-dimensional probability distributions in $\Teff$ in \figref{fig:Teffex}. While our parameters are nicely between the values of the SEGUE follow-up study \cite{Allende08} and SEGUE DR9 (see Sec. \ref{sec:datasets}), the behaviour of our PDF is more interesting: The combined PDF is not even remotely a simple combination of its two contributors. Most interestingly, the expectation value of the combined estimate is not situated between the estimates from each spectroscopy ($6027 \K$) and photometry ($6021 \K$), but significantly higher ($6066 \K$). This complex behaviour can only be accounted for within a full Bayesian approach.

Our final expectation values and uncertainties for this SEGUE test star are $\Teff = 6066\K \pm 44\K, \llg = 3.83 \pm 0.15, \feh = -0.47 \pm 0.07$, for comparison SEGUE DR9 provides $\Teff = 6181\K \pm 19\K, \llg = 3.90 \pm 0.03, \feh = -0.459 \pm 0.006$. Note that we add the reported uncertainties from the SEGUE pipeline just for the sake of completeness. Their formally reported errors cannot be considered realistic. They are severely under-estimated (by about an order of magnitude or more) as shown by the comparisons in \cite{Lee08a, Lee08b} as well the discussion later in this work. The spectral fits in our six standard bands for the best spectroscopic solution are shown in \figref{fig:foutr2p2038f564}. 

This discussion also shows that even a relatively uncertain information can give an improvement to more precise values that is beyond a simple one-dimensional combination. More importantly, mismatches between different sources of information help to flag pathologies in a sample by unexpectedly small overlap of the contributing PDFs.

\section{Application to observations}

Our approach is most needed and also most powerful, when different observations are available for a star and the information content is complementary but limited. With this in mind and to test the stability of our method, we choose both a sample featuring high-resolution spectra ($R \geq 40000$), from observations with VLT, as well as one with low-resolution spectra ($R \sim 2000$) from SDSS/SEGUE. We start this Section with a description of the datasets in use. Then, we first show the performance of the approach when limiting ourselves to photometric data with and without astrometry, followed by the full approach on low- and high-resolution spectra. In the last subsection we compare derived quantities, like distances and ages, and assess the resulting distribution of stars in the temperature-gravity plane.

\subsection{Datasets}\label{sec:datasets}

For the high-resolution sample we obtained a comparison set of stellar parameters from \cite{Jofre13}. Their effective temperatures were derived from the interferometric angular diameters or calibration relations. The gravities stem from astroseismology or Hipparcos parallaxes, and their metallicities are based on the analysis of Fe II lines, which are not significantly affected by non-LTE effects \citep[][]{Bergemann12}.

The high-resolution sample comprises $87$ high-resolution spectra of $40$ nearby stars including the Sun, taken with the HARPS and UVES instruments at VLT, and with NARVAL at the Pic du Midi observatory. This dataset was kindly provided by P. Jofre. In addition, there are Hipparcos parallaxes \citep[][]{vanLeeuwen}, making the sample closely resemble future data from Gaia astrometry combined with Gaia-ESO spectroscopy. The sample is particularly valuable because the spectra were taken on different instruments and there are independent parameter determinations, including interferometric angular diameters and astroseismic surface gravities \citep[][]{Jofre13}. The stars cover a very wide range in metallicities, gravities and temperatures in parameter space \citep[see a complete description in][]{BlancoC14}. Photometry in the $U, B, V, I, J, H, K$ bands, was compiled from the Hipparcos catalogue \citep[][]{Perryman12}, from 2MASS \citep[][]{2MASS}, and from \cite{Johnson66}. $U-$band photometry for HD$22879$ stems from \cite{Koen10}, improved $JHK$-photometry for $\xi$ Hya from \cite{Laney12}. 
Solar photometry was adopted from \cite{BM}, updated with the values of \cite{Ramirez12}. We increased the errors in light of the general uncertainty of the Sun's photometry to $0.03 \mag$.

Our low-resolution sample was selected from SEGUE by \cite{Allende08}, who did an intermediate-resolution follow-up analysis.\footnote{We hence have sets of comparison values, with a mild preference for the SEGUE DR9, since it is very difficult to assess the accuracy and homogeneity of AP08: different parts of the sample were analysed with different methods (equivalent width method for the higher-resolution stars vs spectrum synthesis for the lower-resolution stars). For most of these stars, the spectra were degraded to R $\sim 7000$ from the original R $\sim 15000$ with unclear consequences.} It consists of $78$ stars within the parameter range $-2.5 < \feh < 0.3$, $4000 < \Teff < 7000$ K, and $1.5 < \llg < 4.5$. For these stars, we have low-resolution SEGUE spectra, ($R \sim 2000$), photometry in the SDSS $ugriz$ bands, \cite{Schlegel98} reddening estimates and positional data from SDSS DR9 \citep[][]{SDSSDR9}. One star was removed from the sample, as it was flagged for strongly disagreeing observational information (very low quality measure $Q < -20$, cf. equation \ref{eq:qq}, resulting partly from a strong cosmic in the spectrum). Throughout the text we refer to the parameters from \cite{Allende08} as "AP08" and from SEGUE DR9 as "DR9". 

Two important remarks should be made about these comparison sets: While they are great tools for comparisons, they are subject to uncertainties and systematics that can exceed the quoted errors. Second, the quoted errors are fundamentally different from ours. They just report internal errors from pipelines or spectroscopic fitting routines, which are typically far smaller than realistic error estimates. The most extreme case is the SEGUE parameter pipeline. This is fundamentally different from our error calculations, which attempt to calculate all uncertainties.

\begin{figure*}
\begin{center}
\epsfig{file=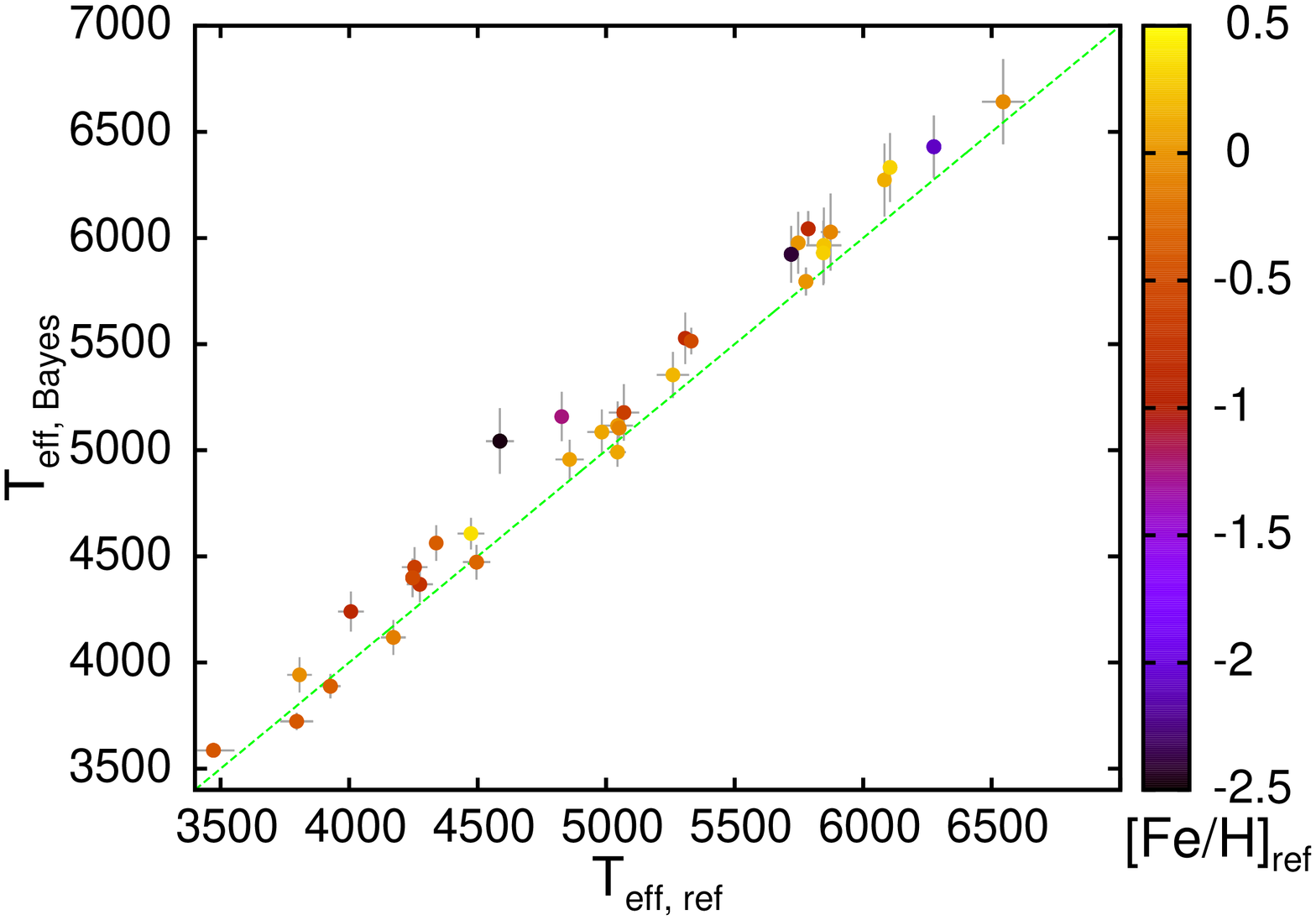,angle=-90,width=0.33\hsize}
\epsfig{file=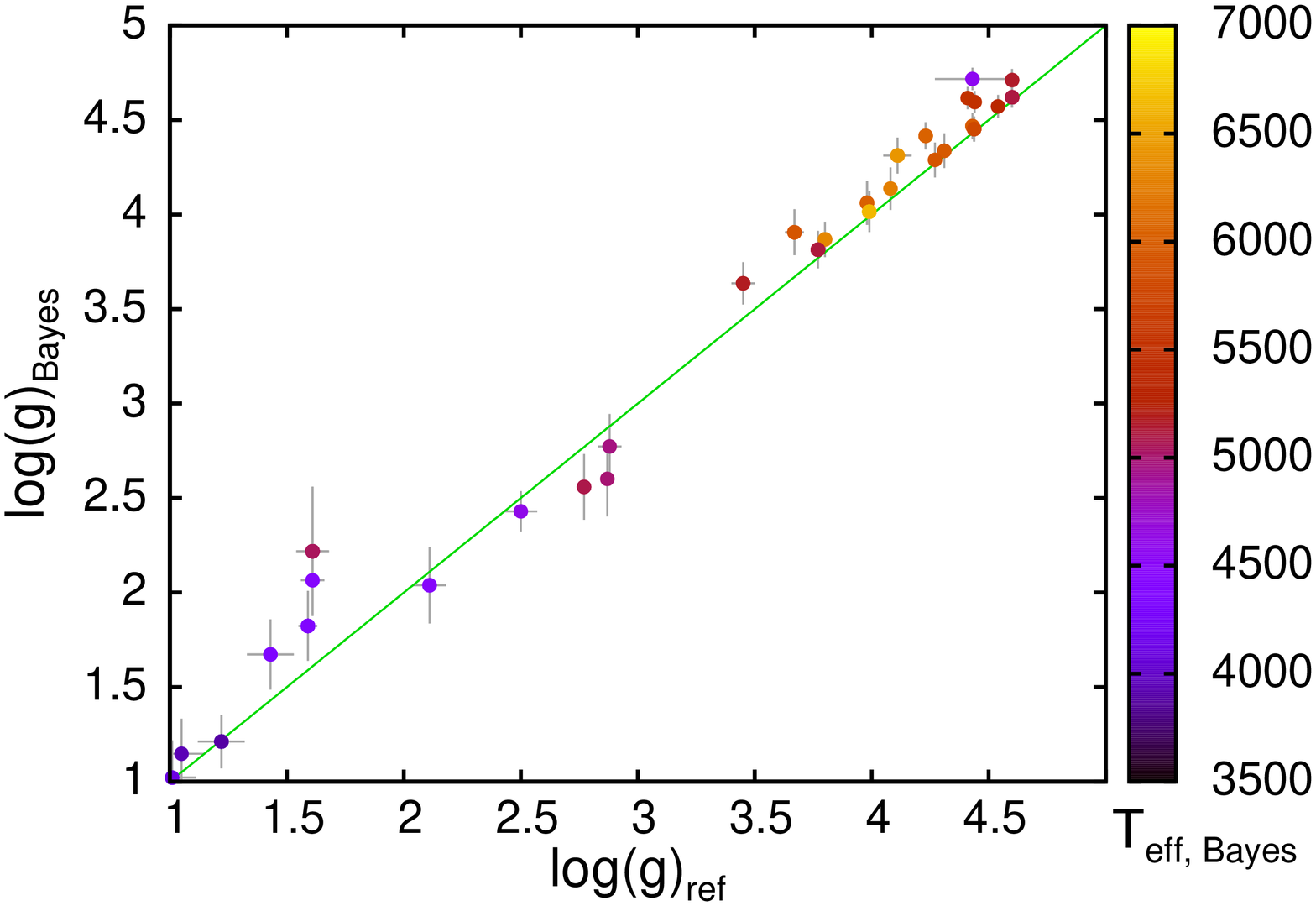,angle=-90,width=0.33\hsize}
\epsfig{file=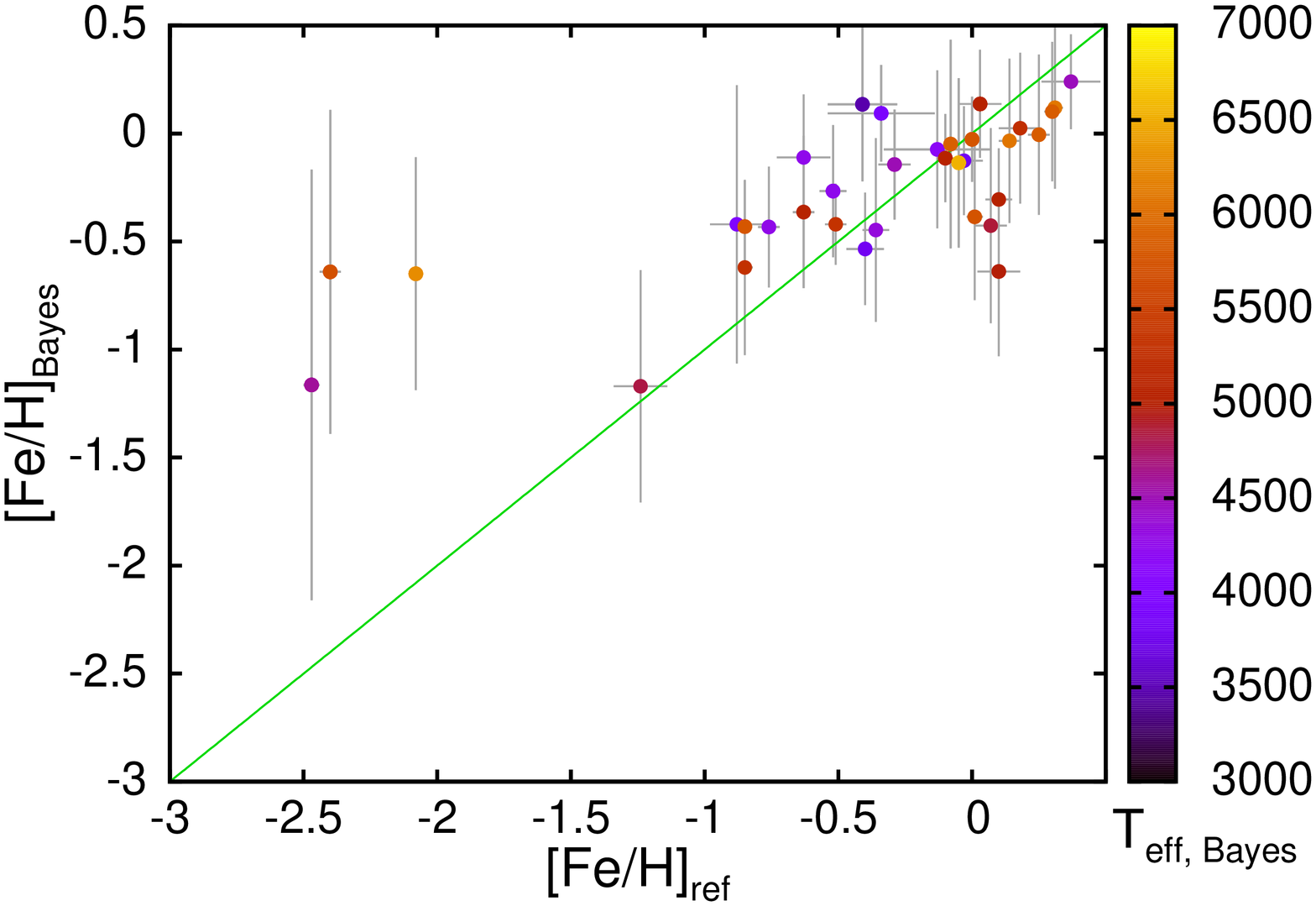,angle=-90,width=0.33\hsize}
\epsfig{file=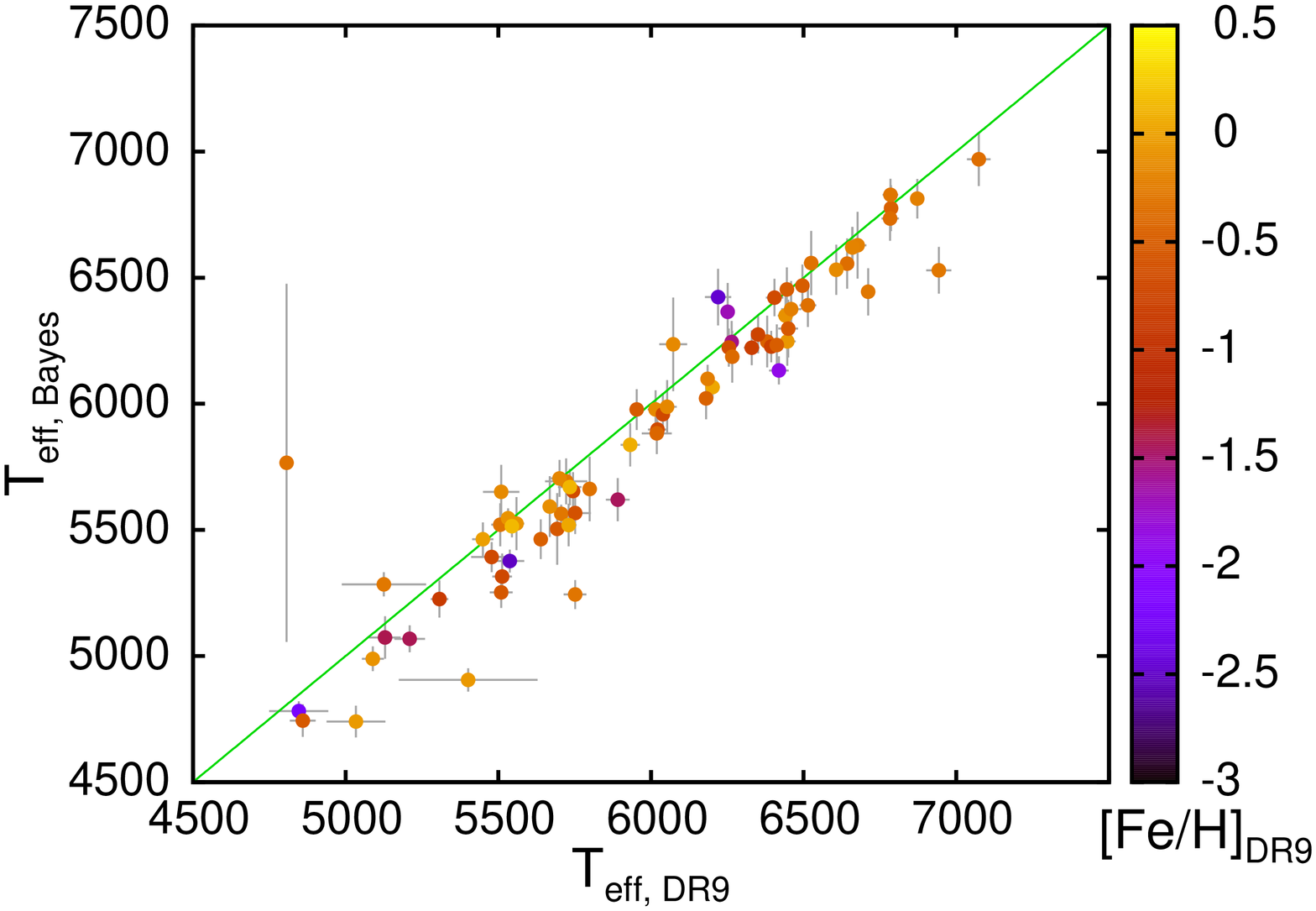,angle=-90,width=0.33\hsize}
\epsfig{file=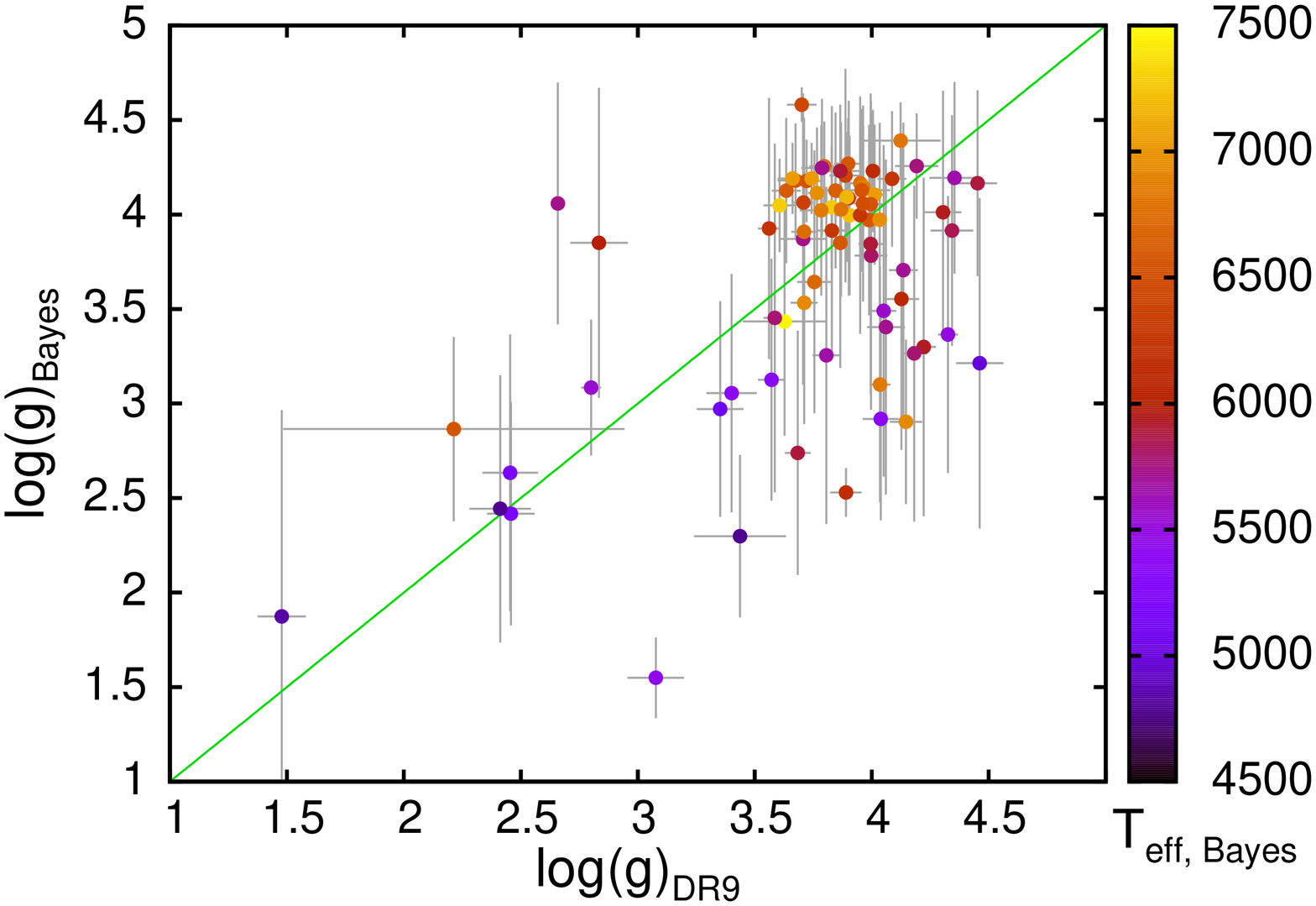,angle=-90,width=0.33\hsize}
\epsfig{file=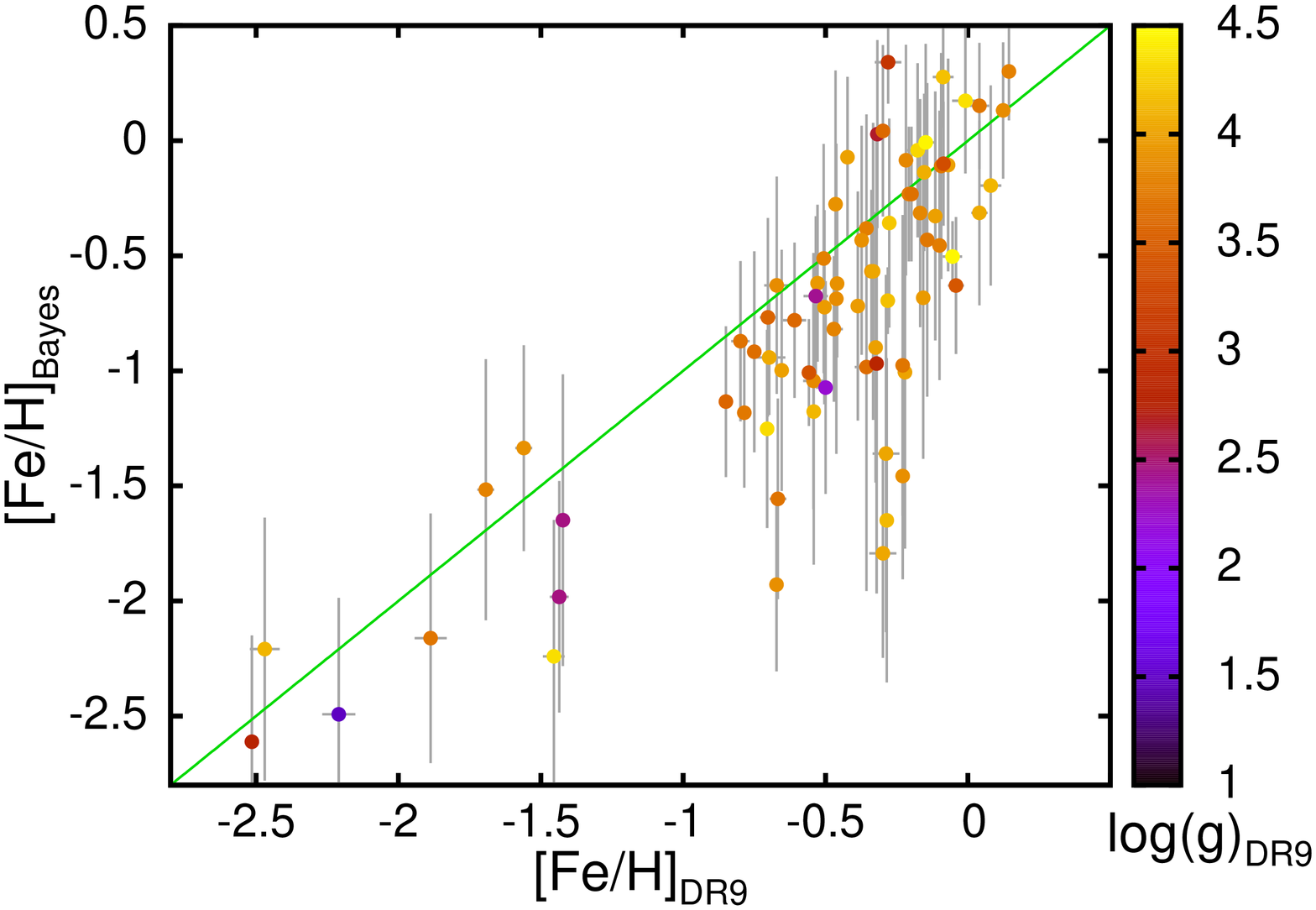,angle=-90,width=0.33\hsize}
\end{center}
\caption{Parameter estimates without using spectroscopic information. The top row shows our high resolution comparison sample, the bottom row shows our comparison with the SEGUE stars. Temperatures are fine in both occasions. In the absence of parallaxes, gravity information is marginal, leading to large uncertainties. Similarly, our high resolution sample has virtually no $U$-band photometry, such that metallicities are weakly determined. This pushes the expectation values strongly towards the middle.}
\label{fig:HRphoto}
\end{figure*}

\subsection{Photometry with and without Astrometry}

Before we explore the performance of the full algorithm against our reference samples, we test it for the simpler case, where spectroscopic information is not available.
For the vast majority of stars in the Galaxy, we will have no or very limited spectral information (e.g. 4MOST will cover only of order $\sim 2\%$ of the stars in the Gaia catalogues). However, we find that the Bayesian method is capable of deriving stellar parameters also when restricted to photometric and astrometric information.

In \figref{fig:HRphoto}, we compare temperatures, gravities and metallicities derived from Johnson photometry and parallaxes only with our high resolution reference sample (top row), and from SDSS photometry alone with values from SEGUE DR9 (bottom row).

For the high-resolution reference sample (top row) the applied photometry is not competitive with what can be expected from modern photometric surveys: For most stars we are restricted to Johnson $B,V,I$ colours at $\sim 0.03 \mag$ precision, and, due to their brightness, very uncertain $2MASS$ photometry. Nevertheless, the photometry gives a good handle on effective temperatures: while our temperatures are mildly higher than the reference, the random scatter is as low as $\sim 110 \K$. The excellent agreement for $\llg$ is a consequence of using Hipparcos parallaxes. We note that for most stars even uncertain parallaxes suffice to fix $\llg$, as they constrain the stellar branch, i.e. whether a star is on the main-sequence or e.g. on the red giant branch. Metallicities are per se very weakly determined with Johnson broad band filters, and particularly without decent $U$-band measurements. The large uncertainties concentrate the values towards the centre of our grid. This underlines the need for intermediate or narrow band photometric surveys to constrain stellar parameters. 

The precise SDSS photometry and the location of SDSS colour bands allow for a better handle on metallicities, as well as for good temperatures. In absence of parallax measurements, photometry alone offers a rough classification of stars, as seen in the bottom row of \figref{fig:HRphoto} with a rms scatter against the SEGUE parameter pipeline of around $0.5 \dex$. While there is significant photometric information, it is not strong enough to be insensitive to the priors. This motivates a closer look at the importance of our assumptions.

\begin{figure}
\begin{center}
\epsfig{file=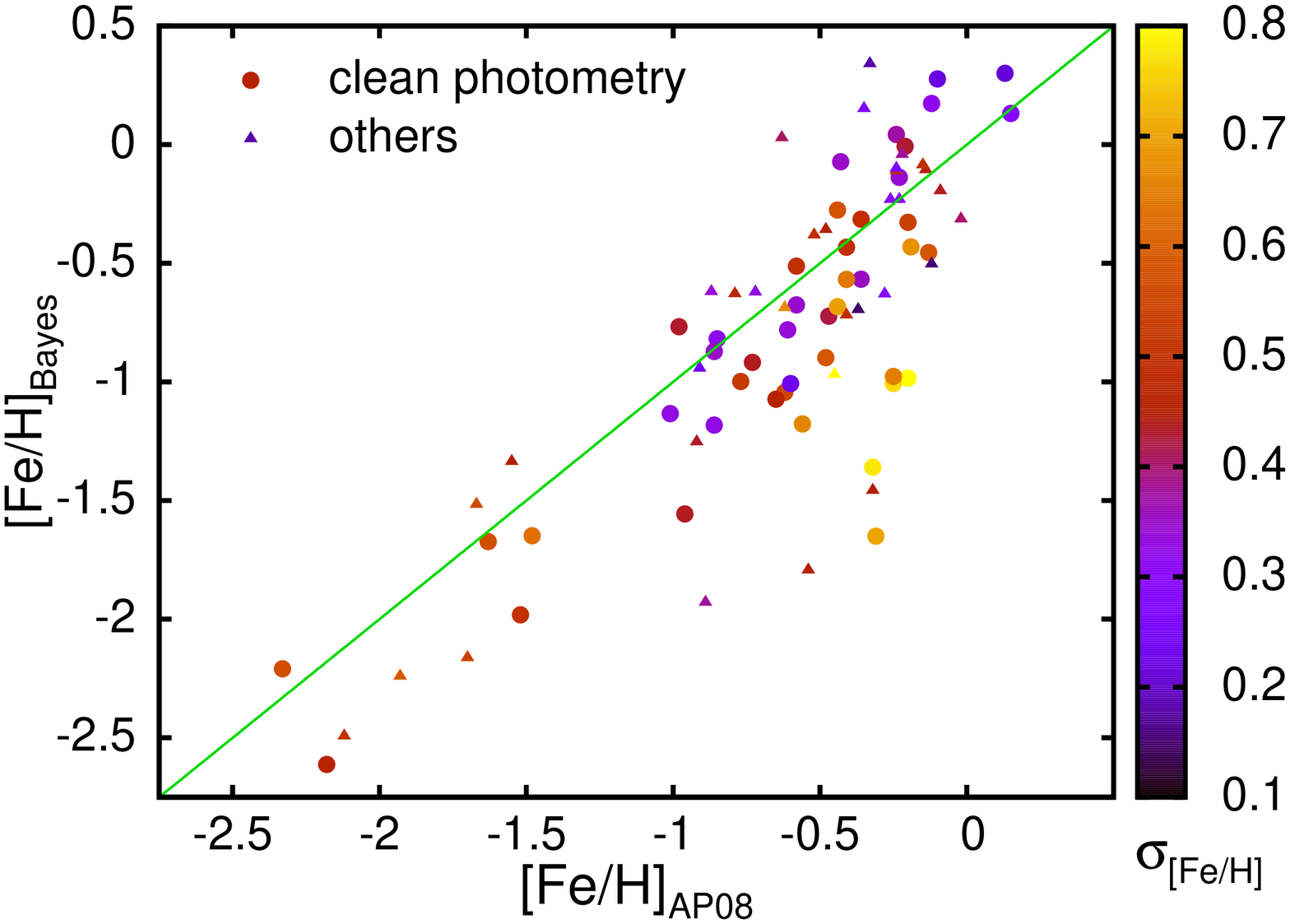,angle=-90,width=\hsize}
\epsfig{file=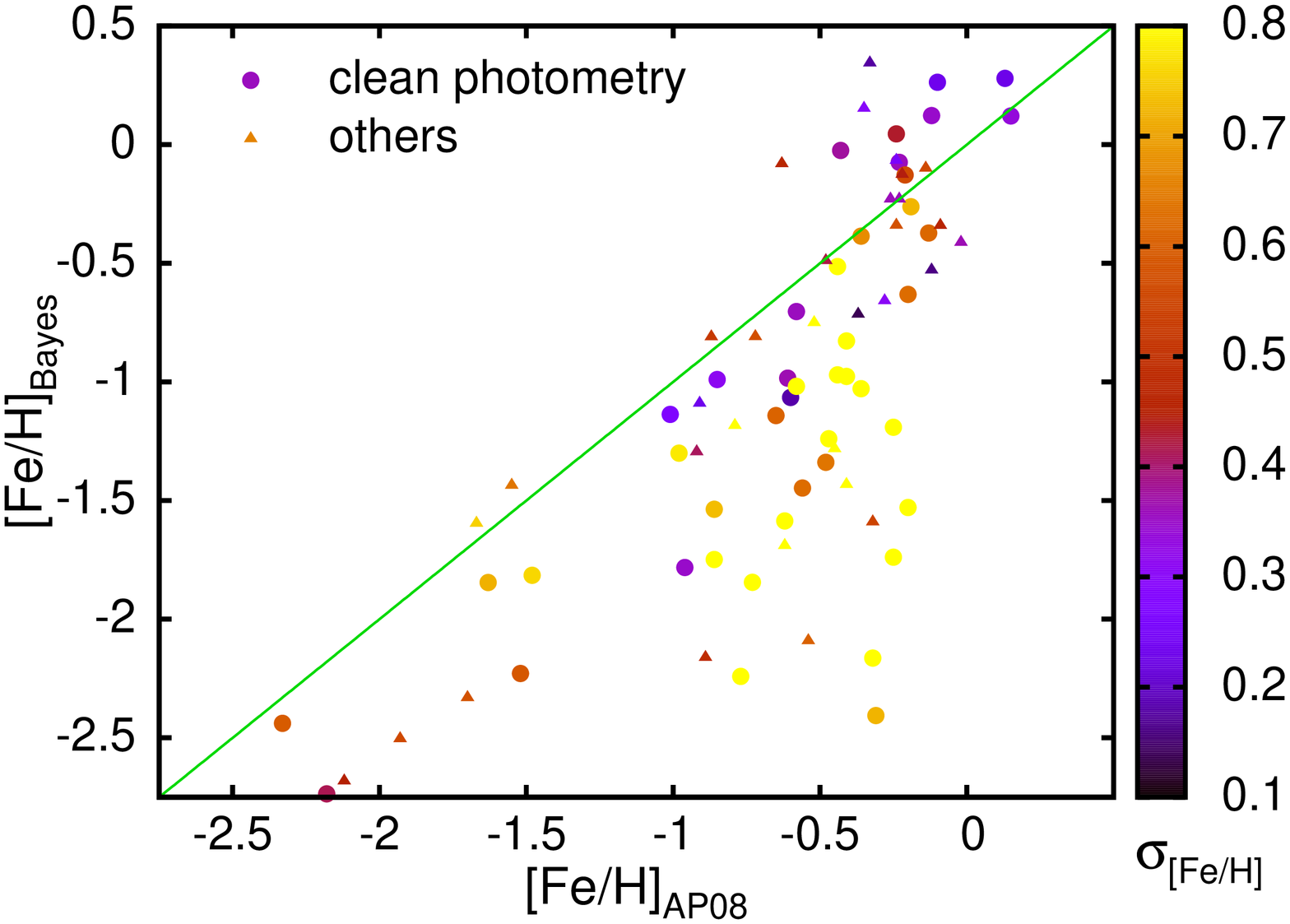,angle=-90,width=\hsize}
\end{center}
\caption{Photometric metallicities (expectation values of the posterior distributions) in the SEGUE/SDSS sample versus the determinations from Allende Prieto et al. (2008). Top panel includes the age prior from eq.\ref{eq:ageprior}, while the bottom panel does not, resulting in a bias towards younger ages. Colours code the standard deviation of the posterior distributions, capped at $0.8 \dex$. Stars with clean photometry according to the SDSS database are depicted with discs, while stars with problematic photometry are shown as smaller triangles.}\label{fig:photmets}
\end{figure}

In \figref{fig:photmets} we compare our photometric metallicities (y-axis) to the metallicities from \cite{Allende08} (x-axis; for the standard SEGUE DR9 comparison, see \figref{fig:HRphoto}) for the SEGUE sample, plotting stars with clean photometry with larger discs and stars with bad photometry with smaller triangles. Colours encode the error estimates from the Bayesian method. Evidently, there is enough information to constrain metallicities at least in the higher metallicity range to an accuracy of about $0.2 \dex$. Contrary to common derivations like \cite{Ivz08}, which fail at metallicities $\ge -0.5$ \citep[cf.][]{Arnadottir10}, our approach is valid throughout the entire metallicity range. However, it is important to realize how important the age prior becomes in this case. In the lower panel we show the same data with a fully flat age prior instead of using eq.~\ref{eq:ageprior}. This flat age prior implies a far larger uncertainty in the gravity of a star, which severely affects objects that cannot be clearly identified as subgiants, or main sequence stars. Via the degeneracy of $u-$band information, their potentially lower gravities allow for a wider range of (mostly lower) metallicities, which lowers the expectation values and boosts the error estimates. Despite this problem, the situation is far better than in the traditional approach: the classical metallicity calibrations like \cite{Ivz08} or \cite{An13} rely on stars falling not only on a fixed age bin, but also onto a single evolutionary sequence. This leads to a metallicity bias and overconfidence concerning the uncertainties. In contrast, the full Bayesian approach makes optimal use of all available colour information, accounts for all sources of uncertainty and allows to explore the effects of prior assumptions.

\begin{figure*}
\begin{center}
\epsfig{file=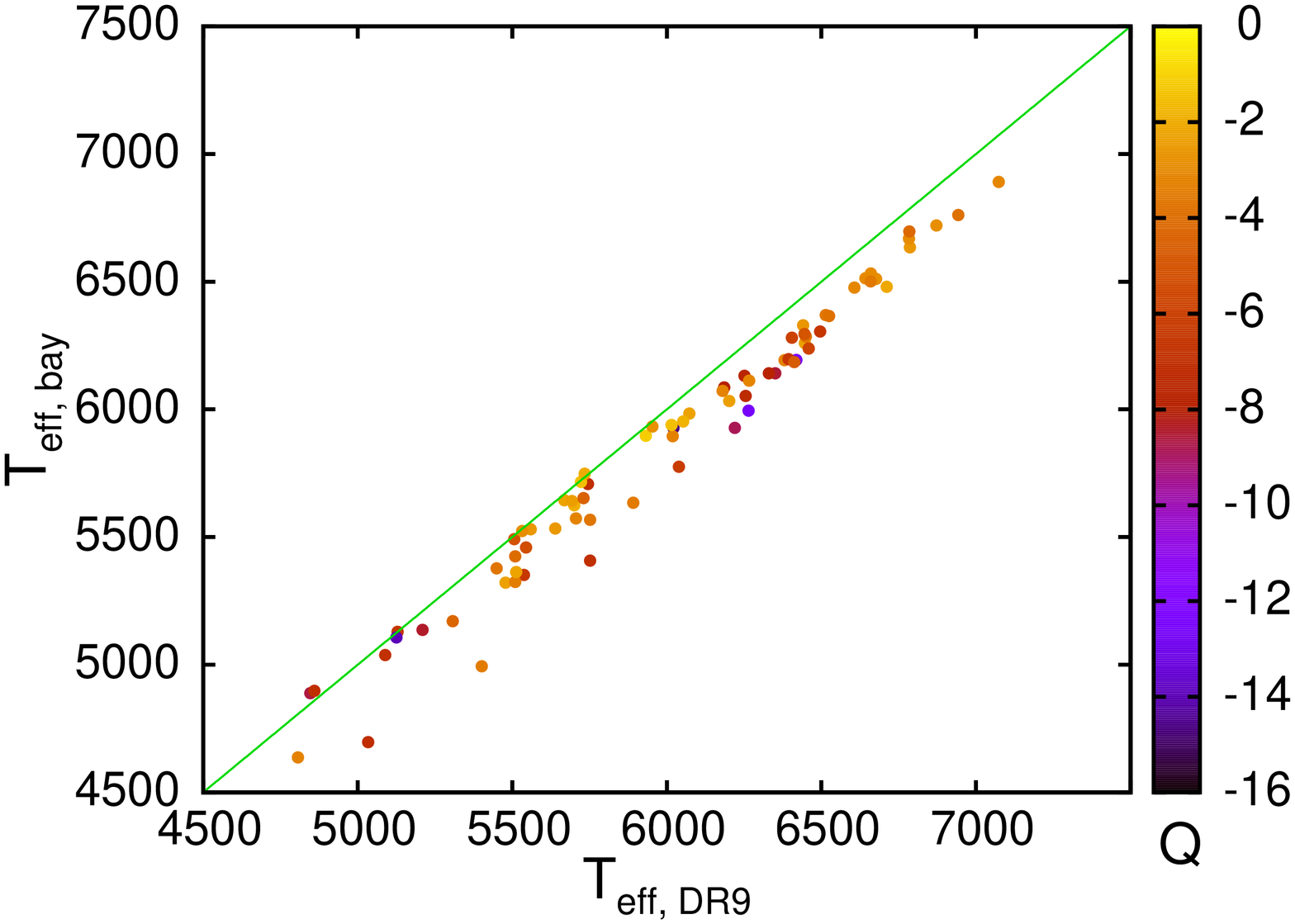,angle=-90,width=0.33\hsize}
\epsfig{file=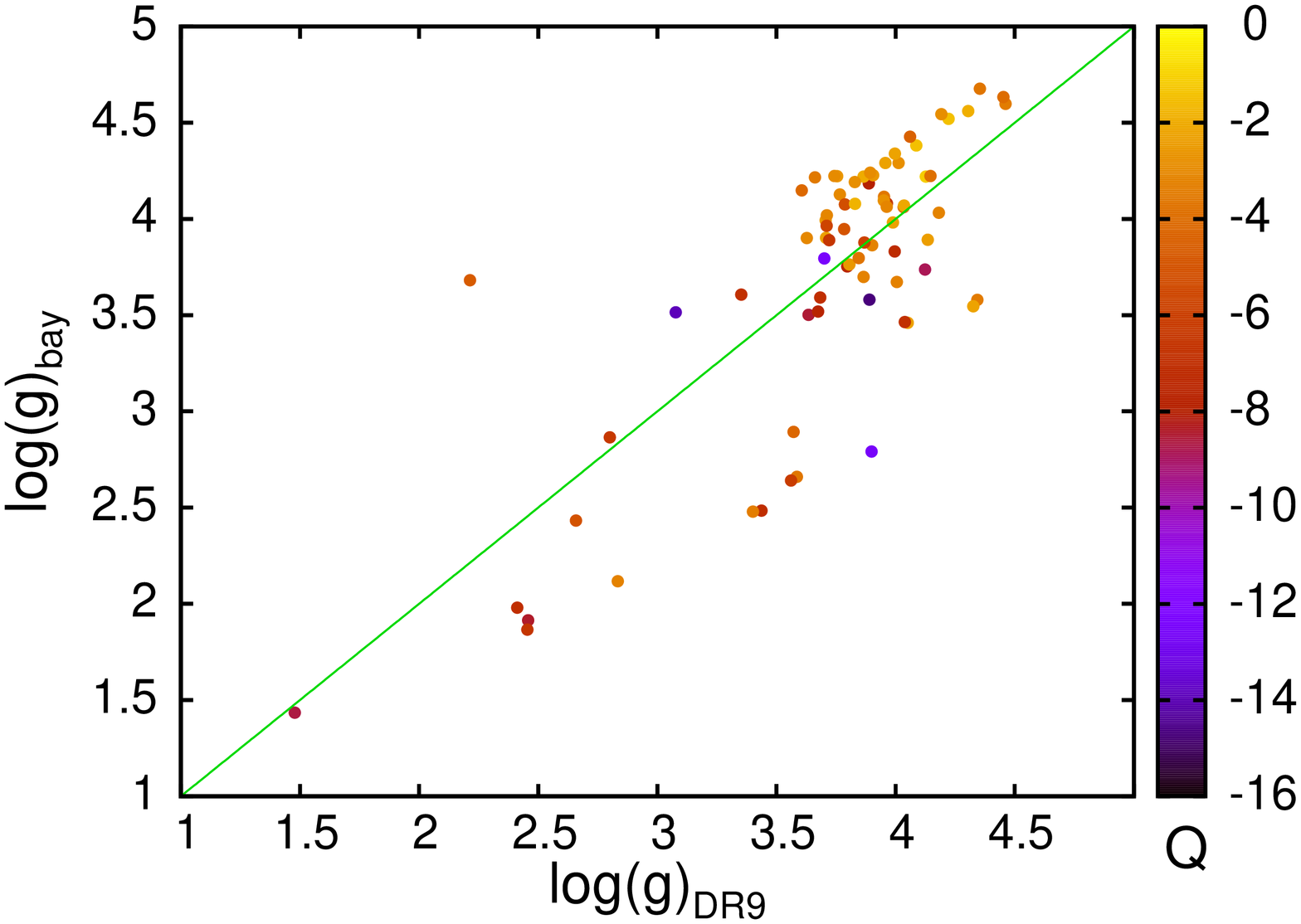,angle=-90,width=0.33\hsize}
\epsfig{file=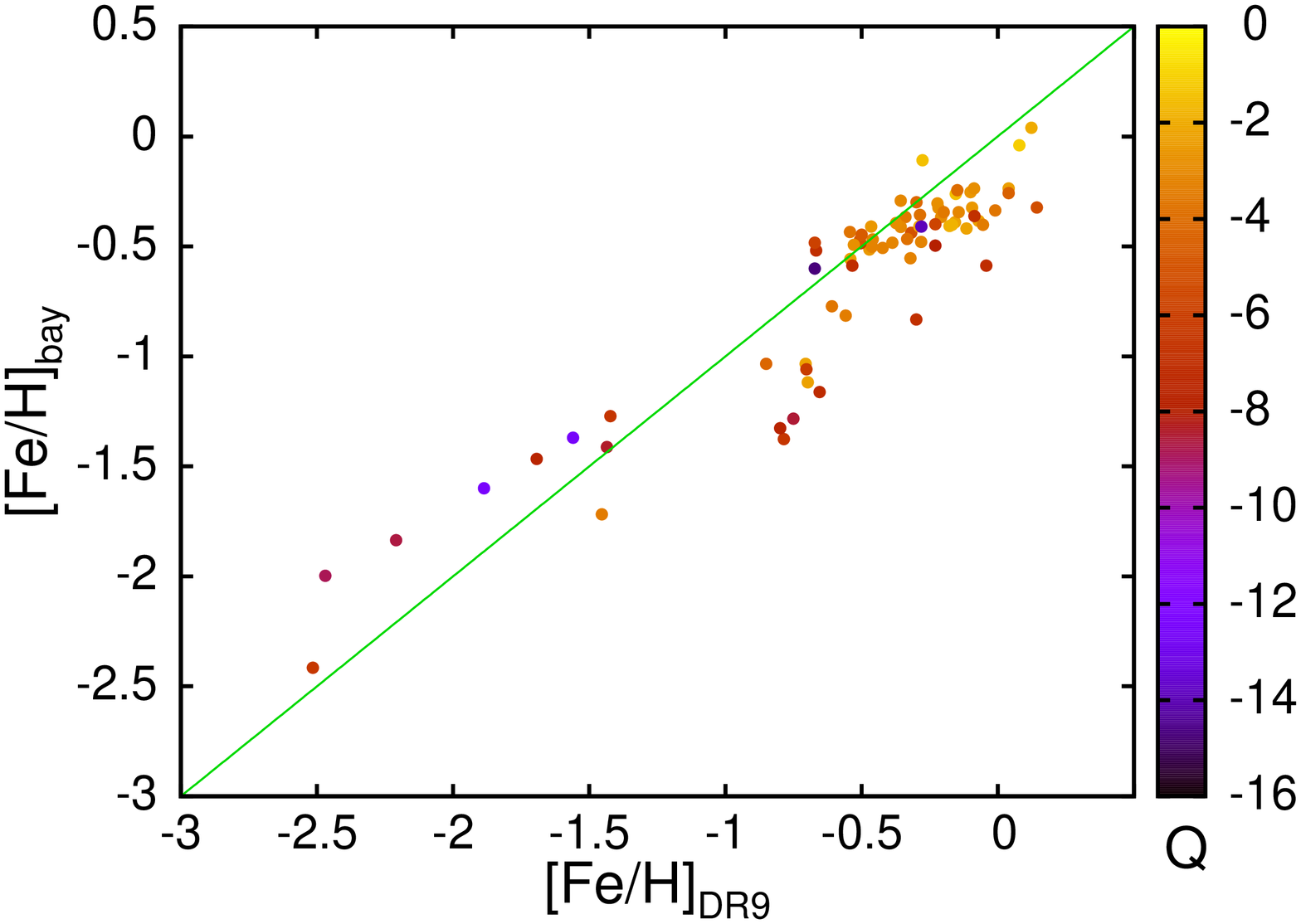,angle=-90,width=0.33\hsize}
\end{center}
\caption{Bayesian expectation values for the SEGUE sub-sample compared to the results from the SEGUE parameter pipeline. The colour codes the quality measure $Q$, as in \figref{fig:res2}.}\label{fig:res3}
\end{figure*}

\subsection{SDSS/SEGUE: Photometry and low-resolution Spectra}\label{sec:SEG}

\figref{fig:res3} shows the comparison of our parameter expectation values with the SEGUE DR9 data release. Colour codes the quality measure 
\begin{equation}\label{eq:qq}
Q = \log_{10}{\frac{\int P_{\rm ph, mod, pr, astr}\cdot P_{\rm sp}d\vX}{2\int P_{\rm ph, mod, pr, astr}^2 d\vX + \int P_{\rm sp}^2 d\vX}} $,$
\end{equation}
which gives a simplified indication on how well the spectroscopic PDF agrees with the remaining information.

Our temperatures are systematically colder than SEGUE DR9 by about $135 \K$. This is a consequence of our spectral and photometric $\Teff$ scales being $160 \K$ and respectively $80 \K$ colder, suggesting that SEGUE DR9 overestimates stellar temperatures. Our spectroscopic temperature scale is nearly identical with that of \citep[][hereafter AP08]{Allende08}, which is on average $\sim 170 \K$ below SEGUE DR9 derivations. The strength of our full approach becomes apparent in the residual scatter of the temperature values after correcting for the systematic offset: while spectroscopic and purely photometric temperatures give a residual rms of $\sim 138 \K$ and $\sim 174 \K$ relative to SEGUE DR9, the full approach excels with $87 \K$. 

The Bayesian gravities are systematically higher for suspected main sequence stars ($\llg > 4$ in both determinations), reflecting the systematic gravity underestimates of SEGUE DR9 in this range (also confirmed by SEGUE not matching expectations for the main sequence). The purely spectroscopic gravities of our method are significantly lower than DR9 and AP08 by $\sim 0.45 \dex$ in the intermediate and lower gravity range (using $\llg < 4.0$ in AP08). This is clearly identified as a bias, since the Bayesian approach reports too young ages, especially for several metal-poor stars. Though the Bayesian approach cannot completely eradicate a systematic bias in one of its inputs, it strongly reduces this problem by systematically increasing the surface gravities by an average of $0.25 \dex$ compared to the purely spectroscopic value. 

The Bayesian metallicity determinations for $\feh$ between $-2$ and $\sim -0.6 \dex$ are robust. However, metal-rich stars have a recognizable metallicity difference between our photometric and our spectroscopic determinations, with the latter being systematically lower. For the open cluster $M67$ \citep[$\feh \sim 0.0$ or $\sim 0.05$][]{Magic10, Gratton00} our spectroscopy alone gives $\feh ~ -0.17$ versus a photometric estimate of $\sim 0.05$. As in the case of the gravities, the Bayesian method partly mitigates this problem: photometric metallicities in this range push the combined estimates towards higher values; however, due to the intrinsic uncertainty of $0.2 \dex$, the corrections are minor. This also shows the importance of fair error assessment: overconfident, i.e. too small, error estimates from spectroscopy prevent a stronger correction of the value by the photometric information, which has intrinsic uncertainties of $\sim 0.25 \dex$ in this range. Tests show stability of our results down to a signal to noise ratio of $\sim 30$ and checks on the continuum placement yielded no conclusive evidence. It is very likely that a finer resolution of the grid of synthetic spectra, its extension to a larger wavelength coverage\footnote{Currently we effectively use less than $1\%$ of the SEGUE spectral range.}, and allowing for $\alpha$-enhancement will solve the problem. This work is in progress and will be presented in a future paper dealing specifically with the analysis of SEGUE spectra.

The most important result is, that even with systematic biases present in the inputs, the Bayesian method itself remains robust, i.e. other parameters are not strongly affected, and the solutions are pushed towards a significantly less biased result.

\subsection{Photometry, Parallaxes and high-resolution Spectra}\label{sec:highres}
\begin{figure*}
\begin{center}
\epsfig{file=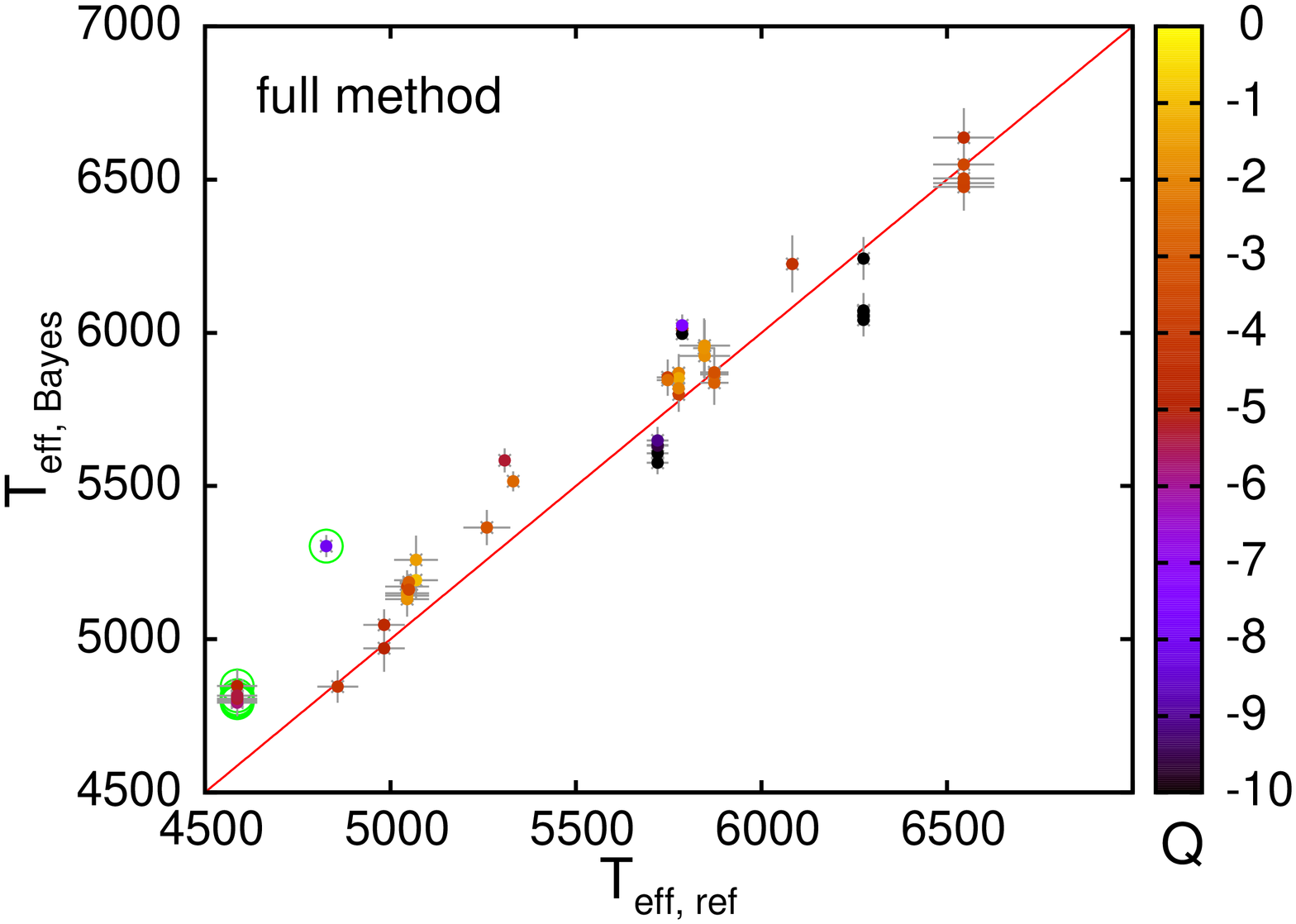, angle=-90,width=0.33\hsize}
\epsfig{file=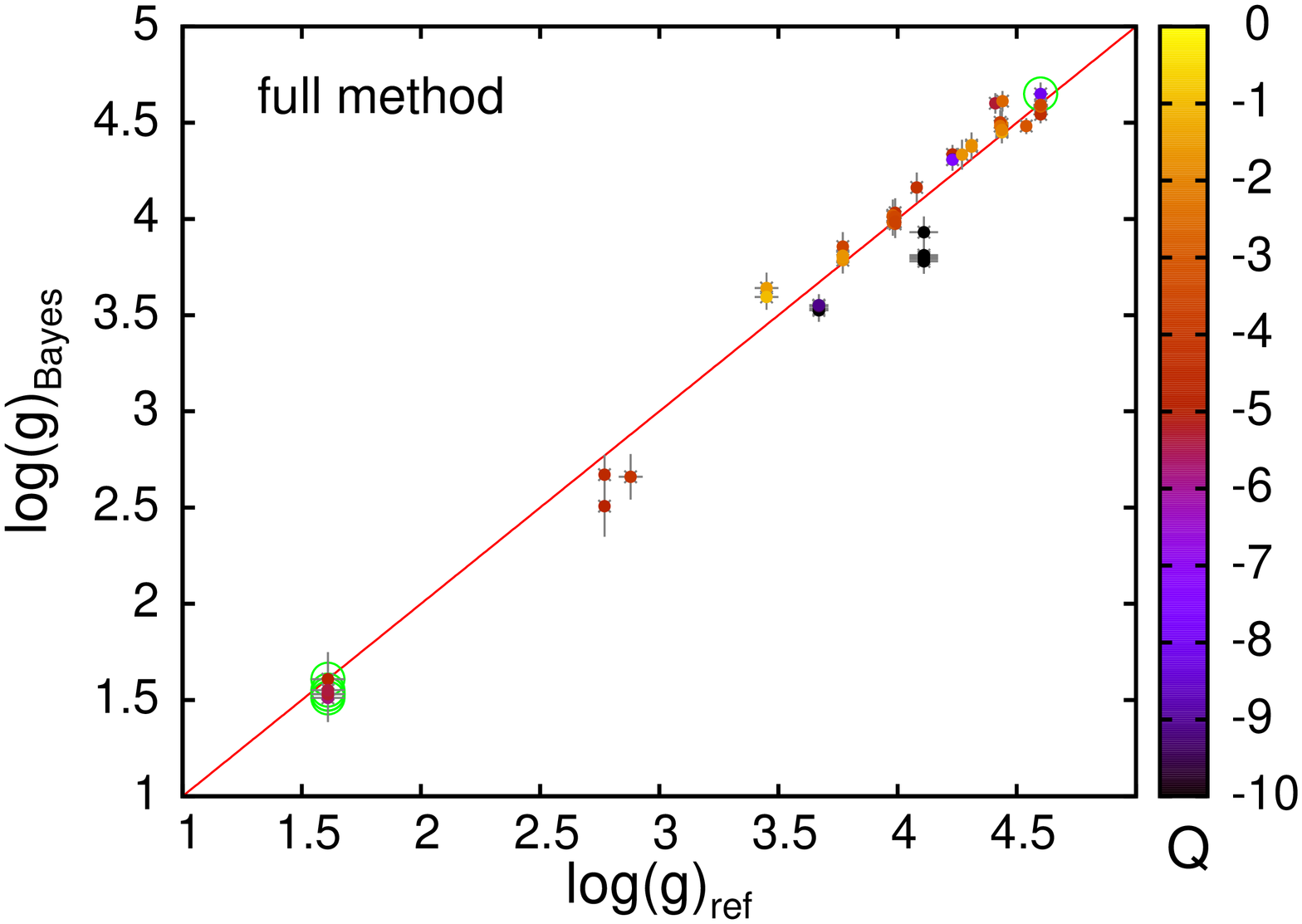,angle=-90,width=0.33\hsize}
\epsfig{file=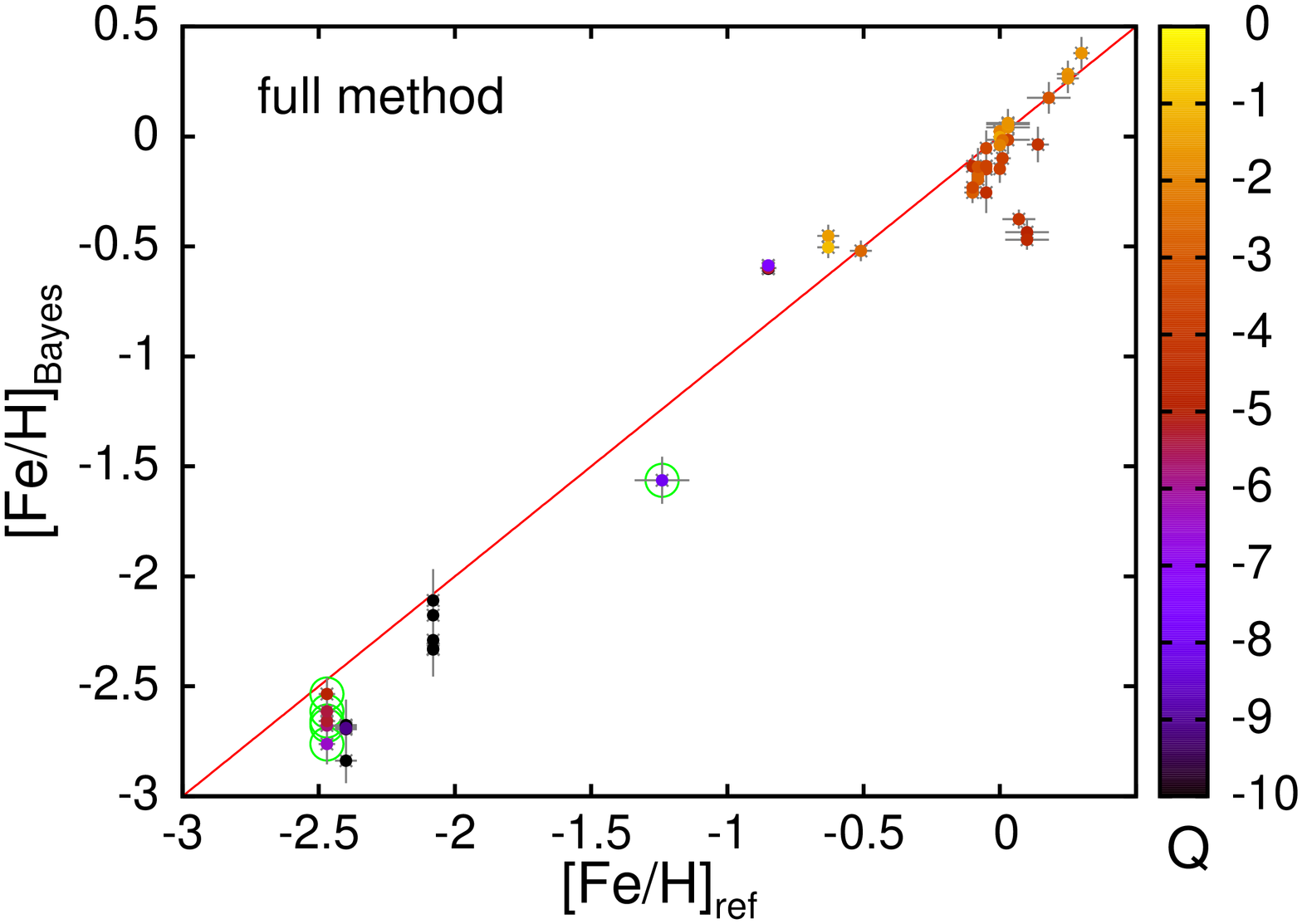,  angle=-90,width=0.33\hsize}
\epsfig{file=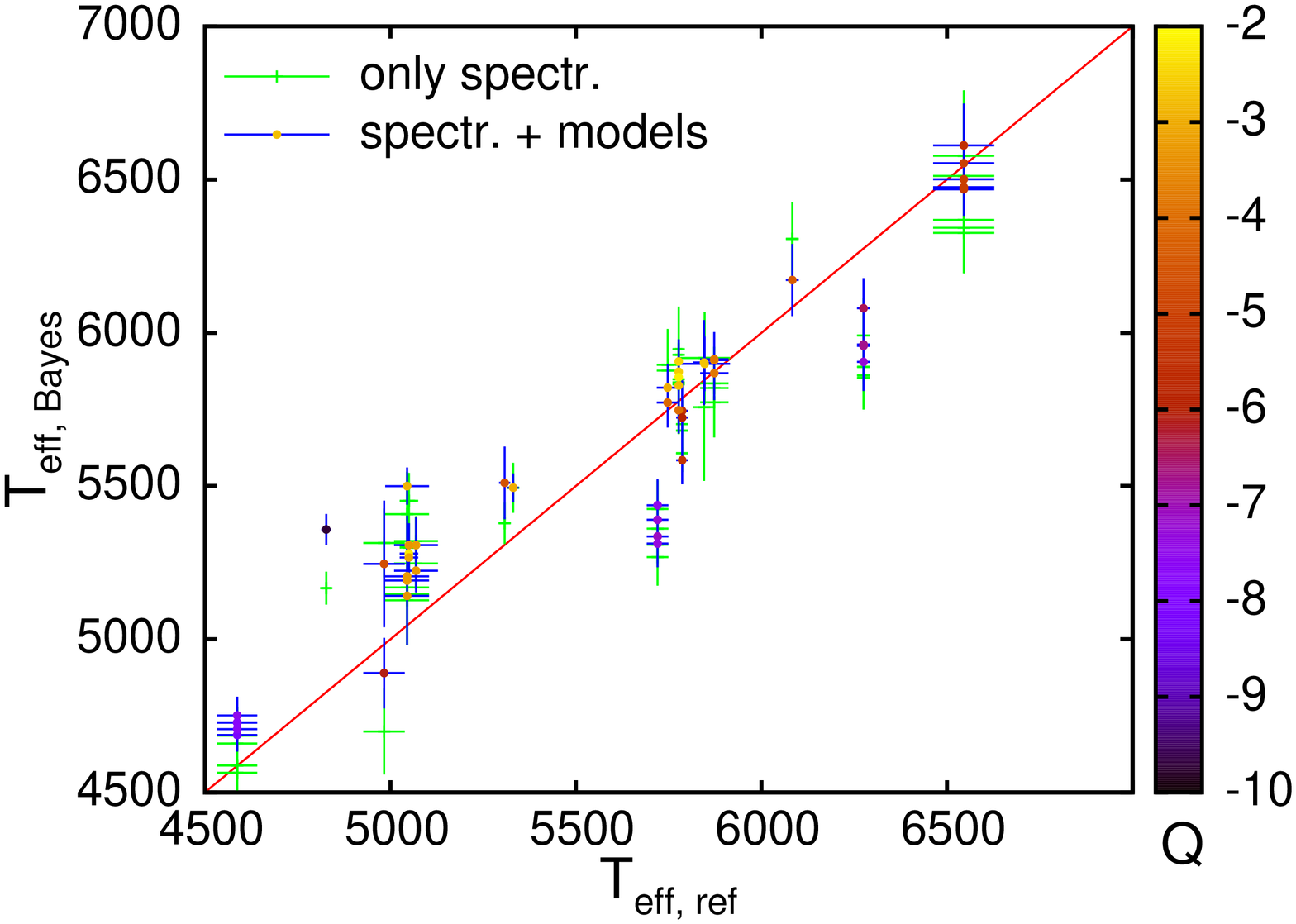, angle=-90,width=0.33\hsize}
\epsfig{file=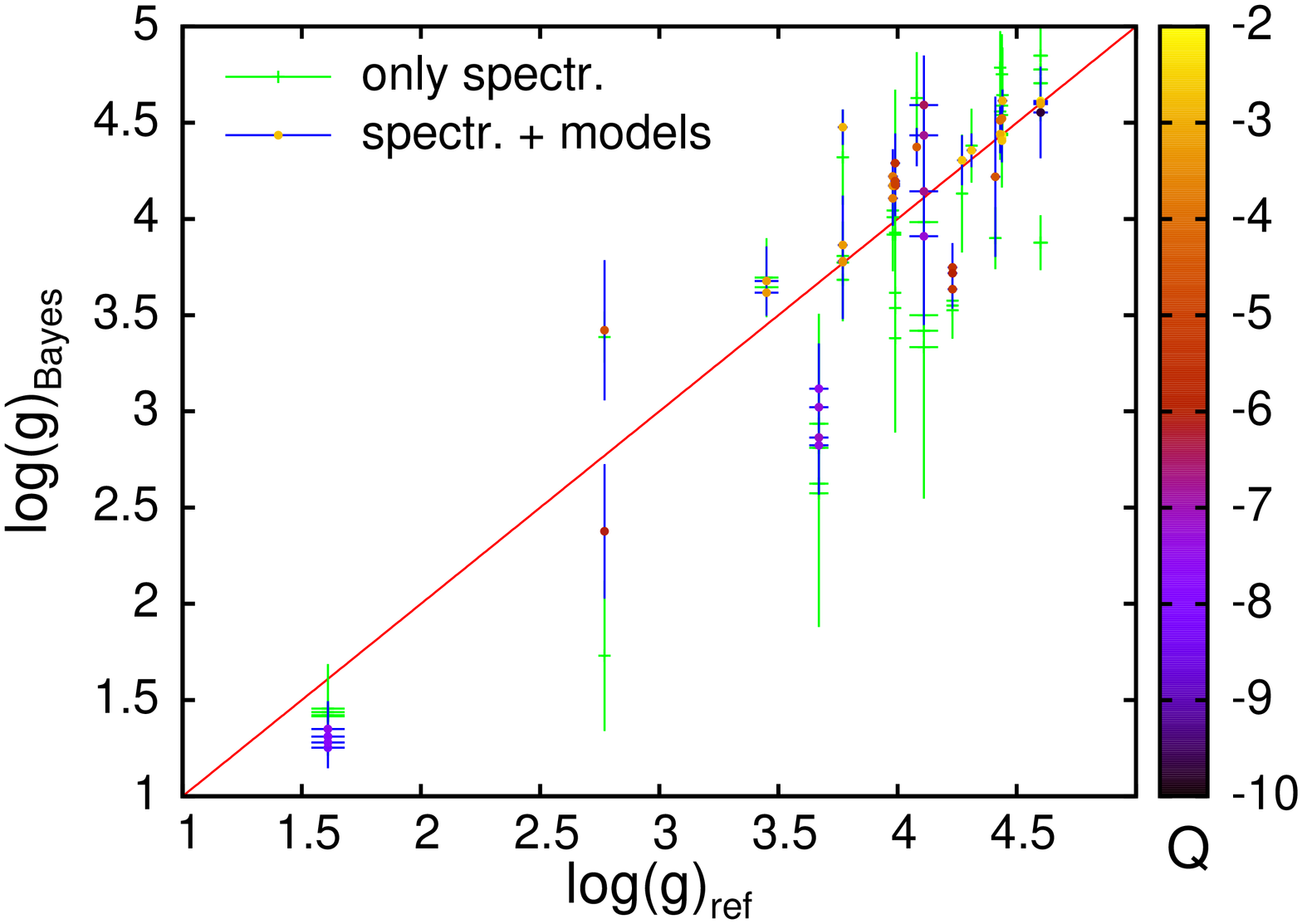,angle=-90,width=0.33\hsize}
\epsfig{file=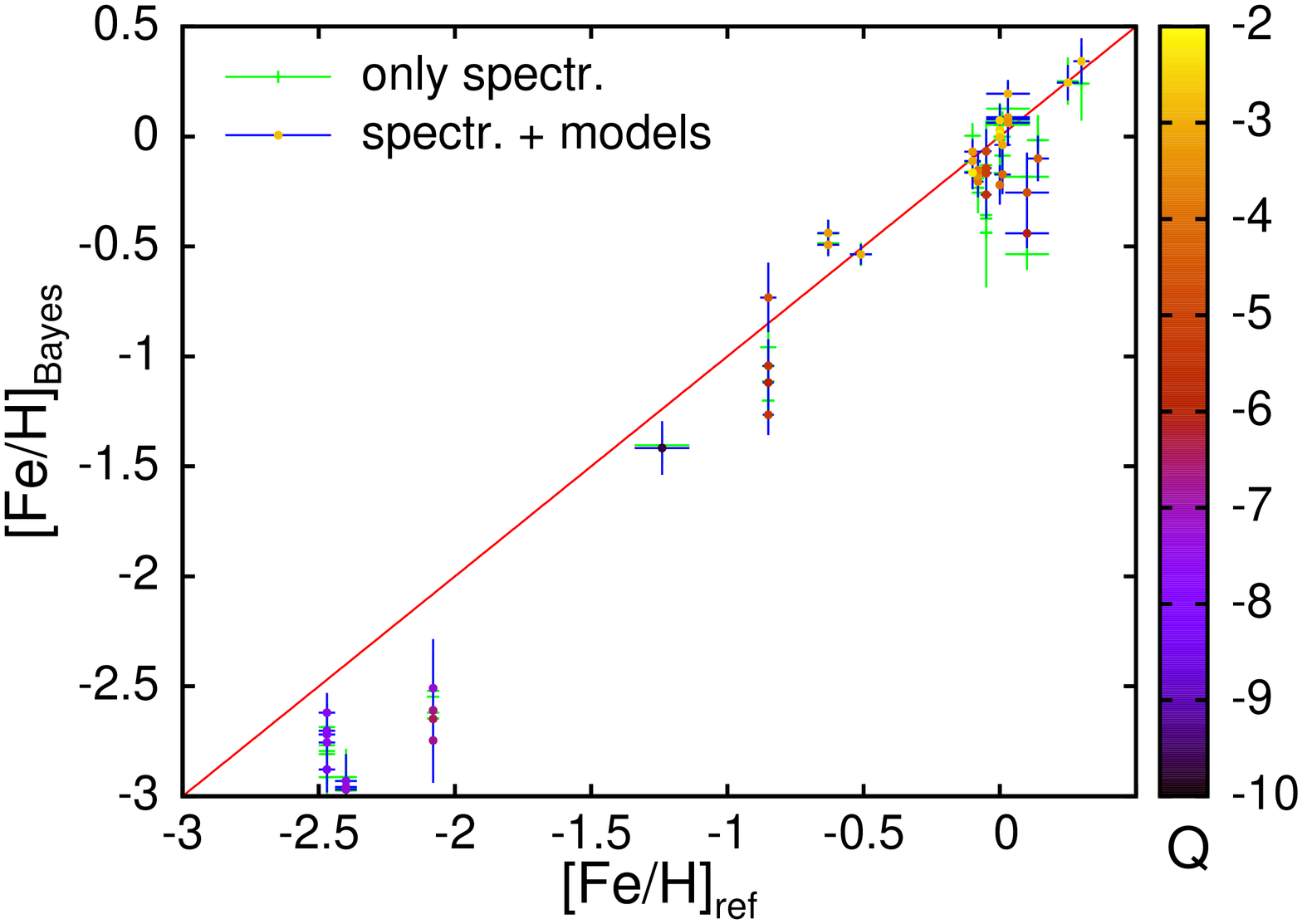,  angle=-90,width=0.33\hsize}
\end{center}
\caption{Expectation values from our Bayesian parameter determinations versus the reference values for stars with high-resolution spectra. In the top row we show the full Bayesian determinations using all available information, i.e. parallaxes, photometry, spectroscopy and stellar model compared to reference values from Heiter et al. (in prep.), which are derived from interferometry, asteroseismology and parallaxes. The bottom row depicts the parameter expectation values from spectroscopy alone (green error bars) and from spectroscopy + the model prior (but no photometry or astrometry).}
\label{fig:res2}
\end{figure*}

Figure \ref{fig:res2} compares the reference parameters (x-axis) to
the expectation values from our full Bayesian analysis (y-axis). Again colours encode the value of the quality measure $Q$ from equation \ref{eq:qq}. We also give statistics in Table \ref{tab:highrescomp}.

Currently, our spectral grids do not cover stars with $\Teff < 4400 \K$, $\logg < 1.4$ and assume a slow stellar rotation of $1 \kms$, which is typical for most G and K stars \citep[][]{Fuhrmann04}. Hence, for the spectroscopic comparison sample we have to exclude stars with $\Teff > 4500 \K$ and drop the fast rotating stars $\eta$ Bootis and $\mu$ Leonis, which have $v \sin i \sim 15 \kms$ and $\sim 5 \kms$. We also remove $\xi$ Hya due to contradictory results from different astroseismic derivations \citep[][]{Stello06}. This leaves $20$ stars with $53$ spectra. The two metallicity outliers at high metallicity in \figref{fig:res2} are $\epsilon Vir$ and $\beta$ Gem. Both have very high
macro-turbulence values \citep[$\sim 5 \kms$][]{Hekker07}, which contradict the
current assumptions of our spectral pipeline.

\begin{table}
\begin{tabular}{l|cc|cc}
\multicolumn{5}{c} {\bf Comparisons to reference sample} \\
parameter & $\Delta \mu$ & $\sigma$ & $\Delta \mu_{\rm astros.}$ & $\sigma_{\rm astros.}$\\
\hline \\
$\Teff / \K$ & $(65 \pm 19)$ & $141$ & $(69 \pm 13)$ & $66$ \\
$\logg / \dex$ & $(-0.024 \pm 0.017) $ & $0.13 $ & $(0.031 \pm 0.009) $ & $0.046 $ \\
$\feh / \dex$ & $(-0.099 \pm 0.026)$ & $0.19 $ & $(-0.049 \pm 0.014) $ & $0.07 $ \\
\end{tabular}
\caption{Differences in the mean expectation values of our sample minus the reference sample $\Delta \mu$ and rms scatter of $\sigma$ for the entire reference sample (left two columns) and the subsample with astroseismic determinations. While the astroseismic subset has competitive accuracy and precision, the remainder of the reference sample strongly scatters against our values.}\label{tab:highrescomp}
\end{table}

From Table \ref{tab:highrescomp}, it is apparent that if we confine the sample to the subset with astroseismic determinations, the random mean scatter in all quantities is reduced, by more than half. This implies that only the astroseismic subset can match or exceed our precision, while the Bayesian method is clearly superior to the traditional analysis on comparable data. Some of the reference gravities were derived from Hipparcos parallaxes, which should make them similar to our results. In this case, our fully Bayesian determinations in gravity appear more reliable than the less sophisticated reference because they also take into account physical information from colours and spectra.  Interferometric $\Teff$, although they are usually taken to be mildly model-dependent, still require an estimate of limb darkening and bolometric fluxes. The former are determined with 1D LTE model atmospheres, while \cite{Chiavassa} showed that 3D hydrodynamical models predict different centre-to-limb variation, which may cause systematic biases in angular diameter estimates. Bolometric fluxes are estimated by interpolating between observed photometric magnitudes with the help of theoretical spectra, giving rise to another systematic uncertainty.

It is instructive to compare the full method results to the spectroscopic results. In the bottom row of \figref{fig:res2} we show expectation values and parameter uncertainty from purely spectroscopic information (green error bars) and when using spectroscopy plus the model prior (coloured points with blue error bars). Spectroscopic surface gravities alone are generally too low by $(0.17 \pm 0.06) \dex$ with a residual scatter of about $0.4 \dex$ compared to the full solution \citep[see][for discussion of similar spectroscopic underestimates]{Ruchti13}. Using spectra in combination with stellar evolution models in the Bayesian framework, but excluding the
parallax and photometric information, improves the residual scatter to $\sim 0.3 \dex$ and fully removes the systematic offset. Hence, while spectroscopic information alone cannot compete with astrometric information, it gives sufficient information on surface gravity to allow for decent values derived by the Bayesian framework.

In Table \ref{tab:allhighres} we provide the stellar parameters and ages from the full Bayesian method. When more than one spectrum is present for a star, we provide the weighted average of the expectation values and errors (we have to assume that the errors between the single determinations are highly dependent) for single spectra. Where no spectral information is available, we fill in the results from the combination of photometry and parallax measurements.

\begin{figure}
\begin{center}
\epsfig{file=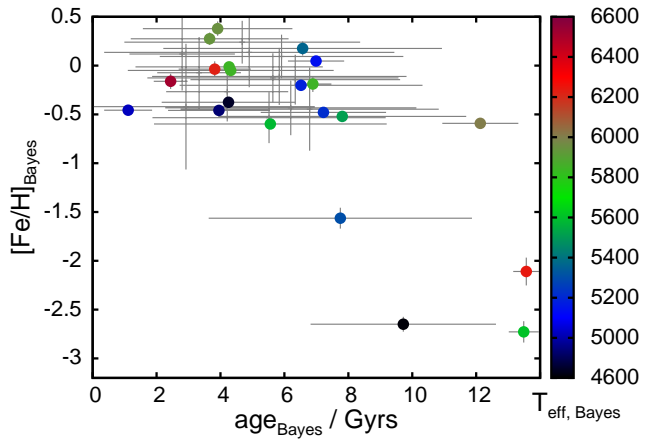,angle=-90,width=\hsize}
\end{center}
\caption{Ages and metallicities for the high-resolution reference sample with Bayesian stellar parameters. Colours encode the temperature estimate. For better visibility we merged the different values of each star according to Table \ref{tab:allhighres}.}
\label{fig:AMR}
\end{figure}

\begin{figure}
\begin{center}
\epsfig{file=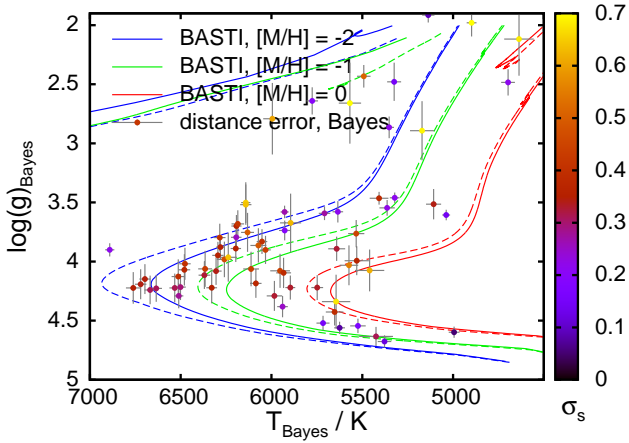,angle=-90,width=\hsize}
\end{center}
\caption{Distance error estimates for the Bayesian method in the temperature-gravity plane. Evidently, distance errors and classification uncertainties are not a clear function of magnitude. While clearly identified main sequence stars offer the best accuracy, even stars with high gravity estimates can have low confidence in their actual classification and distance estimates.}\label{fig:HRdist}
\end{figure}

\subsection{Temperature-gravity Plane, Ages and Distances}\label{sec:adv}

\begin{figure*}
\begin{center}
\epsfig{file=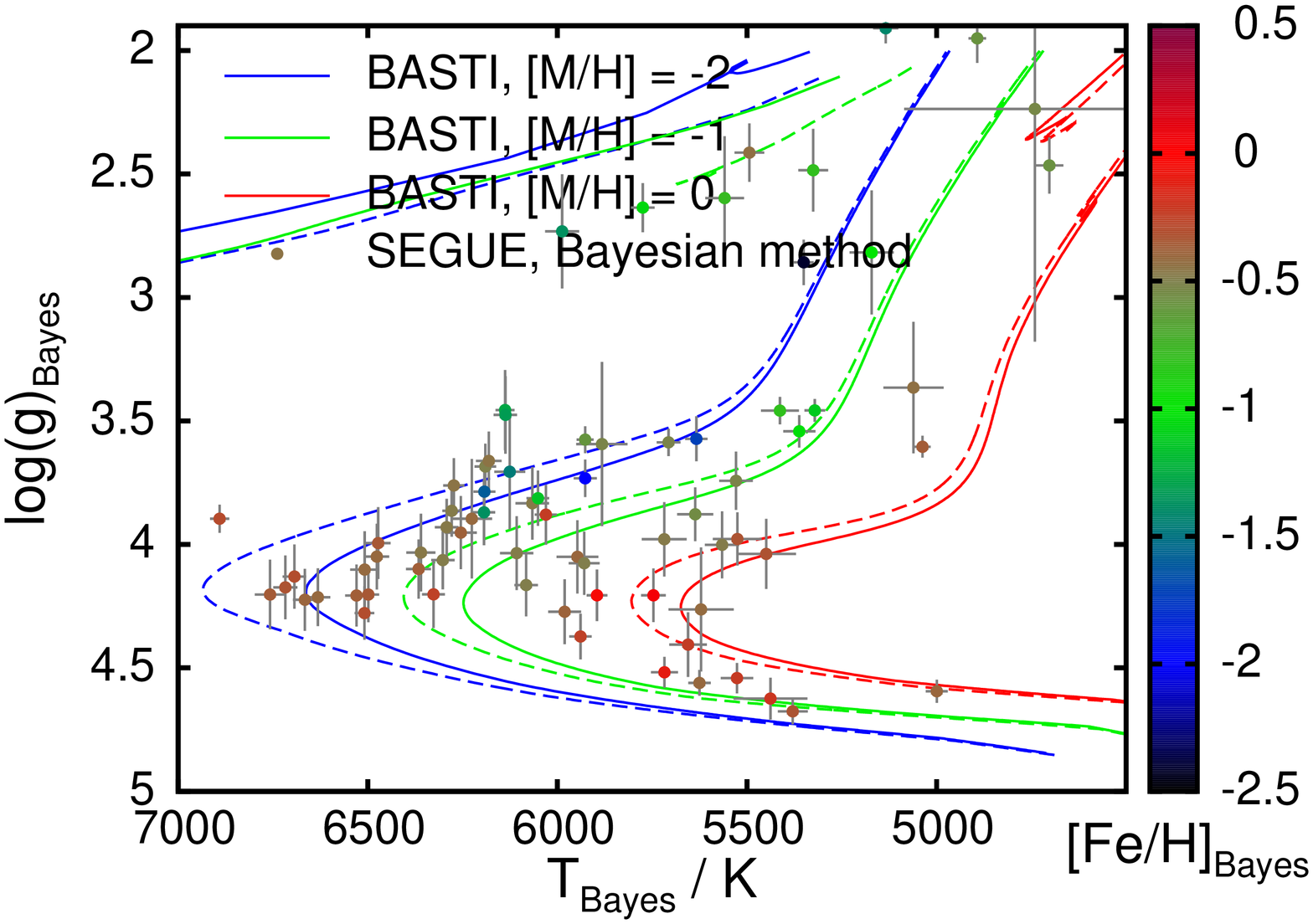,angle=-90,width=0.48\hsize}
\epsfig{file=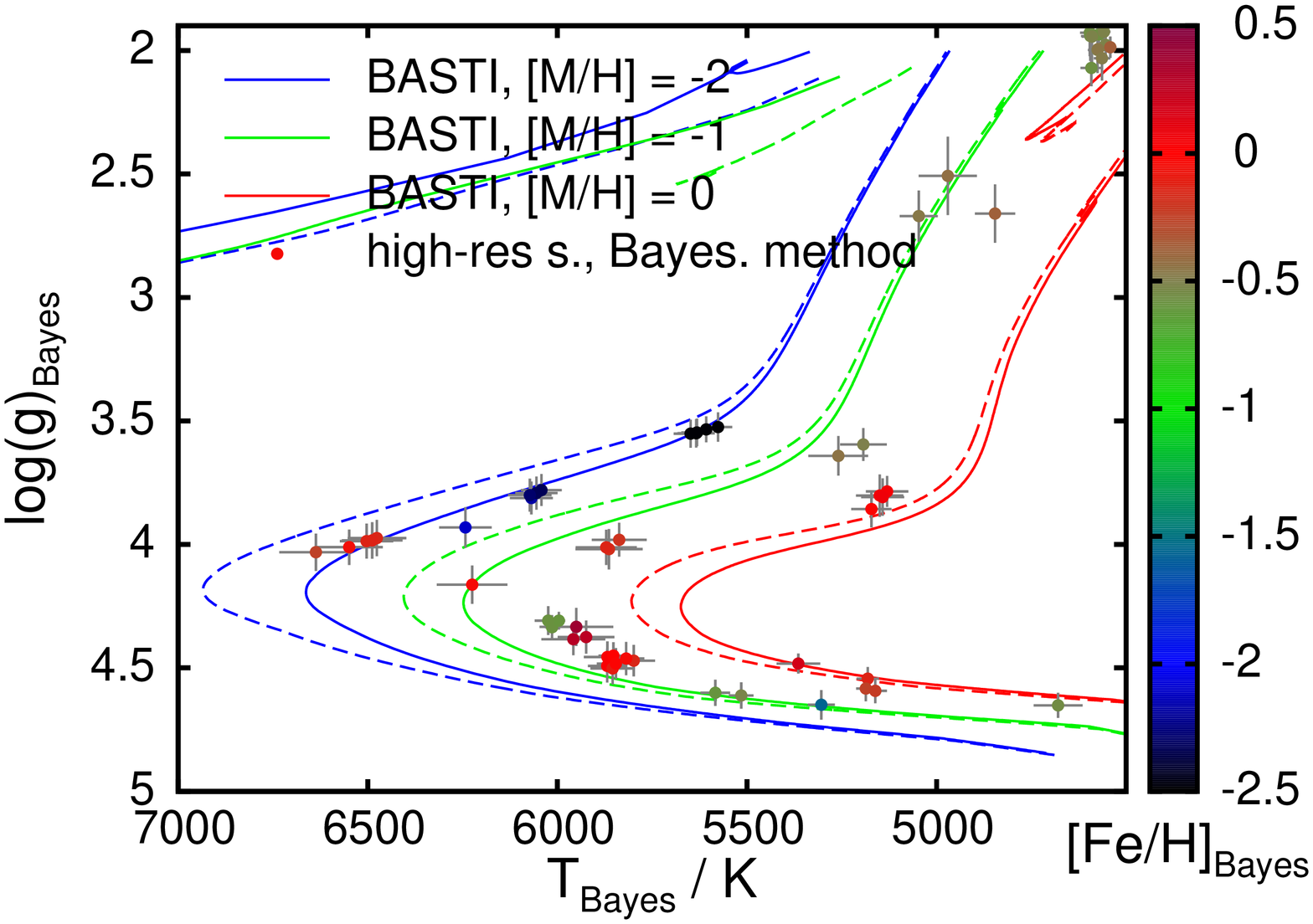,angle=-90,width=0.48\hsize}
\epsfig{file=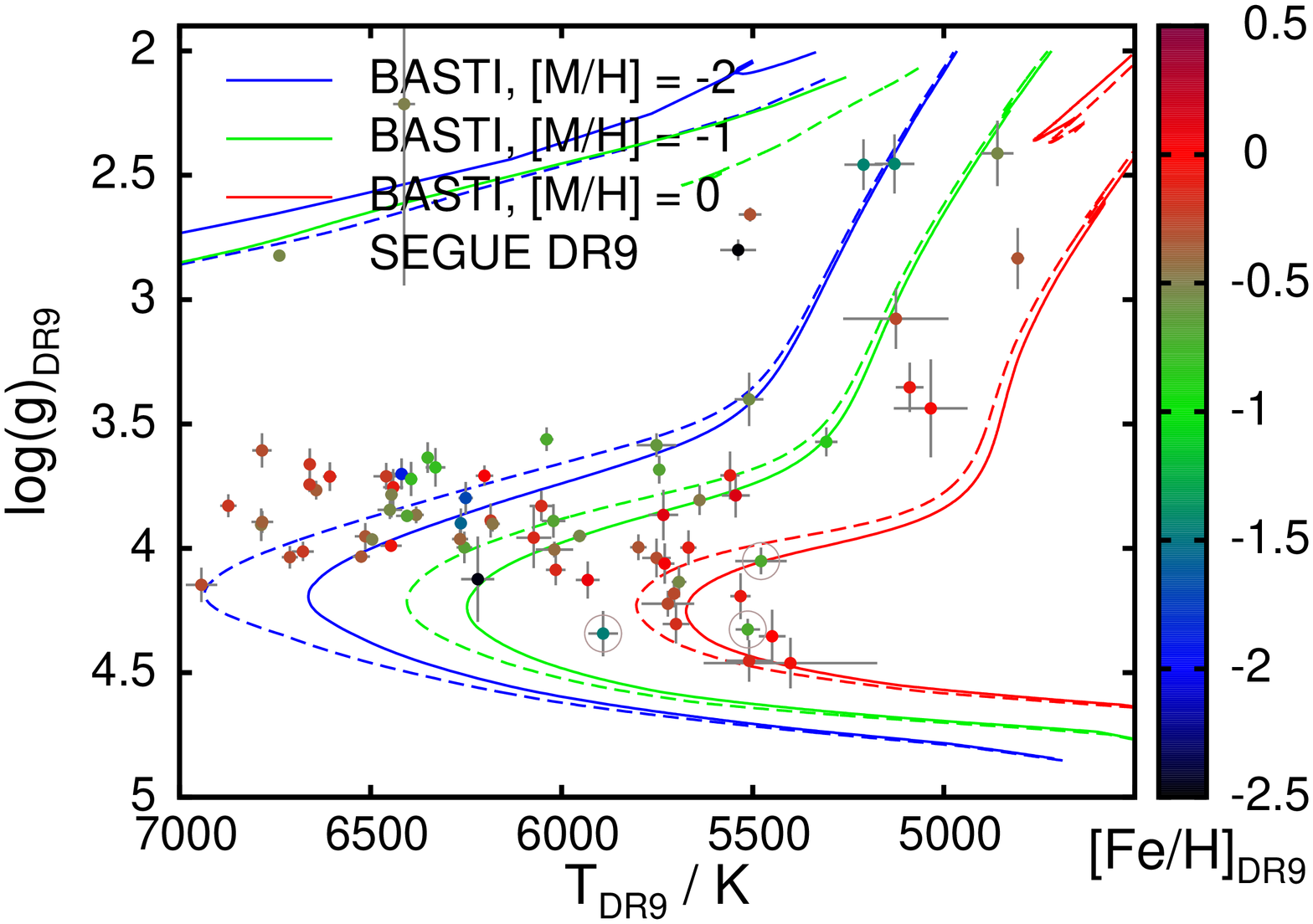,angle=-90,width=0.48\hsize}
\epsfig{file=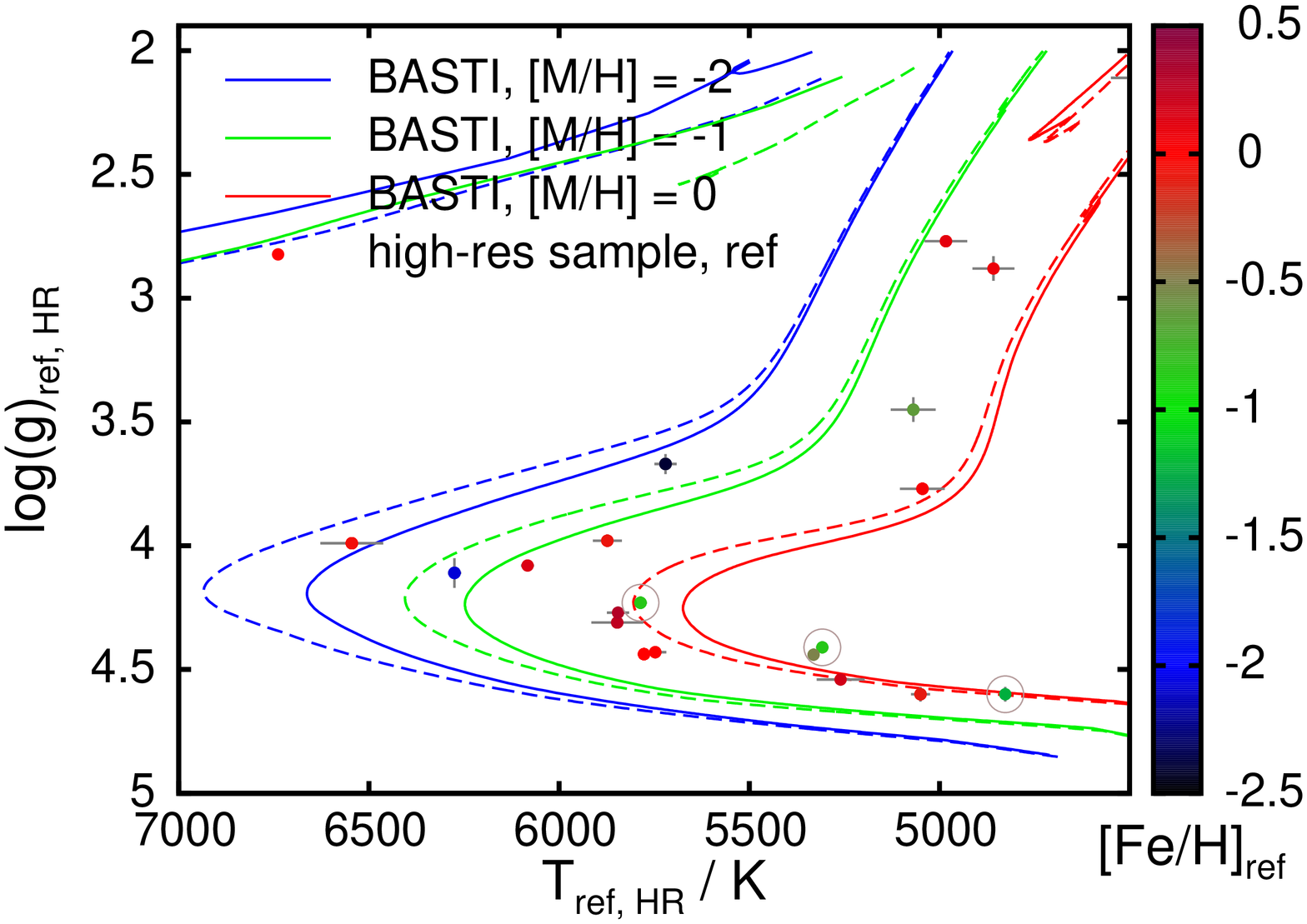,angle=-90,width=0.48\hsize}
\end{center}
\caption{Expectation values from our Bayesian pipeline (top), versus SEGUE DR9 (bottom left) and the high-resolution reference sample from Heiter et al. (in prep., bottom right). Metallicities from each derivation are coded in colours, which are also used for the $10 \Gyr$ (dashed) and $13 \Gyr$ (solid) isochrones for metallicity $[M/H]=-2,-1,0$. Note the disappearance of stars in unphysical positions (right of the turn-off, we mark with grey circles three stars in each the DR9 and high resolution reference samples that are several $\sigma$ in the forbidden region) in the Bayesian method, which are present both in the low and the high resolution reference samples.}\label{fig:HR2}
\end{figure*}

Inspection of sample distributions in parameter space, like in the temperature-gravity planes shown in \figref{fig:HR2}, provides clues about the reliability of each parameter determination. In this figure we show the HR diagrams in the $(\Teff,\logg)$-plane with expectation values from the Bayesian method (top row), versus the reference parameters (bottom), for both the SEGUE sample (left hand side) and the high-resolution reference sample (right). To facilitate the interpretation, we plot isochrones at $10$ and $13$ Gyr at three different metallicities $(-2, -1, 0)$, matching the colour scale of the stars. 

The key differences between our results and those determined by conventional methods are obvious. Despite the mildly biased spectroscopic gravity estimates, our results show a clearly superior performance in this plot. The Bayesian results cover the main sequence, while SEGUE DR9 does not attain main sequence values. Even more striking is the appearance of unphysical stars: Both SEGUE DR9 and the reference sample from Heiter et al. have stars in highly unphysical positions, with the error estimates not even close to the offset from the nearest evolutionary sequence. E.g. both SEGUE DR9 and the high-resolution reference sample place three stars around $\feh \sim -1$ far right of the turn-off or respectively right of the main-sequence. The plot suggests that the gravity offsets between the high-resolution reference values and the Bayesian method track back to a neglected metallicity effect in the reference sample. In principle the Bayesian method could yield stars in between the sequences, since we here give expectation values. A hint of this tendency can be seen, but by construction our errors will correspond to the offset, because the actual likelihood at the unphysical points is near zero.

The resulting age-metallicity relation for the high-resolution sample is displayed in \figref{fig:AMR}. To make the plot easier to read, we merged the entries for different spectra as in Table \ref{tab:allhighres}. The picture very much resembles the results of \cite{Casagrande11}. The younger expectation values for one of the very metal-poor stars corresponds to a larger error estimate, forcing the expectation value away from the hard boundary given by the age of the universe. Further there is no striking trend in metallicity at younger ages.

The importance of a reliable assessment of all stellar parameters in one single approach is demonstrated in \figref{fig:HRdist}. Here we plot the same stars from SEGUE as in the top left panel of \figref{fig:HR2}, but now colour coded with the estimated fractional distance error. It is apparent that even some very high gravity estimates are no guarantee for a good main sequence classification, vice versa stars with lower gravity can have high distance confidence. As expected, these stars are usually cleanly identified subgiants, giants, or even better, red-clump stars. While the distance and its uncertainty are in principle enough to support estimates of mean motion and velocity dispersions in a population, we point out that an investigation of velocity distributions themselves requires accurate estimates of the exact shape of the probability distribution in distance space, which the Bayesian method can deliver.

\begin{figure}
\begin{center}
\epsfig{file=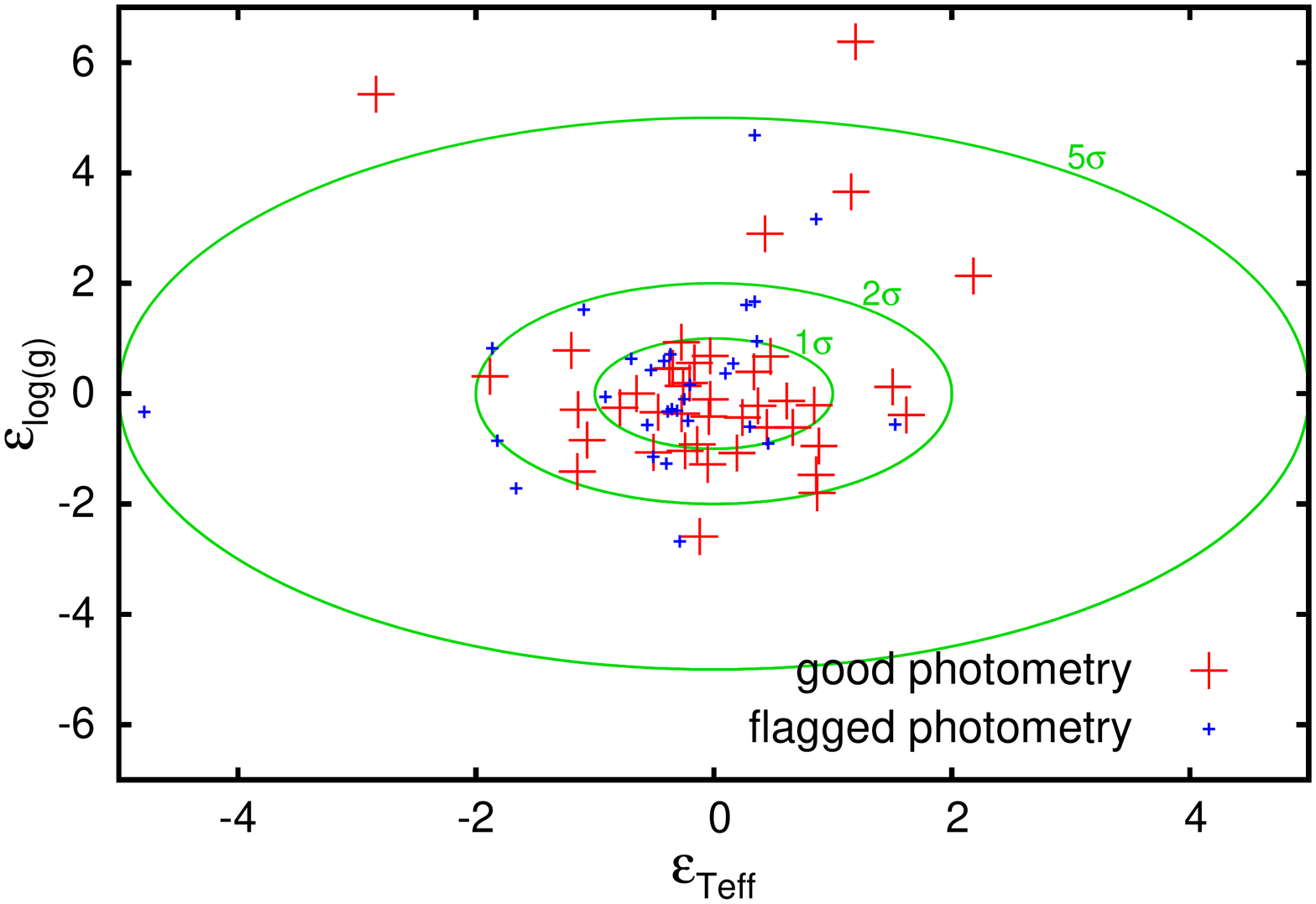,angle=-90,width=\hsize}
\epsfig{file=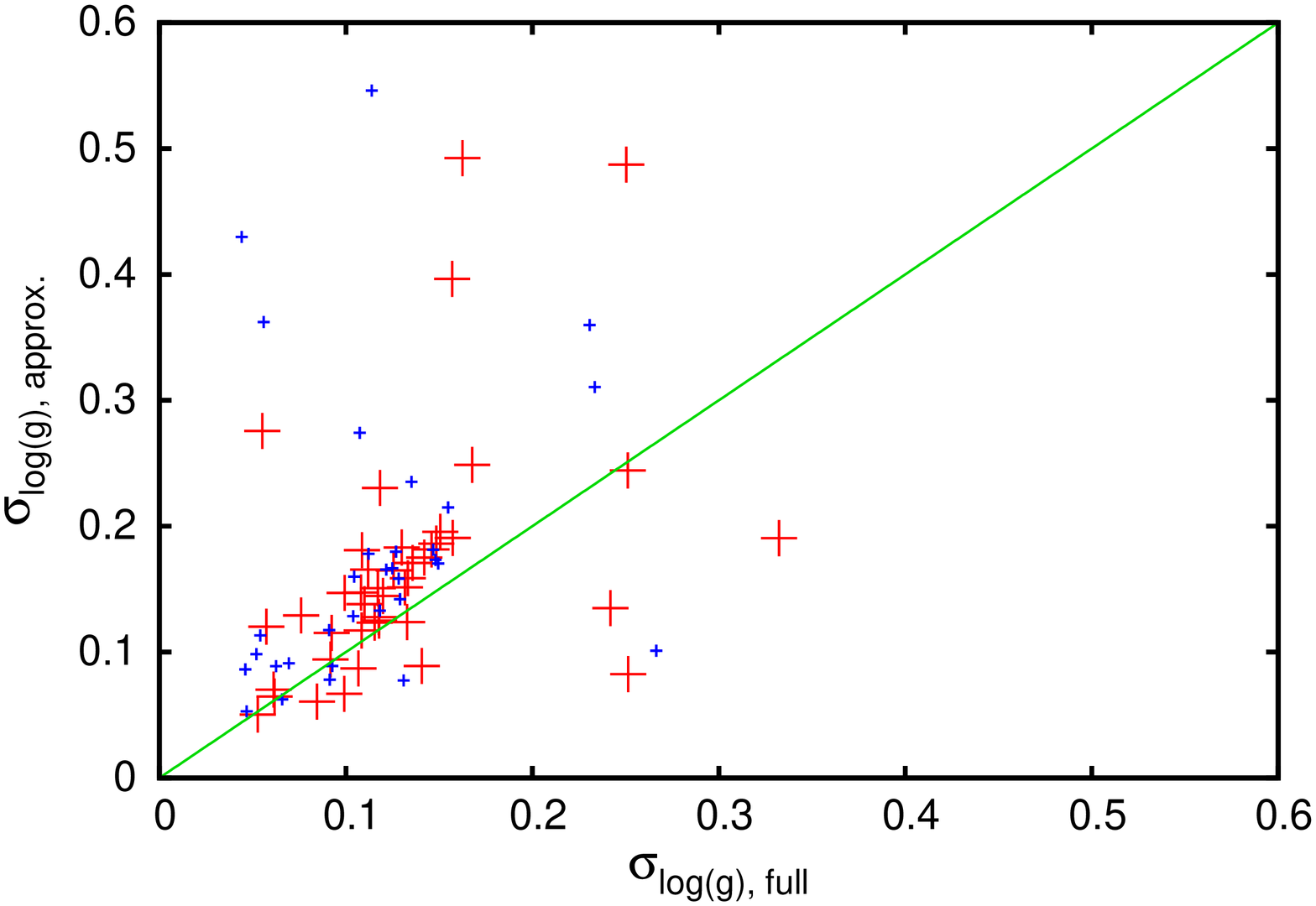,angle=-90,width=\hsize}
\epsfig{file=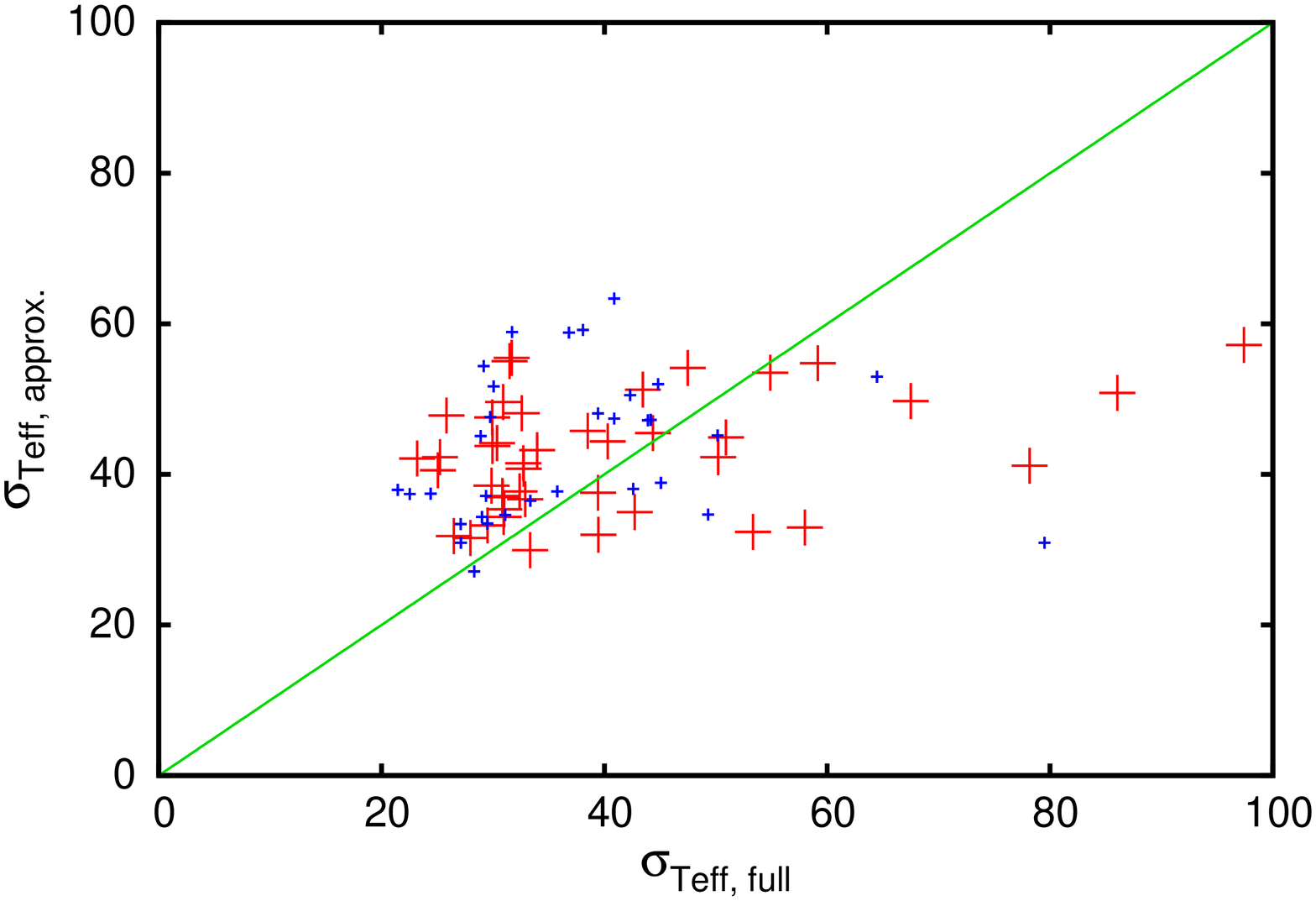,angle=-90,width=\hsize}
\end{center}
\caption{Top panel: Relative deviation of the expectation values in the degraded approach from the expectation values in our standard method. Green lines show the $1,2$ and $5\sigma$ ellipsoids. Centre and bottom panels: Surface gravity and temperature uncertainties for the same stars in the degraded approach (y-axis) versus our standard method (x-axis).}\label{fig:fudge}
\end{figure}

\section{Comparison to Simplified Approach}\label{sec:Gaussian}

In this Section we attempt to compare our implementation of spectroscopic information to an approximation by one-dimensional Gaussian uncertainties. While this is certainly not the only point by which our algorithm differs from other studies in the literature, the simplification of the spectroscopic information is common to the works we are aware of. 

We perform this experiment on the SEGUE sample. To degrade our spectroscopic results, we calculate the uncertainties $\sigma$ and mean/expectation values ($\mu_{\Teff}$, $\mu_{\llg}$, $\mu_{\feh}$) of each quantity separately and then change the spectroscopic PDF to a product of one-dimensional Gaussians in each parameter:
\begin{equation}
P_{sp}(X|O_{sp}) = N \cdot \exp\left(-\sum_i{\frac{(X_i - \mu_i)^2}{2\sigma_i^2}}\right) $,$
\end{equation}
where $i = \left\{\Teff, \llg, \feh \right\}$ and $N$ is the normalisation.

As we see from the spectroscopic PDFs in the top right panels of \figref{fig:casestudy1} and \figref{fig:casestudy2} the spectroscopic information can by no means be described as a product of Gaussian errors in $\Teff$, $\llg$, and $\feh$ separately: the PDF is not even remotely aligned with the coordinate axes and, for most stars we examined, shows a highly irregular shape. 

The top panel in \figref{fig:fudge} shows the relative shifts between our full approach and the more conventional approximation in $\Teff$ and $\llg$ for each star, normalised by the errors derived in our normal method, i.e. 
\begin{equation}
\epsilon_{\Teff} = (\left<\Teff\right> - \left<\Teff\right>_{\rm classic})/\sigma_{\Teff}. 
\end{equation}

\begin{figure*}
\begin{center}
\epsfig{file=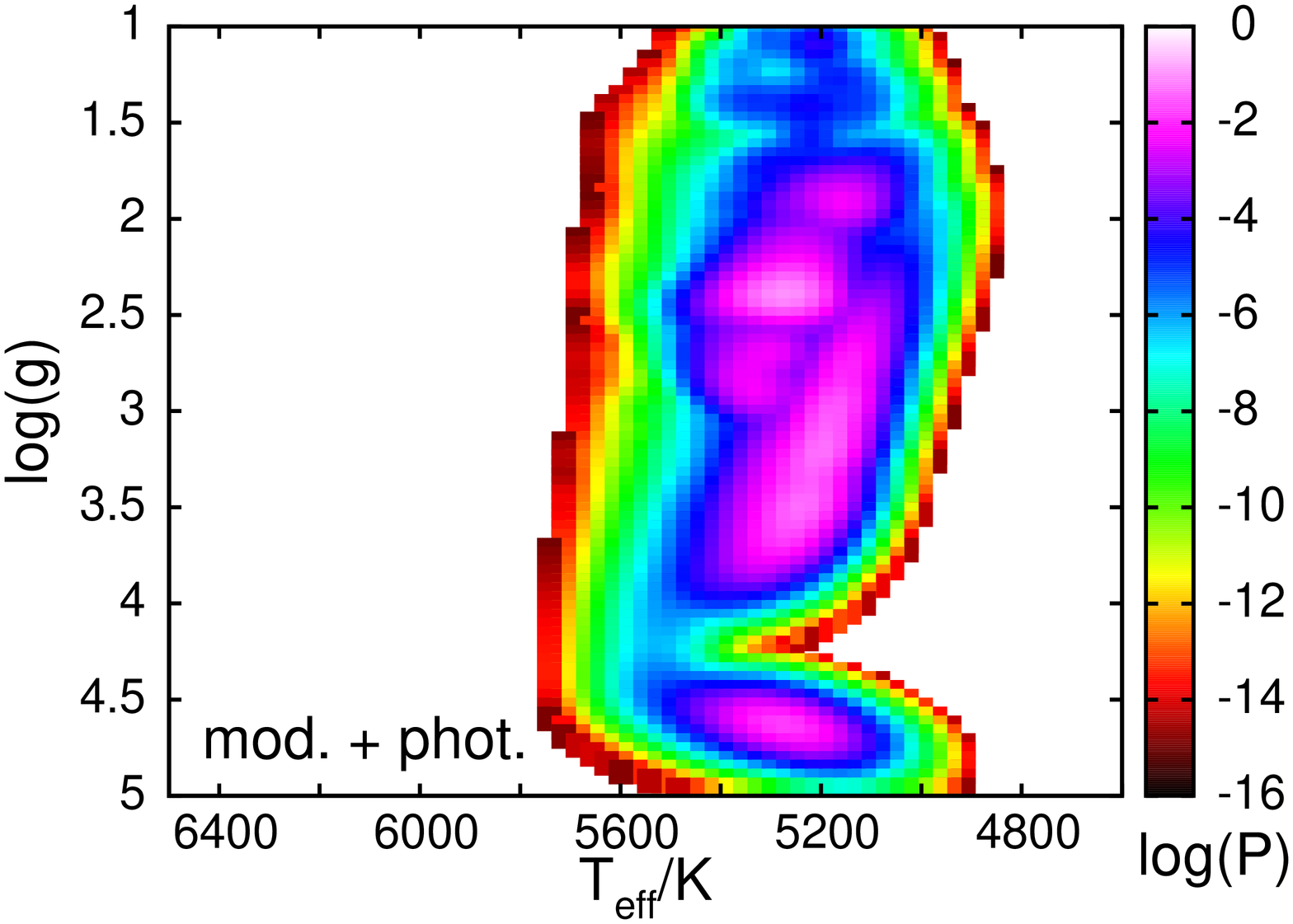,angle=-90,width=0.33\hsize}
\epsfig{file=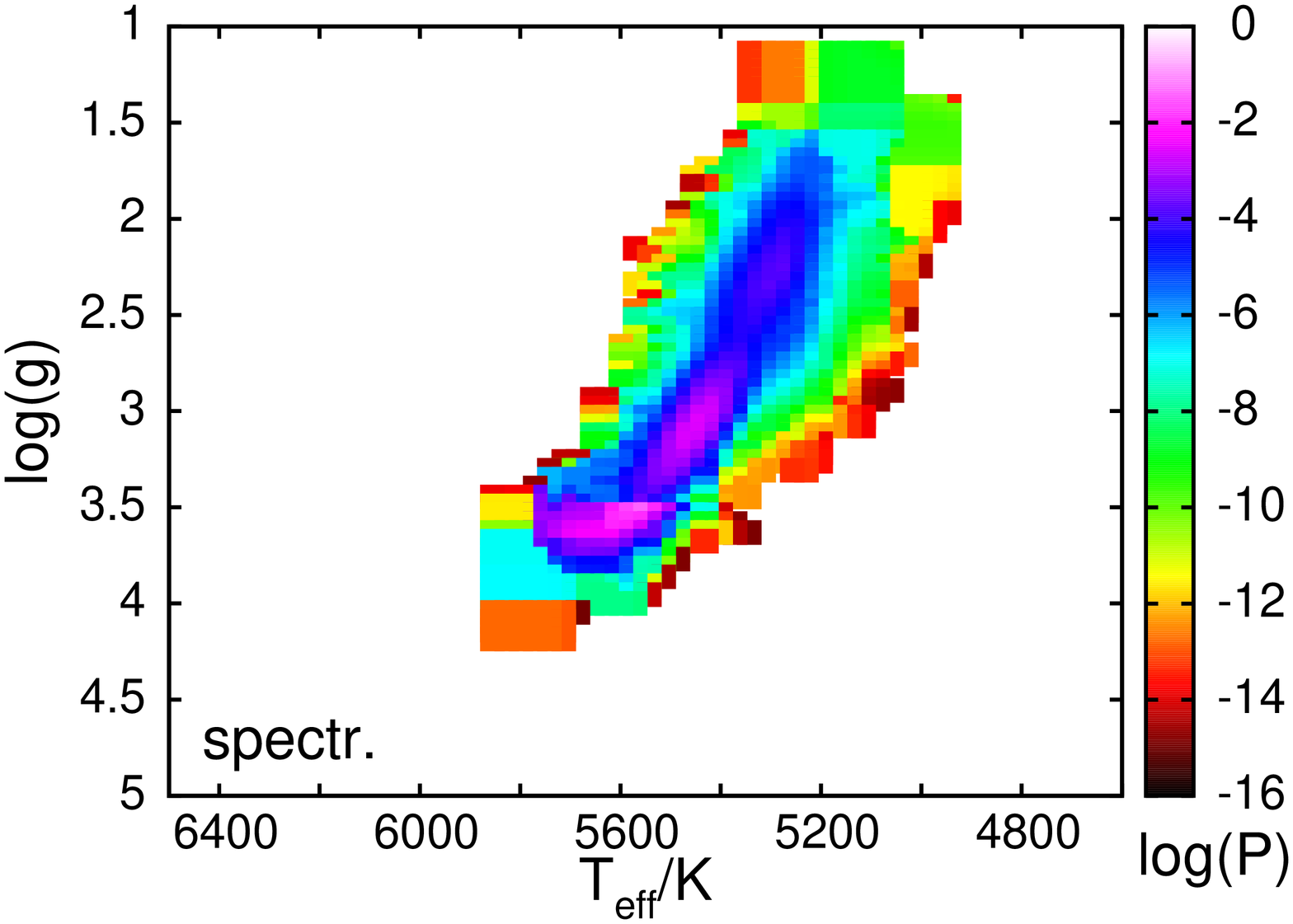,angle=-90,width=0.33\hsize}
\epsfig{file=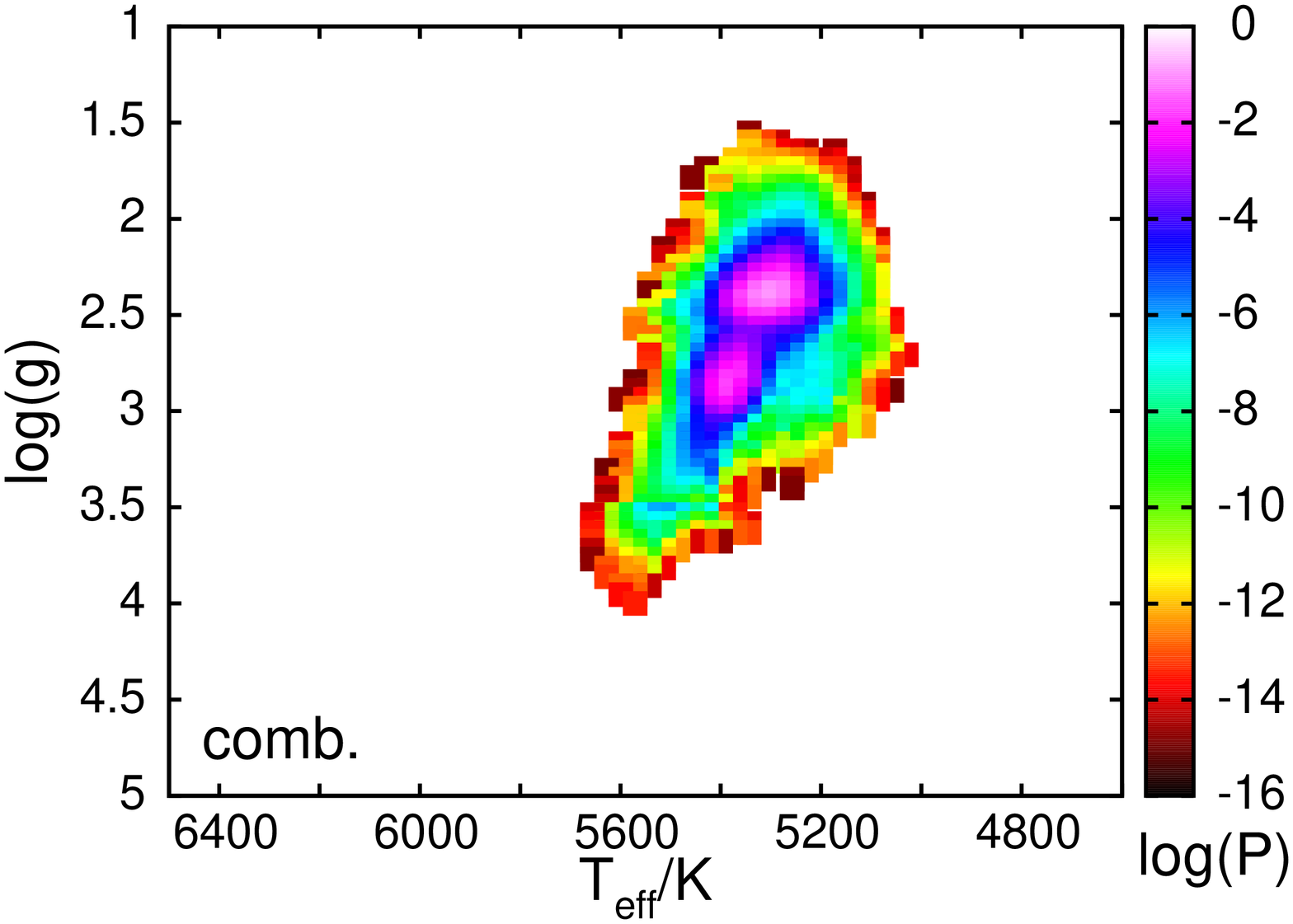,angle=-90,width=0.33\hsize}
\epsfig{file=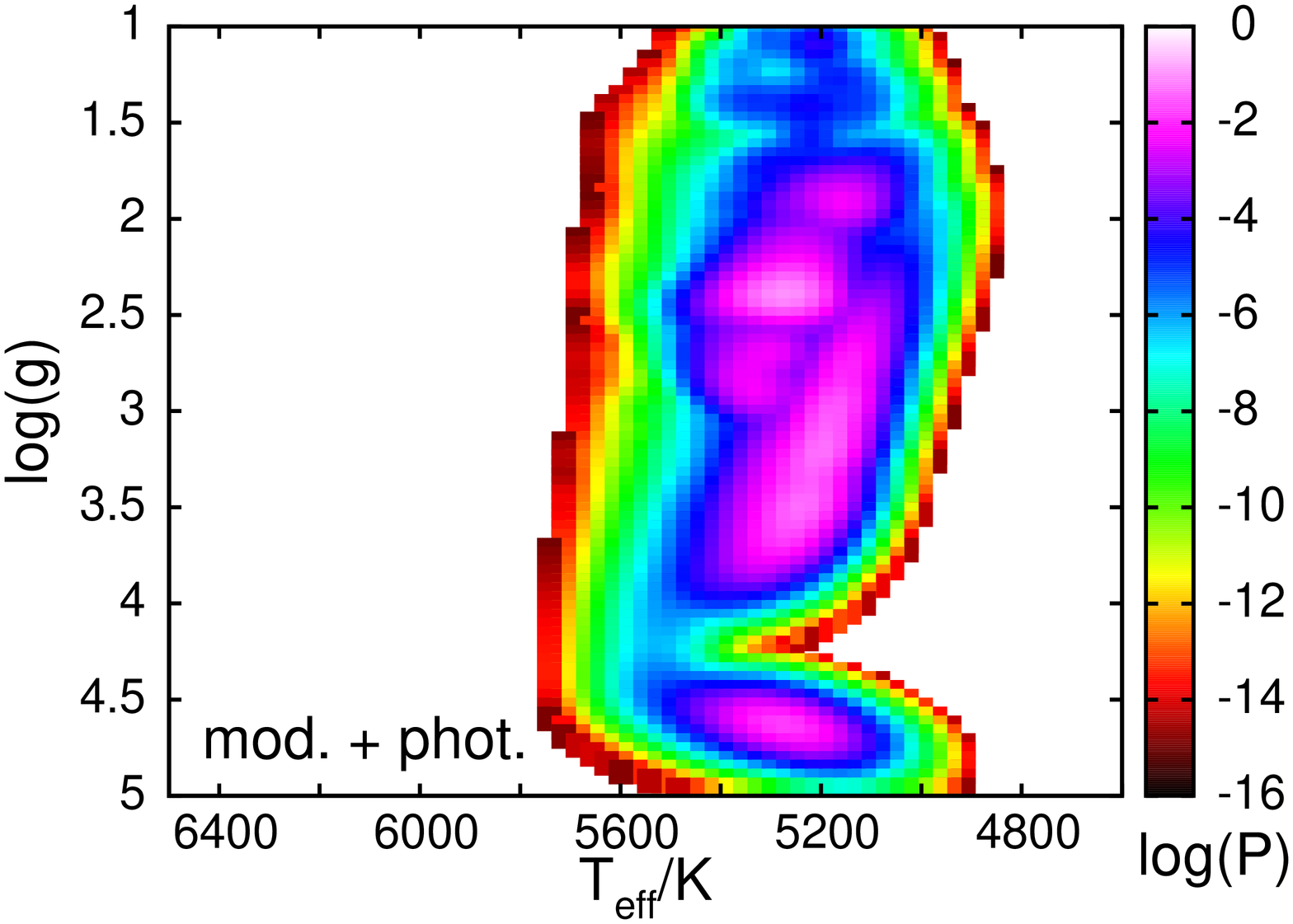,angle=-90,width=0.33\hsize}
\epsfig{file=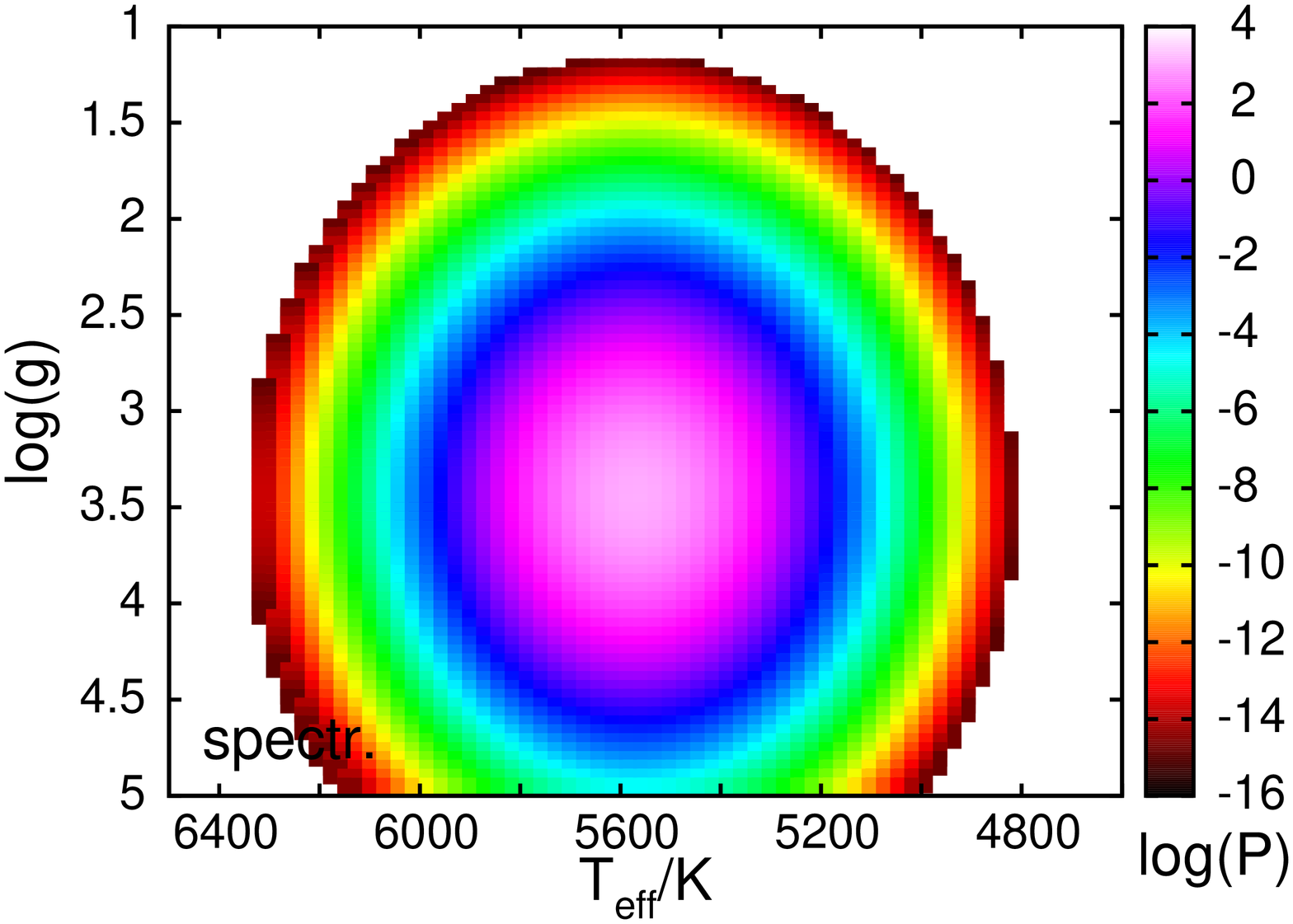,angle=-90,width=0.33\hsize}
\epsfig{file=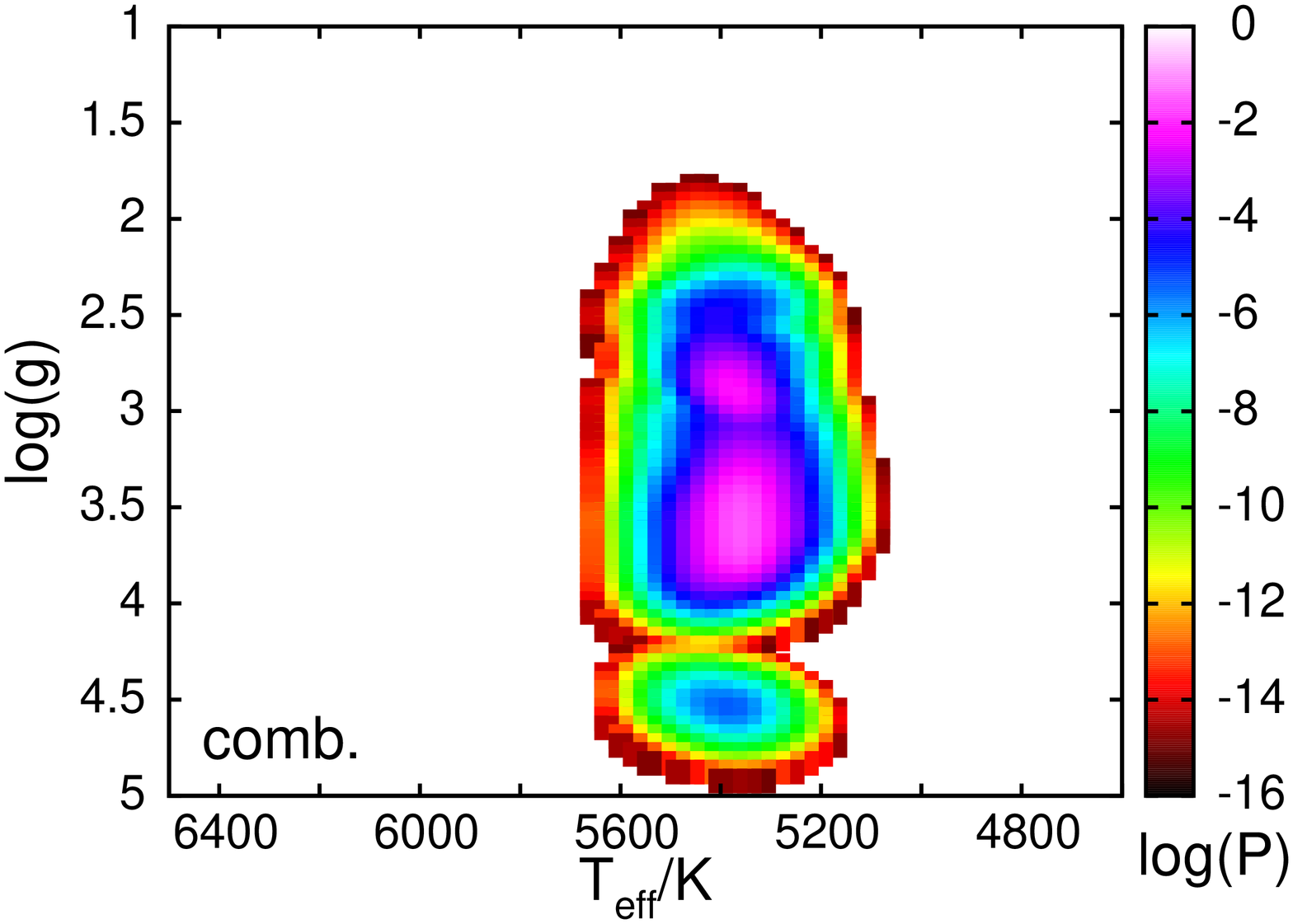,angle=-90,width=0.33\hsize}
\end{center}
\caption{Probability distributions for the SEGUE star on plate $2041$, fibre $8$ from photometry and stellar models (left column), and from spectroscopy only (centre column), once in our full approach (top) and in the Gaussian approximation (bottom) are combined into the final PDF (right). Note that the approximated result on the right hand side is a simple multiplication of the spectroscopic and model/photometric PDFs, while the left hand side is not, because it is a projection of a combination in a higher dimensional space.}
\label{fig:PDFfudge}
\end{figure*}

Intuitively one might expect very small changes, because apart from approximating the spectroscopic PDF, we left all information untouched. The contrary is true, since the shape of the spectroscopic PDF gets distorted and now intersects the other constraints in parameter space at different locations (this problem is aggravated with higher dimensionality of parameter space and a more irregular PDF). Consequently the expectation values of the parameters scatter by more than $1\sigma$. The failure of the "classic" approach can be seen in \figref{fig:PDFfudge}, where we plot the photometric, spectroscopic and photometric probability distributions for our full approach on the left versus the degraded approach on the right hand side. Looking at the invoked difference in the spectroscopic PDF, which, more importantly does not carry any metallicity dependence, helps to understand the stark differences in the resulting parameters.

The bottom panel in \figref{fig:fudge} shows the errors in surface
gravities from each approach. If the classic approximation produced robust error estimates, the values should scatter tightly around the $1:1$ line, instead there is only very weak correlation. The behaviour is a bit more benign in temperatures than in gravities and metallicities. While the
diverse systematics indicate that our results are not perfect anyway, the big
deviations both in the estimated values and their quoted uncertainties show that
the traditional approach to the spectroscopic PDF does not provide a suitable approximation. Thus, use of the full information is mandatory.

\section{Discussion and future developments}\label{sec:discus}

The method presented in this work is essential for accurate determination of astrophysical parameters of stars. Though the demonstrated scheme is essential to obtain accurate and objective error determinations{\footnote The computational cost is affordable. The algorithm is parallelized and without efforts to make it more efficient took about $20$ CPU-minutes per star} way to extract information from the current and upcoming Galactic surveys, several shortcomings need and will be addressed.
\begin{itemize}
\item We are working on extending the grids of stellar spectrum models, i.e. wider wavelength coverage (UV to IR) and finer grid resolution, inclusion of $\alpha$-enhancement and rotation as extra dimensions in the grids.
\item Especially on the low-resolution side, the continuum finding algorithms need to be improved.
\item Parameters, like micro-turbulence, which in fact parametrise the deficits of the current 1D-models in physical realism, must be better constrained or best be made obsolete by the use of more physical models. In the short and intermediate range we will find smoother corrections on a denser grid that allow for more precise evaluation. In the far future, this problem should be solved by better physics, i.e. 3D-NLTE calculations for stars, which are at present still too costly.
\item It is also interesting to include age- and mass- sensitive diagnostics (such as, Ca UV lines), that would in principle allow us to choose spectroscopic models which are more appropriate in a given domain of the HRD. At present, the analysis of OBA stars relies on NLTE model atmospheres, whereas LTE models are standard for FGK stars.
\item The stellar evolution models still apply rather simplistic (frequently grey) atmosphere models. Consequent systematics can be explored via residuals from this Bayesian method e.g. in magnitude space, as well as aberrations of physical parameters. A long-term goal would be to gear the stellar evolution codes with the same atmosphere models used for the spectroscopic modelling to avoid biases by partly contradictory models.
\item The photometric information in our scheme is affected by reddening. Colour distortions and mismatches between the photometric and other information can be directly used to determine reddening, in addition spectral information can be extracted at high resolution e.g. from interstellar Na D lines. Since we simultaneously derive probability distribution functions for stellar distances, the method can be adapted for reddening reconstructions \citep[like the ones by][]{Schlafly13}. 
\end{itemize}

\section{Conclusions}

In this paper we present the first generalised Bayesian approach for stellar parameter determination. 

The essence of the Bayesian method is a combination of several probability distribution functions in the multi-dimensional parameter space, which can be expanded arbitrarily depending on a) the available observational information for a star, and b) the desired physical quantities. The presented framework simultaneously evaluates the spectroscopic informations (gained from comparisons to theoretical spectra) and all other sources of information. This allows to calculate the full probability distributions in parameter space and helps to cut computational costs by pre-constraining the parameter space that has to be searched with the spectroscopic method.

In this work we showed how to combine low or high-resolution spectroscopy, photometry, parallax measurements and reddening estimates to estimate central physical parameters $\Teff, \logg, \meh$ of a star, as well as its mass, age, distance, or detailed chemical composition. The exploitation of theoretical constraints like stellar models, as well as strong mutual dependence or independence of different parameters reduce the complexity and effective dimensionality of the problem and make the computation possible.  The scheme can be easily expanded to other sources of information, in particular to astroseismic e.g. from CoRoT or Kepler.

The presented method has unique advantages compared to other available approaches:
\begin{itemize}
\item It makes an optimal and unbiased use of all observational data and theoretical information for a star, thus providing the parameter estimates that satisfy all observational constraints;
\item The method is robust with respect to missing data, such as low quality or missing spectral or photometric information.
\item The method is vital to gain a grip on derived quantities. E.g. to determine the distance of a star, it is not sufficient to know its best-fit values for surface gravity, temperature, metallicity and their errors; a fair assessment is only possible if we know the full combined PDF in all parameters. We showed that indeed the Bayesian estimates in particular for uncertainties differ from simple expectations.
\item Data from different surveys can be analysed with exactly the same scheme: stellar models are available in most photometric systems and the synthetic spectra grids can be folded with any instrument response function. This avoids systematic offsets caused by applying different analysis methods to different surveys and the Bayesian method can serve as a benchmark for cross-calibration between surveys.
\end{itemize}

We compared our approach to the results of a traditional Bayesian analysis on the SEGUE sample. We use the same photometric input, priors and even spectroscopic analysis, but approximate the spectroscopic PDF by a Gaussian distribution, as usually done in the literature. We find substantial shifts in all parameters, frequently by several standard deviations. This demonstrates that neglect of the full PDFs leads to wrong parameter estimates and unreliable estimates of their errors. Use of our or an equivalent method, which is able to map out the true shape of the full spectroscopic (or any other) PDF, is hence mandatory for any analysis of stellar parameters.

The method requires unbiased assessments from all its sources of information. However, we know that systematic biases (e.g. theoretical atmosphere flaws, stellar evolution uncertainties like convection, nuclear reaction rates, etc.) currently affect these sources. This vulnerability can bias the entire derived parameter set.
To test the performance of our method we compared both to reference samples for low-resolution and for high-resolution spectra. In all cases where we encounter problems, e.g. lower spectroscopic gravities, the Bayesian method remains robust and pushes all values towards the benchmark. Comparisons with each astroseismically and traditionally derived parameters shows that the Bayesian method provides excellent results on the astroseismic sample and clearly superior performance compared to the traditionally derived reference. We provide parameter estimations for these stars in Table \ref{tab:allhighres}.

Similarly the photometric information is affected by reddening. However, this impact can be directly used to determine reddening especially in a larger sample. By the simultaneous determination of distance distributions, the method offers an excellent basis for reddening measurements similar to \cite{Schlafly13}.

Up to the last decade, sample sizes of Galactic surveys determined the scope of model comparisons: at sample sizes of $\sim 1000$ stars, Poisson noise was usually of the same importance as systematic uncertainties and knowledge of the detailed error distributions. In the future we can advance from a more qualitative understanding of best-fit parameters for our Galaxies to full quantitative analysis.
The implies, however, that progress in evaluating the upcoming and present large stellar surveys for the Milky Way critically depends on our ability to cope both with the systematic biases and more importantly derive precise and accurate error distributions, and hence on the development and success of methods like the presented.

\section{Acknowledgements}

We thank the referee for a very helpful report and for suggesting the comparison to the classical Bayesian scheme. It is a pleasure to thank David Weinberg and Sergey Koposov for fruitful discussions and advice and James Binney for helpful comments to the text. We thank U. Heiter, P. Jofre, and S. Cuaresma for providing the observed high-resolution data, and T. Gehren, and F. Grupp for providing stellar atmosphere models used in this work. R.S. acknowledges financial support by NASA through Hubble Fellowship grant HF-$51291.01$ awarded by the Space Telescope Science Institute, which is operated by the Association of Universities for Research in Astronomy, Inc., for NASA, under contract NAS 5-26555. This work was partly supported by the European Union FP7 programme through ERC grant number $320360$. We thank for the great hospitality of the Aspen physics center, where parts of this paper were written.

\section{Appendix}

\subsection{Selection function}\label{sec:SF}

Previous approaches (e.g., \cite{Burnett10}) introduced a selection function. With our choice of symbols, this would read:
\begin{equation}
P(\vX|S,\vO) = P(\vO|\vX)P(\vX)\frac{P(S|\vO,\vX)}{P(S,\vO)}
\end{equation}
where $S$ denotes the selection function.
\cite{Burnett10} then split the selection function  into two parts: the one that depends on the parameters $\vX$ and the other one, that does not and is thus of no importance.
However, there appears to be no reason to introduce the other term: selections of a sample are nearly almost made on observations and not on stellar parameters that are not known a priori. The one example of such a selection function acting on parameter space we could find in the literature, is actually based on a misunderstanding by \cite{Burnett10}: Knowing the available parallax measurement and its error for a star, they try to mimic a typical kinematic quality cut in a sample by zeroing all probability that produces too low parallaxes in proportion to the measured parallax error. However, it is not clear why one should not use the full parallax information here: applying the selection function implies that one has the knowledge necessary to compute the full likelihood, the selection function instead gives an undesirable one-sided constraint against far-away stars, and when pretending not to have the parallax information for testing purposes, the selection function will arbitrarily cut away the tail of effective distance overestimates, leading to wrong confidence and biased error estimates.

\subsection{Details on priors}\label{sec:amprior}
For the metallicity-iron abundance prior we assume a fixed alpha enhancement. It is known that also alpha enhanced stellar models are very well approximated by scaled solar abundance models \citep[cf.][]{Chieffi91, Salaris98}. We use this fact by setting the relation:
\begin{equation}
\meh = \left\{ \begin{matrix}
\feh + 0.1 &$if$& \feh < -1.0 \\
\feh - 0.2 (\feh + 0.5) &$if$& -1.0 \le \feh < -0.5 \\
\feh &$if$& \feh \ge -0.5  \\
\end{matrix} \right.
\end{equation}

The combined prior probability density of age and metallicity is used as::
\begin{equation}\label{eq:ageprior}
P(\tau, \meh) = N \cdot P(\meh) \cdot \left\{ \begin{matrix}
0 &$if$ & \tau > 14 \Gyr  \\ 
1 &$if$ & 11 \Gyr \le \tau \le 14 \Gyr \\
\exp\left(\frac{ \tau - 11 \Gyr}{ \sigma_{\tau}}\right) & $if$ & \tau < 11 \Gyr 
\end{matrix}   \right.
\end{equation}
where 
\begin{equation}
\sigma_{\tau} = \left\{  \begin{matrix}
1.5 \Gyr &$if$& \meh < -0.9 \\
\left(1.5 + 7.5 \cdot \frac{0.9 + \meh}{0.4}\right)\Gyr &$if$& -0.9 \le \meh \le -0.5 \\
9 \Gyr &$else.$&\\
\end{matrix}   \right.
\end{equation}

For the sake of simplicity we give each population the same upper limit of $14 \Gyr$ and allow for a constant density in age down to $11 \Gyr$. Cosmological studies as well as observations in the Milky Way disc \citep[][]{Madau98, AumerB08, SB09} measure a significant decline of star formation rates with time even for Galactic disc stars. Observations and these theoretical models also derive a significantly older age for more metal-poor populations, which motivates the decreasing time constant towards lower metallicities. The high altitude of the SDSS/SEGUE sample additionally favours older ages \citep[cf.][]{Just07}, but in order not to conflict with Cromwell's rule on the other hand, we lean towards a relatively moderate decline with time.

SEGUE measures mostly stars in the high disc, so we describe the spatial distribution for our stars by a primitive thick disc plus halo model, i.e.:
\begin{equation}
\rho(R, z) = e^{-z/z_0}e^{-(R-\Rsun)/R_d} + 0.03 \cdot \left(\frac{r}{\Rsun}\right)^{-2.5}
\end{equation}
where $R$ is the cylindrical galactocentric radial coordinate, $r$ the galactocentric distance, $z$ the altitude above the plane, $z_d = 0.9 \kpc$ the assumed scale height of the Galactic disc, $R_d=2.5\kpc$ the scale length of the Galactic disc, $\Rsun = 8.27 \kpc$ the assumed galactocentric distance of the Sun from \cite{McMillan11, S12}.

\begin{table*}
\begin{tabular}{lccccccccccl}
{\bf name} & {\bf HIP} & {\bf spectra} & {\bf $\feh$} & {\bf $\sigma_{\feh}$} & {\bf $\Teff$} & {\bf $\sigma_{\Teff}$} & {\bf $\llg$} & {\bf $\sigma_{\llg}$} & {\bf $\tau$} & {\bf $\sigma_\tau$} & {\bf remark} \\
\hline \\
HD 107238 & 60172 & 0 & -0.14 & 0.26 & 4473 & 81 & 2.04 & 0.20 & 6.3 & 3.3 & phot.  \\
HD 122563 & 68594 & 4 & -2.650 & 0.076 & 4809 & 47 & 1.54 & 0.13 & 9.7 & 2.9 & comb., bad photometric T \\
HD 140283 & 76976 & 4 & -2.73 & 0.11 & 5608 & 40 & 3.539 & 0.056 & 13.49 & 0.47 & comb. \\
HD 173819 & 92202 & 0 & -0.42 & 0.64 & 4240 & 95 & 1.05 & 0.25 & 2.9 & 4.0 & phot \\
HD 190056  & 98842 & 0 & -0.11 & 0.29 & 4449 & 94 & 2.07 & 0.17 & 5.8 & 4.0 & phot \\
HD 220009 & 115227 & 0 & -0.43 & 0.28 & 4369 & 85 & 1.67 & 0.19 & 6.2 & 3.9 & phot  \\
HD 22879 & 17147 & 3 & -0.592 & 0.024* & 6006 & 19* & 4.316 & 0.044* & 12.1 & 1.2* & comb.$^1$ \\
HD 84937 & 48152 & 1 & -2.11 & 0.14 & 6242 & 70 & 3.931 & 0.082 & 13.57 & 0.41 & comb.$^2$ \\
ksi Hya & 56343 & 1 & -0.458 & 0.032* & 4933 & 35 & 2.476 & 0.080 & 3.9 & 1.6 & comb., metallicity fit questionable \\
Procyon & 37279 & 4 & -0.161 & 0.078 & 6515 & 79 & 3.993 & 0.073 & 2.44 & 0.53 & comb. \\
alpha Cen A & 71683 & 2 & 0.275 & 0.063 & 5939 & 79 & 4.380 & 0.066 & 3.7 & 2.5 & comb. \\
alpha Cen B & 71681 & 1 & 0.175 & 0.072 & 5364 & 58 & 4.482 & 0.041 & 6.6 & 4.4 & comb. \\
Psi Phe & 8837 & 0 & 0.14 & 0.36 & 3586 & 31* & 0.65 & 0.22 & 4.9 & 4.5 & phot.$^3$ \\
Sun & 0 & 4 & -0.013 & 0.046 & 5842 & 49 & 4.464 & 0.063 & 4.3 & 2.9 & comb. \\
18 Sco & 79672 & 2 & -0.050 & 0.059 & 5849 & 54 & 4.492 & 0.064 & 4.3 & 3.2 & comb. \\
61 Cyg A & 104214 & 0 & -0.45 & 0.43 & 4563 & 83 & 4.717 & 0.060 & 6.8 & 4.0 & phot. \\
alpha Tau & 21421 & 0 & 0.09 & 0.22 & 3889 & 57 & 1.21 & 0.14 & 5.9 & 3.8 & phot. \\
Arcturus & 69673 & 0 & -0.27 & 0.31 & 4399 & 91 & 1.82 & 0.19 & 4.2 & 1.9 & phot. \\
alpha Cet & 14135 & 0 & -0.53 & 0.26 & 3723 & 41 & 0.50 & 0.16 & 5.5 & 3.7 & phot. \\
tau Cet & 8102 & 1 & -0.520 & 0.047 & 5515 & 32 & 4.612 & 0.053 & 7.8 & 3.9 & comb. \\
beta Ara & 85258 & 0 & -0.07 & 0.36 & 4118 & 83 & 1.02 & 0.20 & 3.3 & 1.3 & phot.  \\
mu Ara & 86796 & 1 & 0.379 & 0.073 & 5950 & 97 & 4.334 & 0.077 & 3.9 & 2.3 & comb. \\
Pollux & 37826 & 1 & -0.376 & 0.043 & 4846 & 53 & 2.66 & 0.12 & 4.3 & 2.1 & comb., high macroturbulence \\
eps For & 14086 & 2 & -0.479 & 0.049 & 5218 & 68 & 3.614 & 0.072 & 7.2 & 2.0 & comb. \\
eps Vir & 63608 & 2 & -0.457 & 0.050 & 5024 & 58 & 2.62 & 0.12 & 1.11 & 0.75 & comb., high macroturbulence\\
beta Vir & 57757 & 1 & -0.037 & 0.080 & 6225 & 93 & 4.163 & 0.077 & 3.8 & 1.1 & comb. \\
eta Boo & 67927 & 0 & 0.12 & 0.37 & 6332 & 162 & 3.868 & 0.093 & 2.8 & 1.7 & phot., fast rotator \\
delta Eri & 17378 & 2 & 0.047 & 0.050 & 5139 & 59 & 3.791 & 0.071 & 6.99 & 0.88 & comb.$^{4}$ \\
eps Eri & 16537 & 3 & -0.202 & 0.050 & 5184 & 27 & 4.562 & 0.049 & 6.5 & 3.8 & comb. \\
gam Sge & 98337 & 0 & -0.13 & 0.25 & 3942 & 83 & 1.15 & 0.19 & 5.6 & 3.9 & phot.  \\
gmb 1830 & 57939 & 1 & -1.56 & 0.11 & 5304 & 36 & 4.649 & 0.060 & 7.8 & 4.1 & comb. \\
mu Cas & 5336 & 1 & -0.598 & 0.010* & 5584 & 39 & 4.601 & 0.053 & 5.6 & 3.6 & comb.$^1$ \\
mu Leo & 48455 & 0 & 0.24 & 0.22 & 4607 & 74 & 2.43 & 0.11 & 4.7 & 3.7 & phot., low $\Teff$ and rotating \\
beta Hyi & 2021 & 3 & -0.189 & 0.081 & 5848 & 79 & 3.997 & 0.074 & 6.89 & 0,58 & comb. \\
\end{tabular}
\caption{Parameter expectation values and errors for metallicity $\feh$ in $\dex$, temperature $\Teff$ in $K$, surface gravity $\llg$ in $\dex$, and age $\tau$ in $\Gyr$, all values rounded to two significant digits in the formal error. The second column provides the Hipparcos catalogue number for each star, the third column the number of spectra involved. Stars outside the spectral grid or with bad spectra have $0$ used spectra and are denoted with $phot.$ in the last column, as their parameters stem from photometry, stellar models and parallax measurements, while "comb." in the last column denotes a full Bayesian approach. Detailed remarks on single stars: $^1$internal rim solution by $\afe$ step at $-0.6 \dex$, metallicity and errors biased. $^2$UVES and HARPS spectra dropped. $^{3}$Outside model grid (rim solution). $^{4}$NARVAL bad spectral fit. disregarded, though Bayesian values in line with other estimates.}\label{tab:allhighres}
\end{table*}

\label{lastpage}

\begin{thebibliography}{}

\bibitem[Ahn et al.(2012)]{SDSSDR9}
Ahn C.P. et al., 2012, ApJS, 203, 21

\bibitem[Allende Prieto et al.(2008)]{Allende08} 
Allende Prieto, C., Sivarani, T., Beers, T.~C., et al.\ 2008, \aj, 136, 2070 

\bibitem[An et al.(2008)]{An08}
An D. et al., 2008, ApJS, 179, 326

\bibitem[An et al.(2013)]{An13}
An D. et al., 2013, ApJ, 763, 65

\bibitem[{\'A}rnad{\'o}ttir, Feltzing \& Lundstr\"om(2010)]{Arnadottir10}
{\'A}rnad{\'o}ttir A.S., Feltzing S., Lundstr\"om I., 2010, A\&A, 521, 40 

\bibitem[Aumer \& Binney(2009)]{AumerB08}
Aumer M., Binney J., 2009, MNRAS, 397, 1286

\bibitem[Bailer-Jones(2011)]{Bailer10}
Bailer-Jones C., 2011, MNRAS, 411, 435

\bibitem[Bergemann \& Gehren(2008)]{Bergemann08}
Bergemann M., Gehren T., 2008, A\&A, 492, 823

\bibitem[Bergemann et al.(2012)]{Bergemann12}
Bergemann M., Lind K., Collet R., Magic Z., Asplund M., 2012, MNRAS, 427, 27

\bibitem[Bevington \& Robertson(1992)]{Bevington92}
Bevington P.R., Robertson D.K., 1992, {\it Data reduction and error analysis for the physical sciences}, McGraw-Hill, New York

\bibitem[Binney \& Merrifield(1998)]{BM}
Binney J., Merrifield M., 1998, {\it Galactic Astronomy}, PUP, Princeton, NJ

\bibitem[Binney et al.(2013)]{B13}
Binney J. et al., 2013, MNRAS, tmp.2584B, arXiv: 1309.4270

\bibitem[Blanco-Cuaresma et al.(2014)]{BlancoC14}
Blanco-Cuaresma S., Soubiran C., Jofr{\'e} P., Heiter U., 2014, arXiv: 1403.3090

%%\bibitem[Bond et al.(2010)]{Bond10}
%%Bond N.A. et al., 2010, ApJ, 716, 1
%%
\bibitem[Burnett \& Binney(2010)]{Burnett10}
Burnett B., Binney J., 2010, MNRAS, 407, 339 

\bibitem[Casagrande et al.(2011)]{Casagrande11}
Casagrande L., Sch\"onrich R., Asplund M., Cassisi S., Ram\'irez I., Mel\'endez
J., Bensby T., Feltzing S., 2011, A\&A, 530, 138

\bibitem[Chaplin et al.(2011)]{Chaplin11}
Chaplin W.J. et al., 2011, ApJ, 713, 169

\bibitem[Chiavassa et al.(2010)]{Chiavassa}
Chiavassa A, Collet R., Casagrande L., Asplund M., 2010, A\&A, 524, 93

\bibitem[Chieffi et al.(1991)]{Chieffi91}
Chieffi A., Straniero O., Salaris M., in {\it The Formation and Evolution of Star Clusters}, ed. K. Janes, ASPCS, 13, 219

\bibitem[Drell et al.(2000)]{Drell00}
Drell P.S., Loredo T.J., Wasserman I., 2000, ApJ, 530, 593

\bibitem[Fuhrmann(2004)]{Fuhrmann04}
Fuhrmann, K., 2004, AN, 325, 3

\bibitem[Gehren et al.(2004)]{Gehren04}
Gehren T., Liang Y.C., Shi J.R., Zhang H.W. Zhao G., 2004, A\&A, 413, 1045

\bibitem[Gilmore et al.(2012)]{Gilmore12}
Gilmore G. et al., 2012, Msngr, 147, 25

\bibitem[Girardi et al.(2004)]{Girardi04}
Girardi L., Grebel E.K., Odenkirchen M., Chiosi C., 2004, A\&A, 422, 205

\bibitem[Gratton(2000)]{Gratton00}
Gratton R., 2000, ASPC, 198, 225

%\bibitem[Gruberbauer \& Guenther (2013)]{Gruberbauer13}
%Gruberbauer M., Guenther D.B., 2013, MNRAS, 432, 417

\bibitem[Grupp(2004a)]{Grupp04a}
Grupp F., 2004, A\&A, 420, 289

\bibitem[Grupp(2004b)]{Grupp04b}
Grupp F., 2004, A\&A, 426, 309

\bibitem[Hekker \& Mel\'endez(2007)]{Hekker07}
Hekker S., Mel\'endez J., 2007, A\&A, 475, 1003

\bibitem[Ivezi{\'c} et al.(2008)]{Ivz08}
Ivezi{\'c} {\v Z} et al., 2008, ApJ, 684, 287

\bibitem[Jofre et al.(2013)]{Jofre13}
Jofre P. et al., 2013, arXiv:1309.1099

\bibitem[Johnson et al.(1966)]{Johnson66}
Johnson H.L., Mitchell R.I., Iriarte B., Wisniewski W.Z., 1966, CoLPL, 4, 99

\bibitem[J{\o}rgensen \& Lindegren(2005)]{Jorgensen05}
J{\o}rgensen B., Lindegren L., 2005, A\&A, 436, 127

\bibitem[Just \& Jahrreiss(2007)]{Just07}
Just A., Jahrreiss H., 2007, arXiv:0706.3850

\bibitem[Kitaura \& En{\ss}lin(2008)]{Kitaura08}
Kitaura F.S., En{\ss}lin T.A., 2008, MNRAS, 389, 497

\bibitem[Koen et al,.(2010)]{Koen10}
Koen C., Kilkenny D., van Wyk F., Marang F., 2010, MNRAS, 403, 1949

\bibitem[Korn et al.(2003)]{Korn03}
Korn A.J., Shi J., Gehren T., 2003, A\&A, 407, 691

\bibitem[Kurucz(2005)]{Kurucz05}
Kurucz R.L., 2005, Memorie della Societ{\'a} Astronomica Italiana Supplement, 8, 14 

\bibitem[Laney, Joner \& Pietrzy{\'n}ski(2012)]{Laney12}
Laney C.D., Joner M.D., Pietrzy{\'n}ski G., 2012, MNRAS, 419, 1637

\bibitem[Lee et al.(2008a)]{Lee08a}
Lee Y.S. et al., 2008, AJ, 136, 2022

\bibitem[Lee et al.(2008b)]{Lee08b}
Lee Y.S. et al., 2008, AJ, 136, 2050 

\bibitem[Lindley(1982)]{Lindley82}
Lindley D.V., 1982, Academic Press, London, {\it The Bayesian approach to
statistics, in:Some Recent Advances in Statistics}, Eds. J. Tiago de Oliviera
and B. Epstein

\bibitem[Liu et al.(2012)]{Liu12}
Liu C., Bailer-Jones C.A.L., Sordo R., Vallenari A., Borrachero R., Luri X., Sartoretti P., 2012, MNRAS, 426, 2463

\bibitem[Madau, Pozzetti \& Dickinson(1998)]{Madau98}
Madau P., Pozzetti L., Dickinson M., 1998, ApJ, 498, 106

\bibitem[Magic et al.(2010)]{Magic10}
Magic Z., Serenelli A., Weiss A., Chaboyer B., 2010, ApJ, 718, 1378

\bibitem[Magic et al.(2013)]{Magic13}
Magic Z. Collet R. Asplund M., Trampedach R., Jayek W., Chiavassa A., Stein R., Nordlund A., 2013, A\&A, 557, 26

\bibitem[Majewski et al.(2007)]{Majewski07} 
Majewski, S.~R., Skrutskie, M.~F., Schiavon, R.~P., et al.\ 2007, Bulletin of the American 
Astronomical Society, 39, \#132.08 

\bibitem[Marconi et al.(2006)]{Marconi06}
Marconi M., Cignoni M., Di Criscienzo M., Ripepi V., Castelli F., Musella I., Ruoppo A., 2006, MNRAS, 371, 1503

\bibitem[McMillan(2011)]{McMillan11}
McMillan P., 2011, MNRAS

\bibitem[Nordstr{\"o}m et al.(2004)]{Nordstroem04}
Nordstr{\"o}m B. et al., 2004, A\&A, 418, 989

\bibitem[\"Onehag et al.(2011)]{Onehag11}
\"Onehag A., Korn A., Gustafsson B., Stempels E., VandenBerg D.A., 2011, A\&A, 528, A85

\bibitem[Perryman et al.(1997)]{Perryman12}
Perryman M., 1997, A\&A, 323, 49

\bibitem[Pietrinferni et al.(2004)]{Pietrinferni04}
Pietrinferni A., Cassisi S., Salaris M., Castelli F., 2004, ApJ, 612, 168

\bibitem[Pietrinferni et al.(2006)]{Pietrinferni06}
Pietrinferni A., Cassisi S., Salaris M., Castelli F., 2006, ApJ, 642, 797

\bibitem[Pietrinferni et al.(2009)]{Pietrinferni09}
Pietrinferni A., Cassisi S., Salaris M., Percival S., Ferguson J.W., 2009, ApJ,
697, 275

\bibitem[Plez(2012)]{Plez12}
Plez B., 2012, ascl:1205.004 

\bibitem[Pont \& Eyer(2004)]{Pont04}
Pont F., Eyer L., 2004, MNRAS, 351, 487

\bibitem[Ram\'irez et al.(2012)]{Ramirez12}
Ram\'irez I. et al., 2012, ApJ, 752, 5

\bibitem[Reetz(1999)]{Reetz99}
Reetz J., 1999, Ap\&SS, 265, 171

\bibitem[Rix \& Bovy (2013)]{Rix13}
Rix H.-W., Bovy J., 2013, A\&ARv, 21, 61 

\bibitem[Ruchti(2013)]{Ruchti13}
Ruchti G., Bergemann M., Serenelli A., Casagrande L., Lind K., 2013, MNRAS, 429, 126

\bibitem[Salaris \& Weiss(1998)]{Salaris98}
Salaris M., Weiss A., 1998, A\&A, 335, 943

\bibitem[Salpeter(1955)]{Salpeter55}
Salpeter E., 1955, ApJ, 121, 161

\bibitem[Schlafly, Green \& Finkbeiner(2013)]{Schlafly13}
Schlafly E., Green G., Finkbeiner D.P., 2013, AAS, 22114506

\bibitem[Schlegel, Finkbeiner \& Davis(1998)]{Schlegel98}
Schlegel D.J., Finkbeiner D.P., Davis M., 1998, ApJ, 500, 525 

\bibitem[Sch\"onrich(2012)]{S12}
Sch\"onrich R., 2012, MNRAS, 427, 274

\bibitem[Sch\"onrich \& Binney(2009)]{SB09}
Sch\"onrich R., Binney J., 2009, MNRAS, 396, 203

\bibitem[Serenelli et al.(2013)]{Serenelli13}
Serenelli A., Bergemann M., Ruchti G., Casagrande L., 2013, MNRAS, 429, 3645

\bibitem[Shi et al.(2014)]{Shi14}
Shi J.R., Gehren T., Zeng J.L., Mashonkina L., Zhao G., 2014, ApJ, 782, 80

\bibitem[Shkedy et al.(2007)]{Shkedy07}
Shkedy Z., Decin L., Molenberghs G., Aerts C., 2007, MNRAS, 377, 120

\bibitem[Skrutskie et al.(2006)]{2MASS}
Skrutskie et al., 2006, AJ, 131, 1163

\bibitem[Sneden(1973)]{Sneden73}
Sneden C., 1973, ApJ, 184, 839

\bibitem[Steinmetz et al.(2006)]{Steinmetz06}
Steinmetz M. et al., 2006, AJ, 132, 1645

\bibitem[Stello et al.(2006)]{Stello06}
Stello D., Kjeldsen H., Bedding T.R., Buzasi D., 2006, A\&A, 448, 709

\bibitem[van Leeuwen(2007)]{vanLeeuwen}
van Leeuwen F., 2007, A\&A, 474, 653

\bibitem[Yanny et al.(2009)]{Yanny09}
Yanny B. et al., 2009, AJ, 137, 4377

\end{thebibliography}
\end{document}